\documentclass[a4paper,10pt]{article}
\usepackage{cite,amsmath,amsfonts,amsthm,fullpage}
\usepackage{youngtab}
\newcommand{\Pf}{\mathop\mathrm{Pf}\nolimits}
\newcommand{\sgn}{\mathop\mathrm{sgn}\nolimits}
\newcommand{\Pa}{\mathop\mathrm{P}\nolimits}
\newcommand{\SCP}{\mathop\mathrm{SCP}\nolimits}
\newcommand{\DP}{\mathop\mathrm{DP}\nolimits}
\newcommand{\FP}{\mathop\mathrm{FP}\nolimits}
\newcommand{\FDP}{\mathop\mathrm{FDP}\nolimits}
\newcommand{\SCDP}{\mathop\mathrm{SCDP}\nolimits}
\newcommand{\bt}{\mathbf{t}}

\theoremstyle{plain}

\newtheorem{Lemma}{Lemma}
\newtheorem{Proposition}{Proposition}

\theoremstyle{remark}
\newtheorem{Remark}{Remark}

\def\hp{{\hat \phi}}
\def\p{{\phi}}
\def\pd{\psi^\dag}
\def\l{\langle}
\def\r{\rangle}
\def\g{\Gamma}
\def\t{\texttt{T}}

\def\FF {\mathcal{F}}

\def\Tr{\mathrm {Tr}}

\def\det{\mathrm {det}}

\def\ln{\mathrm {ln}}

\def\res{\mathop{\mathrm {res}}\limits_}

\def\Hf{\mathop\mathrm{Hf}\nolimits}

\def\bp{\begin{Proposition}\rm}
\def\ep{\end{Proposition}}
\def\bc{\begin{corollary}}
\def\ec{\end{corollary}}
\def\bl{\begin{Lemma}\em}
\def\el{\end{Lemma}}
\def\be{\begin{equation}}
\def\ee{\end{equation}}
\def\br{\begin{Remark}\rm\small}
\def\er{\end{Remark}}
\def\brs{\begin{remarks}.\\ \rm\
\begin{enumerate}}
\def\ers{\end{enumerate}\end{remarks}}
\def\bea{\begin{eqnarray}}
\def\eea{\end{eqnarray}}

\def\Tr{\mathrm {Tr}}

\def\det{\mathrm {det}}

\def\sgn{\mathrm {sgn}}
\def\ln{\mathrm {ln}}

\def\res{\mathop{\mathrm {res}}\limits}

\def\&{&{\hskip -20pt}}

\newcount\YDcount\YDcount=0
\def\YDsize{10pt}

\def\YD#1{%
\ifnum#1=0
 \ifnum\YDcount=0 \ifx\varnothing\undefined\emptyset\else\varnothing\fi
 \else\vskip1.4pt\egroup\YDcount=0\fi
\else
 \ifnum\YDcount=0 \YDcount=1\vcenter\bgroup\vskip1pt
 \else\nointerlineskip\fi
 \vbox{\hrule\hbox{\vrule height\YDsize
 \loop\hskip\YDsize\vrule\ifnum\YDcount<#1\advance\YDcount1\repeat}\hrule
 \kern-0.4pt}\expandafter\YD
\fi}

\begin{document}
\author{ A. Yu.
Orlov\thanks{Institute of Oceanology, Nahimovskii Prospekt 36,
Moscow 117997, Russia, email: orlovs55@mail.ru}\and T. Shiota\thanks{Kyoto University, Kyoto, Japan, Email address: shiota@math.kyoto-u.ac.jp
} \and K.
Takasaki\thanks{Graduate School of Human and Environmental Studies, Kyoto University, Yoshida,
Sakyo, Kyoto, 606-8501, Japan
E-mail address: takasaki@math.h.kyoto-u.ac.jp
}}
\title{Pfaffian structures and certain solutions to
  BKP hierarchies I. Sums over partitions}

\maketitle

\begin{abstract}

We introduce a useful and rather simple class of 
BKP tau functions which which we shall call ``easy tau functions''.
We consider two versions of BKP hierarchy, one we will call ``small
BKP hierarchy'' (sBKP) related to $O(\infty)$ introduced in \cite{DJKM-B1} and ``large BKP
hierarchy'' (lBKP) related to $O(2\infty +1)$ introduced in \cite{KvdLbispec}  
(which is closely related to the large $O(2\infty)$ DKP hierarchy (lDKP) introduced in \cite{JM}).
Actually ``easy tau functions'' of the sBKP hierarchy were already considered in \cite{HLO}, here
we are more interested in the lBKP case and also the mixed small-large BKP tau functions \cite{KvdLbispec}.
Tau functions under consideration are equal to certain sums over partitions and to certain multi-integrals 
over cone domains. In this way they may be applicable  in models of random partitions and models of random matrices.
Here is the first part of the paper where sums of Schur and projective Schur functions over partitions are considered.

\end{abstract}

\bigskip


\section{Introduction}

In the seminal papers of Kyoto school KP hierarchies related to
different symmetry groups were introduced. In such a way DKP and
BKP hierarchies appeared as KP hierarchies related to $D_\infty$ and $B_\infty$ type
root systems, while the original KP hierarchy of Dryuma-Zakharov-Shabat was assigned to the root system of type $A_\infty$. 
However different realizations of these hierarchies are possible. The BKP and DKP hierarchies presented in \cite{DJKM-B1} are
subhierarchies of the standard KP one: it is related to $O(\infty)$
subgroup of $GL(\infty)$ symmetry group of KP. Authors of \cite{KvdLbispec} refer these BKP and DKP hierarchies as
respectively neutral BKP and DKP hierarchies. We shall call them respectively small BKP (sBKP) and small DKP (sDKP) hierarchies. 
There is also different DKP hierarchy presented in the paper \cite{DJKM-B2}
\footnote{We mean the hierarchy referred in \cite{DJKM-B2},\cite{JM} as the $D_\infty'$ one.}
 and related to $O(2\infty) \supset GL(\infty)$  which contains
KP as the subhierarchy. We shall this hierarchy as large DKP one (lDKP). At last in \cite{KvdLbispec} the BKP hierarchy related to 
$O(2\infty+1) \supset GL(\infty)$ which also contains KP as a subhierarchy was introduced. We shall refer it as 
large BKP hierachy\footnote{In \cite{KvdLbispec} the large DKP and the large BKP hierarchies were called fermionic (also charged)
 DKP and BKP ones, and the small DKP and BKP hierarchies were called neutral DKP and BKP ones; 
we found these names a little bit misleading}  These "large" hierarchies are
rather interesting and much less studied than the "small" ones. 
In \cite{DJKM-B2} and \cite{KvdLbispec} the fermionic representation for sDKP, sBKP, lDKP and lBKP tau
function was written down and bilinear equation ('Hirota
equations') were presented. The tau function of these hierarchies
appeared in a number of various problems. In the paper \cite{L1} it
was shown that under certain restrictions lBKP (and lDKP) tau
functions coincide with the so-called Pfaff lattice tau function
\cite{AMS} which in particular describes the orthogonal ensemble of random
matrices. In \cite{L1} nice fermionic representations for
orthogonal and for symplectic ensembles were found and in this way it was
shown that the partition function of these ensembles are examples of lBKP tau function.
In the recent paper \cite{T-09} the coupled
"large" 2-DKP hierarchy was introduced and the quasi-classical
limit of the lDKP hierarchy and of the 2-lDKP hierarchy was
studied.

 General solutions of lDKP and lBKP hierarchy may be found as solutions to
Hirota-type equations  \cite{JM},\cite{KvdLbispec}\footnote{ There is an interesting remark
by J.Harnad that the Hirota equation for lDKP and lBKP may be
treated as analogues of the Plucker relations for isotropic
Grassmannians called Cartan relations \cite{Harnad-private}, see \cite{HS} for more details 
on the topic of Cartan relations.}; Hirota equations for 2-lBKP hierarchy were written down in \cite{T-09}.

In the present paper we shall study certain 'simple' classes of
solutions of the lBKP and 2-lBKP hierarchies (``easy tau functions'') singled out by
equations \eqref{simplification-g-},\eqref{simplification-g+},\eqref{simplification-g-+} as
it is explained in Section \ref{large-BKP-sDKP}. Actually these tau functions on the one hand generalize two examples presented 
by J. van de Leur in \cite{L1} on the other hand generalize tau functions considered in \cite{OSch1} and called tau functions
of hypergeometric type. We believe that such tau functions
will have various applications. In addition we find its natural to
consider certain solutions of the lBKP hierarchy coupled to sBKP
one. Let us mark that special solutions of  sBKP were studied in
\cite{You},\cite{Nimmo},\cite{Q}. sBKP was used in studies of
various random processes, see
\cite{Foda},\cite{ShoMatsumoto},\cite{Serbian},\cite{LO}.

A lBKP tau function depends on the same set of higher times
$\bt=(t_1,t_2,\dots)$ as a KP tau function and on two discrete parameters (discrete times)
$l$ and $l'$ (instead of one parameter in the KP case), and may be written in form of Schur function
expansion
 \be\label{generalDKPtau}
\tau_{ll'}(\bt)=\sum_{\lambda\in \Pa}\,
s_\lambda(\bt)\,\Pi_\lambda(l,l')
 \ee
 where $\Pi_\lambda(l,l')$ are certain Pfaffians. In
 \eqref{generalDKPtau} sum runs over the set of all partitions denoted by
 $\Pa$. In case of 2-lBKP hierarchy a typical series has the
 following form
 \be\label{general-coupled-DKPtau}
\tau_{ll'}(\bt,{\bar \bt})=\sum_{\lambda,\mu\in \Pa}\,\,
s_\lambda(\bt)\,\Pi_{\lambda\mu}(l,l')s_\mu({\bar \bt})
 \ee
 which is an analog of the Takasaki series for 2D TL hierarchy,
 which is
  \be\label{Takasaki-Schur}
\tau_{l}(\bt,{\bar \bt})=\sum_{\lambda,\mu\in \Pa}\,\,
s_\lambda(\bt)\,\pi_{\lambda\mu}(l)s_\mu({\bar \bt})
  \ee
  where $\pi_{\lambda\mu}(l)$ are certain determinants.

 In the present paper we shall derive \eqref{generalDKPtau} from
 the fermionic representation of the lBKP tau function given in
 \cite{KvdLbispec}  and consider a set of examples and applications.
In particular we will introduce a certain class of lBKP tau
functions which may be considered as a generalization of the
hypergeometric function (compare to \cite{OSch2} and \cite{Q})
which depends on the lBKP higher times $\bt=(t_1,t_2,\dots)$ and a
set of parameters denoted by $U=\{U_m,m\in\mathbb{Z}\}$,
\[
\tau(\bt)=\sum_{\lambda\in\Pa} \, e^{-U_\lambda}
s_\lambda(\bt)
 \]
 and more general tau functions, see
Section \ref{large-BKP-sDKP}. Here $s_\lambda$ are the Schur functions, $\bt$ are
lBKP higher times. The sum ranges over the set of all partitions
denoted by $\Pa$.

 An example of such tau functions is 
 \[
\tau(\bt)=\sum_{\lambda\subset (m^N)} \, s_\lambda(\bt)
 \]
 where the sum ranges over all Young diagrams $\lambda$ which can be arranged into a $m$
 by $N$ rectangular   where $m,N$ are given numbers.
 At first sight one can think that it is just a particular case of the well-known
 series for the KP tau functions \cite{SS},\cite{JM}
  \[
\tau(\bt)=\sum_\lambda \,\, s_\lambda(\bt)\,\pi_\lambda
  \]
  This guess is not right: for simplicity take $m=N=\infty$; then in KP case the numbers $\pi_\lambda$ should
  solve Plucker relations while $\pi_\lambda$ all equal to $1$  do not solve.

Another interesting example of the lBKP tau function is as follows.
Consider a subset of all partitions $\Pa$ denoted
$\FP$ (``fat partitions'')  which consists of all partitions of even length of 
form
$(\lambda_1,\lambda_1,\lambda_2,\lambda_2,\lambda_3,\lambda_3,\dots)=\lambda\cup\lambda$.
  Then
 \[
\tau(\bt)=\sum_{\lambda\in\FP} \, e^{-U_{\lambda}}
s_{\lambda}(\bt)
 \]
which is also an example of the lBKP tau function and which we will 
 relate to a discrete version of $\beta=4$ ensembles. This tau function will be used in 
Section \ref{partition for random-section} in a problem related to random motion.

We also present different examples of tau functions which are written as
multiple integrals over a cone domain. Such integrals appear in the theory of random
matrices \cite{Mehta},\cite{Forr1}. Let us point out the pioneer paper \cite{AvM-Pfaff},\cite{AMS} 
which relates the orthogonal ensemble to Pfaff lattice and also the very helpful paper by J/ van de Leur
who have shown that both ensembles of real $\beta=1$ and $\beta=4$ are examples of lBKP theory. 
 In \cite{OST-II} we modify some results
of \cite{L1} to the case of sBKP and considered also the cases of three different $\beta=1$, $\beta=2$ , 
and $\beta=4$ ensembles

We shall explain what is the meaning of "independent variables" $U=\{
U_i\}$ and what are equations with respect to these variables. It
is suitable to parameterize $U$ by new variables $\bt^*$, see
\eqref{U_n(bt^*,{bar bt^*})} and \eqref{U_n(bt^*,{bar bt^*},q)}, then for certain
specifications of $\bt$ (see \eqref{t-infty}-
\eqref{t(a;q)}) we find that
\[
\tau(\bt,U(\bt^*))=\sum_{\lambda\in\Pa} \, e^{-U_\lambda(\bt^*)}
s_\lambda(\bt)
 \]
are again lBKP tau functions now with respect to parameters $\bt^*$ which
play the role of higher times. Let us mark that this tau function turned out to be
a partition function for a model of random turn motion of vicious walkers introduced by
M.Fisher \cite{F}. This problem is considered in
Section \ref{partition for random-section}. 

These times may be also considered
as group times for convolution flows \cite{HO-convol} and related
to the action of vertex operators. Hamiltonians of
these flows act in a diagonal way on the basis of Schur functions
 \footnote{First similar Hamiltonians were considered in the study of generalized
 Kontsevich model in \cite{Mironov}.}.

These convolution flows on arbitrary lBKP tau function may be also
interpreted in terms of 'dual' multisoliton lBKP tau functions
whose higher times $\bt^*$ are related to parameters $U=U(\bt^*)$
mentioned above (see Section \ref{interlinks-section}). This link between two lBKP tau
functions is quite similar to the case studied in KP where such
link between two tau functions was used for technical purposes in
papers \cite{NOk} and \cite{ABW} and was 
clarified in \cite{hypsol} and in \cite{hypsolHO}.

We found it is pertinent to present certain small BKP tau
functions such as
 \[
\tau(\bt',U)=\sum_{\lambda\in \DP}\, e^{-U_\lambda}
Q_{\lambda}({\tfrac 12\bt}')
  \]
  and also
lBKP tau functions coupled to sBKP tau functions (Section ref), the
simplest example
 \[
\tau(\bt,\bt')=\sum_{\lambda\in \Pa\atop \ell(\lambda)\le N}\,
s_\lambda(\bt)\,Q_{\lambda^-}({\tfrac 12\bt}')
 \]
 where $Q_{\lambda^-}$ are projective Schur functions, $\lambda^-$ denotes a strict
 partition whose parts are shifted parts of a partition $\lambda$. This
 expression is a lBKP tau function with respect to the set $\bt=(t_1,t_2,\dots)$ of higher
 times. At the same time it is sBKP tau function with respect to the times $\bt'$.

\section{Sums of Schur functions\label{Sums of Schur functions}}

\paragraph{Subsets of partitions.}
In the following, we consider sums over partitions and strict
partitions , which will
 be denoted by Greek letters $\alpha$, $\beta$. Recall  \cite{Mac} that a
strict partition $\alpha$ is a set of integers (parts)
$(\alpha_1,\dots,\alpha_k)$ with $\alpha_1>\dots
>\alpha_k\ge 0$. The length of a partition $\alpha$, denoted
$\ell(\alpha)$, is the number of non-vanishing parts, thus it is
either $k$ or $k-1$.

Let $\Pa$ be the set of all partitions. 
We shall need two special subsets of $\Pa$.

The first one consists of all partitions $\lambda=(\lambda_1,\dots,\lambda_{2n})$,
$0\le n\in\mathbb Z$, $\lambda_{2n}\ge0$, which satisfy
$$
\lambda_i+\lambda_{2n+1-i}\ \ \hbox{is independent of\,\ $i$}\,,\quad i=1,\dots,2n\,,
$$
or equivalently
 \be\label{SCP}
h_i+h_{2n+1-i}=2c\ \ \hbox{is independent of\,\ $i$\,\ (hence }=h_1+h_{2n}\ge2n-1),\quad i=1,\dots,2n\,,
 \ee
where $h_i=\lambda_i-i+ 2n\,$, and $2c$ is a natural number conditioned by $2c\ge 2n$.
This subset consists of all partitions $\lambda$ of length $l(\lambda)\le2n$
whose Young diagram satisfies the property that its complement in the
rectangular Young diagram $\overline Y$ corresponding to
$(\lambda_1+\lambda_{2n})^{2n}$ coincides with itself rotated 180 degrees
around the center of $\overline Y$.
This set of partitions will be denoted by $\SCP(c)$ or simply SCP, for
``self-complementary partitions''.
 If we introduce 
\be\label{y-h}
 y_i:=h_i - {c}\,,\quad {c} = \frac{h_1+h_{2n-1}}2 \,,
 \ee
then relation \eqref{SCP} may be rewritten as
\be\label{y+y=0}
y_i + y_{2n+1-i}=0\,.
 \ee

The second subset we need consists of the partitions $\lambda$ which satisfy,
equivalently,
 \be\label{FP}
\lambda_{2i}=\lambda_{2i-1}\,, \quad i=1,2,\dots,
 \ee
or $\lambda=\mu \cup \mu:=(\mu_1,\mu_1,\mu_2,\mu_2,\dots,\mu_{k},\mu_{k})$
($\exists\,\mu=(\mu_1,\mu_2,\dots,\mu_k)\in\Pa$), or that the conjugate
partitions of $\lambda$ are even, i.e., the ones whose parts are even numbers.
This set of partitions will be denoted by FP, for ``fat partitions''.

Following \cite{Mac} we will denote by DP
the set of all strict partitions (partitions with distinct parts),
namely, partitions
$(\alpha_1,\alpha_2,\dots,\alpha_k)$, $1\le k\in\mathbb Z$  with the strict
inequalities
$\alpha_1>\alpha_2>\cdots>\alpha_k>0$.

Strict partitions $\alpha$ with the property
 \be
\alpha_{2i}=\alpha_{2i-1}+1 \quad\hbox{for}\quad 2i-1\le l(\alpha)\,,
 \ee
where we set $\alpha_{2i}=0$ if $l(\alpha)=2i-1$,
will be called fat strict partitions. The set of all fat strict partitions will be denoted by 
FDP.\footnote{This subset was used in \cite{HLO} where it was denoted by $\DP'$.}

The set of all self-complementary strict partitions will be denoted by SCDP.

Let $R_{NM}$ denote the set of all partitions whose Young diagram
may be placed into the rectangle $N \times M$, namely, $R_{NM}$ is
the set of all partitions $\lambda$ restricted by the conditions
$\lambda_1\le M$ and $ \ell(\lambda) \le N$.

\paragraph{Sums over partitions.} Consider the following sums (for $\bt := (t_1, t_3, \dots )$, $\bt^*:=(t_1^*, t_3^*,\dots )$,
  $\bar{\bt} := (\bar{t}_1, \bar{t}_3, \dots )$, $\bt := (t_1, t_3, \dots )$,
  $\bt^*:=(t_1^*, t_3^*, \dots )$,   $\bar{\bt} := (\bar{t}_1, \bar{t}_3, \dots )$), $N$).

 \be
 \label{S^{(1)}}
S^{(1)}(\bt,N;U,{\bar A})\,:=\,\sum_{\lambda\in\Pa \atop \ell(\lambda)\le N}\,{\bar A}_{h(\lambda)}e^{-U_{\{h\}}}
\,s_\lambda(\bt)
 \ee 
where $h(\lambda)=\lambda_i-i+N$. The factors ${\bar A}_h$  on the  right-hand side of \eqref{S^{(1)}} are
   determined in terms a pair $(A, a)=:{\bar A}$ where $A$ is an infinite skew symmetric matrix and $a$
an infinite vector.  For a strict partition $h=
(h_1,\dots,h_N )$, the numbers
${\bar A}_h$ are defined as the Pfaffian of an antisymmetric $2n
\times 2n$ matrix ${\tilde A}$  as follows:
  \be
  \label{A-c}
{\bar A}_{h}:=\,\Pf[{\tilde A}]
  \ee
where for $N=2n$ even
  \be
  \label{A-alpha-even-n}
{\tilde A}_{ij}=-{\tilde A}_{ji}:=A_{h_i,h_j},\quad 1\le
i<j \le 2n
  \ee
and for $N=2n-1$ odd
 \be \label{A-alpha-odd-n} {\tilde
A}_{ij}=-{\tilde A}_{ji}:=
\begin{cases}
A_{h_i,h_j} &\mbox{ if }\quad 1\le i<j \le 2n-1 \\
a_{h_i} &\mbox{ if }\quad 1\le i < j=2n  .
 \end{cases}
  \ee
In addition we set ${\bar A}_0 =1$.

Then 
 \be
U_{\{h\}}\,:=\,\sum_{i=1}^N\,U_{h_i}
 \ee
where $U_n$, $n=0,1,2,\dots$ is a set of given complex numbers. This set is denoted by $U$.

As we see the factor $e^{-U_{\{h\}}}$ can be included into the factor ${\bar A}_{h}$ by redefinition of the data ${\bar A}$
as follows:
 \[
  A_{nm}\to A_{nm}e^{-U_n-U_m}\,,\quad a_n\to a_ne^{-U_n}
 \]
However we prefer to keep $U$ as a set of parameters.

 {\bf Example 0}
We choose the following matrix $A$ is given by
 \be \label{ExA0}
A_{ik}=\,(A_0)_{ik}\,:=\begin{cases} \sgn(i-k) & \hbox{if}\ 1\le i,k \le L\\[3pt] 0 
& \hbox{otherwise}
                              \end{cases}\,,
 \qquad a_k\,=\begin{cases} 1 &\hbox{if}\ k\le L \\[3pt] 0 & \hbox{otherwise}\end{cases}\,.
 \ee 
\br \em The matrix $A_1$ is infinite. However if in series \eqref{S^{(1)}} we put
  $U_n =+\infty$  for  $ n>L$, it will be the same as if we deals with the finite $L$ by $L$ matrix $A$, given by \eqref{ExA0}.
\er

 {\bf Example 1}

 \be \label{ExA1}
A_{ik}=\,(A_1)_{ik}\,:=\, 1 \,,\quad i<k\,, \qquad a_k= 1
 \ee
Then
 \be
({\bar A}_1)_{\{h\}}=1
 \ee

 {\bf Example 2}

The matrix $A$  is a finite $2n$ by $2n$ matrix, and $a=0$, thus the sum \eqref{S^{(1)}} ranges only  
partitions with even number of non-vashing parts. We put
 \be\label{ExA2}
A_{ik}=(A_2)_{ik}\,:=\,-\delta_{i,2c-i}\,,\quad i<k
 \ee
Then 
\be
\label{transweight}
({\bar A}_2)_{\{h\}}=
\begin{cases}1   &\mbox{ iff }\ \lambda\in\SCP(c) \\
0 &\mbox{ otherwise }\ 
\end{cases}
\ee
where $h=(h_1,\dots,h_N)$ is related to $\lambda=(\lambda_1,\dots,\lambda_{N})$ as $ h_i=\lambda_i-i+N\,,i=1,\dots,N=2n$.

 {\bf Example 3} Given set of additional variables $\bt'=(t_1',t_3',t_5',\dots)$
 where we take
 \be\label{ExA3}
 A_{nm}=\,(A_3)_{nm}\,:=\,\frac 12\,
e^{-U_m-U_n}\,Q_{(n,m)}(\tfrac 12{\bt'}), \qquad
a_n=(a_3)_n\,:=\,e^{-U_n}Q_{(n)}(\tfrac 12{\bt'})
 \ee 
 Here, the {\em  projective Schur functions} $Q_\alpha$  are weighted polynomials in the variables $t'_m$,
 $\deg t_m' =m$, labeled by strict partitions
(See \cite{Mac} for their detailed definition.)

\br \em\label{Q(t-infty)}
Let us introduce notation $\bt'_\infty=(1,0,0,\dots)$. It is known that
 $Q_{h}(\tfrac 12{\bt'_\infty})=\Delta^{*}(h)\prod_{i=1}^N\frac{1}{h_i!}$ where 
 \be\label{Delta^{*}}
\Delta^{*}(h)\,:=\,\prod_{i<j}\frac{h_i-h_j}{h_i+h_j}
 \ee
Thus for this choice of $\bt'$ we obtain
 \be
({\bar A}_3)_{\{h\}}=\Delta^{*}(h)\prod_{i=1}^N\frac{1}{h_i!}
 \ee
One may compare it with Example 5 where $f(n)=n$.

\er

  {\bf Example 4}
 \be\label{ExA4}
 A_{nm}=\,(A_4)_{nm}\,:=\,\delta_{n+1,m}-\delta_{m+1,n}.
 \ee

Then  
\be
\label{transweight4}
({\bar A}_4)_{\{h\}}=
\begin{cases}1   &\mbox{ iff }\ \lambda=(\lambda_1,\dots,\lambda_{2n})\in\FP \\
0 &\mbox{ otherwise }\ 
\end{cases}
\ee
where $h=(h_1,\dots,h_N)$ is related to $\lambda=(\lambda_1,\dots,\lambda_{N})$ as $ h_i=\lambda_i-i+N\,,i=1,\dots,N=2n$.

\,

\,

 \br \em\label{Examples-5-6-7} For some applications we may need further examples.
In Examples 5-7  ${\bar A}$ depends on a given function on the lattice denoted by $f$. In particular one can choose $f(n)=n$. 
Below are examples of matrices $A$ whose  Pfaffians are well-known (see \cite{Ishikawa} and references there).

 {\bf Example 5} 
\be\label{ExA5}
 A_{nm}=\,(A_5)_{nm}\,:=\,\frac{f(n)-f(m)}{f(n)+f(m)}
 \ee
Then for $ h_i=\lambda_i-i+N\,,i=1,\dots,N$, we have
\be
({\bar A}_5)_{\{h\}}=\Delta^{(5)}_N\left(f(h)\right)
 \ee
where
 \be
\Delta^{(5)}_N\left(f(h)\right):=\prod_{i<j\le N}\frac{f(h_i)-f(h_j)}{f(h_i)+f(h_j)}
 \ee

    {\bf Example 6}
 \be\label{ExA6}
 A_{nm}=\,(A_6)_{nm}\,:=\,\frac{f(n)-f(m)}{1-f(n)f(m)}
 \ee
Then for $ h_i=\lambda_i-i+N\,,i=1,\dots,N$, we have
 \be
({\bar A}_6)_{\{h\}}=\Delta^{(6)}_N(f(h))
 \ee 
where
 \be\label{Delta^{(6)}}
\Delta^{(6)}_N\left(f(h)\right)\,:=\,\prod_{i<j\le N}\frac{f(h_i)-f(h_j)}{1-f(h_i)f(h_j)}
 \ee

  {\bf Example 7}
 \be\label{ExA7}
 A_{nm}=\,(A_7)_{nm}\,:=\,\frac{f(n)-f(m)}{(f(n)+f(m))^2}.
 \ee
Then for $ h_i=\lambda_i-i+N\,,i=1,\dots,N$, we have
 \be
({\bar A_7})_{\{h\}}=\Delta^{(7)}_N\left(f(h)\right)
 \ee 
where
 \be\label{Delta^{(7)}}
\Delta^{(7)}_N\left(f(h)\right)\,:=\,\Biggl(\prod_{i<j\le N}\frac{f(h_i)-f(h_j)}{(f(h_i)+f(h_j))^2}\Biggr) \,\Hf\left( \frac{1}{f(h_i)+f(h_j)}\right)
 \ee

\er

\,

Having these examples we introduce the notation
  \be
 \label{S^{(1)}i}
S^{(1)}_i(\bt,N;U)\,:=S^{(1)}(\bt,N;U,{\bar A}_i)\,=\,\sum_{\lambda\in\Pa \atop \ell(\lambda)\le N}\,({\bar A_i})_{h(\lambda)}\,e^{-U_\lambda}
\,s_\lambda(\bt)\,,\quad i=1,\dots,6
 \ee 

In particular we obtain 
 \bea
 \label{S^{(1)}0}
S^{(1)}_{0}(\bt,N;M,U)&:=&\sum_{\lambda\in R_{N,M}}\,e^{-U_\lambda}\,s_\lambda(\bt)
\\
 \label{S^{(1)}1}
S^{(1)}_{1}(\bt,N;U)&:=&\sum_{\lambda\in\Pa \atop \ell(\lambda)\le N}\,e^{-U_\lambda}s_\lambda(\bt)
\\
 \label{S^{(1)}2}
S^{(1)}_{2}(\bt,N;U,c)&:=&\sum_{\lambda\in\SCP(c) \atop \ell(\lambda)\le N}\,e^{-U_\lambda}s_\lambda(\bt)
\\
 \label{S^{(1)}3}
S^{(1)}_{3}(\bt,N,\bt';U)&:=&\sum_{\lambda\in\Pa\atop \ell(\lambda)\le N}\,e^{-U_\lambda} 
Q_{\alpha(\lambda)}(\tfrac 12\bt') \,s_\lambda(\bt)
\\
 \label{S^{(1)}4}
S^{(1)}_{4}(\bt,N=2n,U)&:=&\sum_{\lambda\in\FP\atop \ell(\lambda)\le N}\,e^{-U_\lambda}s_\lambda(\bt)
\\
 \label{S^{(1)}i=567}
S^{(1)}_{i}(\bt,N;U,f)&:=&\sum_{\lambda\in\Pa\atop \ell(\lambda)\le N}\,\Delta^{(i)}_N\left(f(h)\right)\,e^{-U_\lambda}s_\lambda(\bt)\,,\quad i=5,6,7
\eea
The coefficients $U_{\{\alpha\}}$ are defined as \be
 U_{\{\alpha\}} := \sum_{i=1}^k U_{\alpha_i},
 \ee
The notation $U_\lambda$ serves for 
 \be
U_\lambda\,:=\,U_{\{h\}}\,,\quad h_i=\lambda_i-i+\ell(\lambda)
 \ee
\begin{Proposition} \em \label{Prop1}
Sums \eqref{S^{(1)}},\eqref{S^{(1)}0}-\eqref{S^{(1)}i=567} are tau functions of the ``large'' BKP hierarchy introduced in 
\cite{KvdLbispec} with respect to the time variables $\bt$. Sums \eqref{S^{(1)}3} are tau functions of the
BKP hierarchy introduced in \cite{JM} with respect to the time variables $\bt'$.

\end{Proposition}

\paragraph{Sums over pairs of strict partitions.}

In the Frobenius notations \cite{Mac} we write $\lambda=(\alpha|\beta)=(\alpha_1,\dots,\alpha_k|\beta_1,\dots,\beta_k)$.
where $\alpha=(\alpha_1,\dots,\alpha_k)$, $\alpha_1>\cdots>\alpha_k\ge 0$ and $\beta=( \beta_1,\dots,\beta_k)$ may be 
viewed as strict partitions. It is clear that $\ell(\alpha)=\ell(\beta),\ell(\beta)\pm 1$, and we imply this restriction
in sums over pairs of strict partitions below.

Now we consider
 \be
 \label{S^{(2)}}
S^{(2)}(\bt; U,{\bar A},{\bar B})\,:=\,1+\sum_{\alpha ,\beta\in\DP}\,
e^{U_{\{-\beta-1\}}-U_{\{\alpha\}}}\,{\bar A}_{\alpha}
\,s_{(\alpha|\beta)}(\bt)\,{\bar B}_{\beta}
  \ee
where given infinite skew matrices $A$ and $B$ and given vectors $a$ and $b$, the  factors ${\bar A}_{\alpha}$ and 
${\bar B}_{\alpha}$ are defined in the same way as before. 
 \be
U_{\{\alpha\}}=\sum_{i=1}^k\,U_{\alpha_i}\,,\quad 
U_{\{-\beta-1\}}=\sum_{i=1}^k\,U_{-\beta_i-1}
 \ee

We introduce the following notation
\be
 \label{S^{(2)}ij}
S^{(2)}_{ij}(\bt;U)\,:=\,S^{(2)}(\bt; U,{\bar A}_i,{\bar A}_j)
  \ee
where $i,j=1,\dots,7$ and matrices $A_i$ are taken from the Examples 1-7 above.
In particular we obtain series
 \bea
 \label{S^{(2)}11}
S^{(2)}_{11}(\bt;U)&:=&\sum_{\lambda\in\Pa}\,e^{-U_\lambda}s_\lambda(\bt)
\\
 \label{S^{(2)}22}
S^{(2)}_{22}(\bt;U)&:=&
1+\sum_{\alpha,\beta\in\SCDP}\,
e^{U_{\{\beta \}}-U_{\{\alpha \}}} s_{(\alpha|\beta)}(\bt)
\\
\label{S^{(2)}24}
S^{(2)}_{24}(\bt;U)&:=&1+\sum_{\alpha\in\SCDP
,\beta\in\FDP}\,e^{U_{\{\beta \}}-U_{\{\alpha \}}}
\,s_{(\alpha|\beta)}(\bt)
\\
 \label{S^{(2)}31}
S^{(2)}_{31}(\bt,\bt';U)&:=&1+\sum_{\alpha
,\beta\in\DP}\,e^{U_{\{\beta \}}-U_{\{\alpha \}}}
\,Q_{{\bar\alpha}(\alpha)}(\tfrac 12\bt')
\,s_{(\alpha|\beta)}(\bt)
\\
 \label{S^{(2)}41}
S^{(2)}_{41}(\bt;U)&:=&1+\sum_{\alpha\in\FDP
,\beta\in\DP}\,e^{U_{\{\beta \}}-U_{\{\alpha \}}}
\,s_{(\alpha|\beta)}(\bt)
\\
 \label{S^{(2)}44}
S^{(2)}_{44}(\bt;U)&:=&
1+\sum_{\alpha,\beta\in\FDP}\,
e^{U_{\{\beta \}}-U_{\{\alpha \}}} s_{(\alpha|\beta)}(\bt)
\\
 \label{S^{(2)}33}
S^{(2)}_{33}(\bt,\bt',\bt'';U)&:=&1+\sum_{\alpha
,\beta\in\DP}\,e^{U_{\{\beta \}}-U_{\{\alpha \}}}
\,Q_{{\bar\alpha}(\alpha)}(\tfrac 12\bt')
\,s_{(\alpha|\beta)}(\bt)\,Q_{{\bar\beta}(\beta)}(\tfrac 12\bt'')
\\
 \label{S^{(2)}34}
S^{(2)}_{34}(\bt,\bt';U)&:=&1+\sum_{\alpha\in\DP
,\beta\in\FDP}\,e^{U_{\{\beta \}}-U_{\{\alpha \}}}
Q_{{\bar\alpha}(\alpha)}(\tfrac 12\bt')
\,s_{(\alpha|\beta)}(\bt)
\\
 \label{S^{(2)}ij567}
S^{(2)}_{ij}(\bt;U,f)&:=&1+\sum_{\alpha\in\DP
,\beta\in\DP}\,e^{U_{\{\beta \}}-U_{\{\alpha \}}}
\Delta^{(i)}(f(\alpha)) \,s_{(\alpha|\beta)}(\bt)\, \Delta^{(j)}(f(\beta))\,,\quad i,j=5,6,7
 \eea

Each
 $Q_\alpha(\tfrac 12\bt')$ is known to be a BKP 
 \cite{DJKM-B1},\cite{JM} tau function. (This was a nice observation of
 \cite{You},\cite{Nimmo}).  The fact that only
 odd subscripts appear in the BKP higher times $t_{2m-1}$ is
 related to the reduction from the KP hierarchy.

\begin{Proposition} \em \label{Prop3}
Sums \eqref{S^{(2)}},\eqref{S^{(2)}ij} are tau functions of the ``large'' BKP hierarchy introduced in 
\cite{KvdLbispec} with respect to the time variables $\bt$. Sums \eqref{S^{(2)}31} are tau functions of the
BKP hierarchy introduced in \cite{JM} with respect to the time variables $\bt'$. Sums \eqref{S^{(2)}33} are tau functions of the
two-component BKP hierarchy introduced in \cite{JM} with respect to the time variables $\bt'$ and $\bt''$.

\end{Proposition}

\br \em

Let us remind that for the small BKP hierarchy obtained from KP we have the following \cite{HLO} 
 \be
S=\sum_{\alpha\in\DP}\,{\bar A}_\alpha\,Q_\alpha(\bt')
 \ee
By specification of the data ${\bar A}$ we obtain
 \be
 \label{smallBKPoutproduct1}
\sum_{\alpha\in\DP}\,e^{-U_{\{\alpha\}}}Q_\alpha(\tfrac 12 \bt') \,,\quad \sum_{\alpha\in\DP}\,
e^{-U_{\{\alpha\}}}Q_\alpha(\tfrac 12\bt')Q_\alpha(\tfrac 12{\bt''})\,,\quad \sum_{\alpha\in\DP'}\,
e^{-U_{\{\alpha\}}}Q_\alpha(\tfrac 12 \bt')
 \ee

The sums \eqref{smallBKPoutproduct1} are particular examples (see \cite{HLO}) of  BKP tau  functions,
 as introduced in \cite{DJKM-B1}, defining solutions to what was called the small BKP
 hierarchy in \cite{KvdLbispec}.

 The coupled small BKP yields series
 \be
 \label{S5}
S_5({\bt'},{\bt''},D):=\sum_{\alpha,\beta\in\DP \atop
\ell(\alpha)=\ell(\beta)}\,Q_\alpha(\tfrac
12\bt')D_{\alpha,\beta}Q_\beta(\tfrac 12{\bt''})
 \ee

The coefficients $D_{\alpha,\beta}$ in (\ref{S5}) are defined as
determinants:
 \be\label{D-alpha-beta}
D_{\alpha,\beta}=\det\,\left( D_{\alpha_i,\beta_j}\right)
 \ee
 where $D$ is a given infinite matrix. Taking $D_{nm}=e^{U_m-U_n}s_{(n|m)}(\bt)$ we reproduce \eqref{S^{(2)}33}.

\er

\,

\subsection{Action of $\Psi DO$ algebra on sums}

Here we shall discribe certain group transformation properties of sums $S^{(1)}$ and $S^{(2)}$. 
We shall also present some Virasoro invariant sums $S^{(1)}$.
 
Consider the following operator acting on the space of functions of infinitely many variables $\bt=(t_1,t_2,\dots)$
\be\label{{hat W}{(g)}k}
{\hat W}^{(g)}_k: = \res_x \, x^k \left( g(D_x)\cdot Z(y,x)\right)|_{y=x}
 \ee
where   $Z(x,y)$ is the vertex operator \cite{JM} 
 \be\label{vertexZ}
  Z(y,x):= \left(e^{\sum_{n=1}^\infty \left(y^n-x^n\right) t_n} 
e^{\sum_{n=1}^\infty \left(\frac{1}{nx^n}-\frac{1}{ny^n}\right) \frac{\partial}{\partial t_n}}\,-\,1\right)
\sum_{n=0}^{+\infty}\frac{x^n}{y^{n+1}}
 \ee
and where  $ D_x:=x\partial_x+\frac 12$ is the Euler operator and $g$ is a given function of one variable. We assume that
$g(D_x)\cdot x^n=g(n+\frac 12)x^n$.

The operators ${\hat W}^{(g)}_k$ act as symmetry transformation generators on tau functions, and this action may be
embedded into the algebra of infinite matrices with the central extension, see \cite{JM} and references therein.
The matrix which corresponds to ${\hat W}^{(g)}_k$ is as follows
  \be\label{{W(g)}k}
W^{(g)}_{k}: = \,\Lambda^k g(D) \,,\qquad \left( \Lambda \right)_{nm}=\delta_{n,m-1}\,,\quad 
\left(D\right)_{nm}=\left(n+\frac 12\right)\delta_{n,m}
 \ee

  \br \em
 Operators ${\hat W}^{(g)}_k$ may be also viewed as the element $x^kg(D_x)$ of the  algebra of pseudodifferential 
operators ($\Psi DO$) on the circle with a central extension, see Appendix \ref{PsiDO} 

 \er
 
In the Preposition below we shall use the notation $(A)_-$ to denote the antisymmetric part of a matrix $A$.

 \bp \em
For any data ${\bar A}=(A,a)$, ${\bar B}=(B,b)$ where $A,B$ are (infinte) antisymmetric matrices and $a,b$ are 
(infinite) column vectors, consider the following one-parameter family of 
\be
 {\bar A}(U,\texttt{t}):=\left(\left(e^{\texttt{t} W^{(g)}_{k}}e^U A e^U e^{\texttt{t} W^{(g)}_{k}}\right)_-,
e^{\texttt{t} W^{(g)}_{k}}e^Ua \right)\,,
\quad
{\bar B}(U,\texttt{t}):=\left(\left(e^{\texttt{t} W^{(g)}_{k}}e^UBe^Ue^{\texttt{t} W^{(g)}_{k}}\right)_-,
e^{\texttt{t} W^{(g)}_{k}}e^Ub \right)
 \ee
where $U=diag(U_n)$ and where we assume that matrices and vectors in the right hand sides of equalities do exist as formal 
series in a parameter $\texttt{t}$. Then
 \be\label{flowsS{(1)}}
e^{\texttt{t}{\hat W}^{(g)}_{k}}\cdot S^{(1)}(\bt,N,U,{\bar A})\,=\,S^{(1)}(\bt,N,0, {\bar A}(U,\texttt{t}))\,,\quad k > 0
 \ee
 \be
e^{\texttt{t}{\hat W}^{(g)}_{k}}\cdot S^{(2)}(\bt,N,U, {\bar A}, {\bar B})\,=
\,S^{(2)}(\bt,N,0,{\bar A}(U,\texttt{t}), {\bar B}(U,\texttt{t}))\,,\quad k > 0
 \ee
where ${\hat W}^{(g)}_k$ and ${ W}^{(g)}_k$ are given respectively by \eqref{{hat W}{(g)}k} and \eqref{{W(g)}k}, and
the exponential $e^{\texttt{t}{\hat W}^{(g)}_{k}}$ is considered as formal Taylor series in the parameter $\texttt{t}$.

\ep

\subsection{ Pfaffian representations \label{Pfaffian representations-1}}

For 
 \be
\bt=\bt({\bf x}^{(M)})=:[x_1]+\cdots +[x_M]
 \ee
we have for any $N\ge M=1$
 \be
S^{(1)}(\bt(x_i);N,U,\bar{A})\,=\,\sum_{n=0}^\infty\,a_ne^{-U_n}x_i^n
 \ee
and for any $N\ge M=2$ we have
 \be
S^{(1)}(\bt(x_i,x_j);N,U,\bar{A})\,=\,\frac{1}{x_i-x_j}
\sum_{m>n\ge 0}^\infty\,A_{nm}e^{-U_n-U_m}\left(x_i^mx_j^n-x_i^n x_j^m \right)
 \ee

\bp \em\label{Pfaff-Miwa}

For $M=N$ we have
\be
S^{(1)}(\bt({\bf x}^{(M)});N,U,\bar{A})\,=\,
\frac{1}{\Delta_N(x)}\Pf [{\tilde S}]
\ee
where for $N=2n$ even
  \be
  \label{S-alpha-even-n}
{\tilde S}_{ij}=-{\tilde S}_{ji}:=(x_i-x_j)S^{(1)}(\bt(x_i,x_j),N,U,\bar{A}),\quad 1\le
i<j \le 2n
  \ee
and for $N=2n-1$ odd
 \be \label{S-alpha-odd-n} {\tilde
S}_{ij}=-{\tilde S}_{ji}:=
\begin{cases}
(x_i-x_j)S^{(1)}(\bt(x_i,x_j),N,U,\bar{A}) &\mbox{ if }\quad 1\le i<j \le 2n-1 \\
S^{(1)}(\bt(x_i),N,U,\bar{A}) &\mbox{ if }\quad 1\le i < j=2n  
 \end{cases}
  \ee
and where
 \be\label{true-sign-Delta_N(x)}
\Delta_N(x)\,:=\,\prod_{0\le i < j \le N}\,(x_i-x_j)
 \ee
 
\ep

We shall omit more spacious formulae for the case $M \neq N$.

 \br \em\label{AppendixS^{(1)}(bt(x_i,x_j))}
 Let us write down the entries of ${\tilde S}$ to express 
$S^{(1)}_i,\,i=0,\dots,6$

 \be
S^{(1)}_1(\bt(x_i,x_j),N,U)=(x_i-x_j)^{-1}\sum_{m>n\ge 0} e^{-U_n-U_m}(x_i^mx_j^n-x_j^mx_i^n)
 \ee
 \be
S^{(1)}_1(\bt(x_i),N,U)=\sum_{n=0}^\infty e^{-U_n}x_i^n
 \ee
\,
 \be
S^{(1)}_2(\bt(x_i,x_j),2n,U)=(x_i-x_j)^{-1}\sum_{n = 0}^\infty e^{-U_n-U_{c-n}}(x_i^{c-n}x_j^n-x_j^{c-n}x_i)
 \ee
\,
 \be
S^{(1)}_3(\bt(x_i,x_j),N,U)=(x_i-x_j)^{-1}\sum_{m>n\ge 0}^\infty e^{-U_n-U_{m}}Q_{(n,m)}(\bt')(x_i^{n}x_j^m-x_j^{n}x_i^m)
 \ee
 \be
S^{(1)}_3(\bt(x_i),N,U)=\sum_{n\ge 0}^\infty e^{-U_n}Q_{(n)}(\bt') x_i^{n}
 \ee
\,
 \be
S^{(1)}_4(\bt(x_i,x_j),N`,U)=(x_i-x_j)^{-1}\sum_{n\ge 0}^\infty e^{-U_n-U_{n+1}}(x_i^{n}x_j^{n+1}-x_j^{n}x_i^{n+1})
 \ee
 \be
S^{(1)}_4(\bt(x_i),N,U)=\sum_{n\ge 0}^\infty e^{-U_n} x_i^{n}
 \ee

 \er

In particular substituting \eqref{ExA1},\eqref{ExA4}  we obtain
 \be
S^{(1)}_1(\bt,N,U=0)=\frac{1}{\Delta_N(x)}\Pf\,\frac{x_j-x_i}{(1-x_i)(1-x_j)(1-x_ix_j)}
 \ee
 \be
S^{(1)}_4(\bt,N,U=0)=\frac{1}{\Delta_N(x)}\Pf\,\frac{x_j-x_i}{1-x_ix_j}
 \ee
Then it follows that
 \be\label{sum-Schur-Pa}
\sum_{\lambda\in\Pa}\,s_\lambda(\bt({\bf x}^N))\,=\,\prod_{i=1}^N(1-x_i)^{-1}\,\prod_{i<j\le N}(1-x_ix_j)^{-1}
 \ee
and
 \be\label{sum-Schur-FP}
\sum_{\lambda\in\Pa}\,s_{\lambda\cup\lambda}(\bt({\bf x}^N))\,=\,\prod_{i<j\le N}(1-x_ix_j)^{-1}
 \ee
Formulae \eqref{sum-Schur-Pa} and \eqref{sum-Schur-FP} are known, see Ex-s 4-5 in I-5 of \cite{Mac}.

It is convenient to re-write these formulae in a way independent of the choice of $N$:

 \bp \label{exponentials}\em

 \be\label{SchurSum1}
\sum_{\lambda\in\Pa}\,s_\lambda(\bt)\,=\,
e^{\frac 12\sum_{m=1}^\infty\,mt_m^2\,+\,\sum_{m=1}^\infty \, t_{2m-1}}
 \ee
and
 \be\label{SchurSum4}
\sum_{\lambda\in\Pa}\,s_{\lambda\cup\lambda}(\bt)\,=\,
e^{\frac 12\sum_{m=1}^\infty\,mt_m^2\,+\,\sum_{m=1}^\infty \, t_{2m}}
 \ee
Relations \eqref{SchurSum1} and \eqref{SchurSum4} will be used later in Section{} to solve
certain combinatorial problem.
From
 \be
s_{\lambda^{tr}}(\bt)=(-1)^{|\lambda|}s_\lambda(-\bt)
 \ee
we obtain
\be\label{SchurSum1-}
\sum_{\lambda\in\Pa}\,(-1)^{|\lambda|} s_\lambda(\bt)\,=\,
e^{\frac 12\sum_{m=1}^\infty\,mt_m^2\,-\,\sum_{m=1}^\infty \, t_{2m-1}}
 \ee
and
 \be\label{SchurSum-even}
\sum_{\lambda\in\Pa_{even}}\, s_{\lambda}(\bt)\,=\,
e^{\frac 12\sum_{m=1}^\infty\,mt_m^2\,-\,\sum_{m=1}^\infty \, t_{2m}}
 \ee

\ep

  By the simple re-scaling $t_m\to z^mt_m$ in equations \eqref{SchurSum1}-\eqref{SchurSum-even}
  and equating factors before same powers of $z$ we obtain

\bp \label{polynomials}\em

\bea
\label{areafixedsum1}
\sum_{\lambda\in\Pa \atop
|\lambda|=\t}\,s_\lambda(\bt) &=& s_{(\t)}({\tilde\bt})
\,\qquad\qquad\qquad \left[\,{\tilde t}_{2m-1}= t_{2m}\,,\,\, {\tilde t}_{2m}=\frac m2 t_m^2\,\right]
\\
\label{areafixedsum1-}
\sum_{\lambda\in\Pa \atop
|\lambda|=\t}\,(-1)^{|\lambda|} s_{\lambda}(\bt) &= &s_{(\t)}({\tilde\bt}) 
\,\qquad\qquad\qquad \left[\,{\tilde t}_{2m-1}= -t_{2m}\,,\,\, {\tilde t}_{2m}=\frac m2 t_m^2\,\right]
  \\
 \label{areafixedsum4}
\sum_{\lambda\in\Pa \atop
|\lambda|=\t}\,s_{\lambda\cup\lambda}(\bt) & = & s_{(\t)}({\tilde\bt})\,\qquad \qquad\qquad
 \left[\,{\tilde t}_m=\frac m2 t_m^2 +t_{2m}\,\right]
 \\
\label{areafixedsum-even}
\sum_{\lambda\in\Pa_{even} \atop
|\lambda|=\t}\,s_{\lambda}(\bt) & = & s_{(\t)}({\tilde\bt}) 
\,\qquad\qquad \qquad \left[\,{\tilde t}_m=\frac m2 t_m^2 - t_{2m}\,\right]
  \eea
where auxilary sets of times ${\tilde\bt}=({\tilde t}_1,{\tilde t}_2,\dots)$ are specified in the brackets to the right of
equalities. 

\ep

For instance we get \eqref{areafixedsum1} from
 \eqref{SchurSum1} using the equality
  \[
  \sum_{\lambda\in\Pa}\,z^{|\lambda|}s_\lambda(\bt)=e^{\sum_{m=1}^\infty
  \,\frac {z^{2m}}{2} mt_m^2+\sum_{m=1}^\infty\,
  z^{2m-1}t_{2m-1}}=\sum_{\t=0}^\infty\,z^{\t}\, s_{(\t)}({\tilde\bt})
  \]
 where ${\tilde\bt}=
 \left(t_1,\frac{1\cdot t_1^2}{2},t_3,\frac{2\cdot t_2^2}{2},t_5,\frac{3\cdot t_3^2}{2},
 \dots \right)$.

Formula \eqref{areafixedsum1} in case $\bt=(1,0,0,\dots)$ has an interpretation in terms of total numbers of standard tableaux 
of weight $(1^\t)$ and numbers of involutive permutations of $S_\t$, see Ex 12 I.5 of \cite{Mac}, 
\cite{BaikRains-involution}, \cite{BF} (see also 
\eqref{standard-tableaux-number}).

We get from Proposition \ref{Pfaff-Miwa}

\bp \em

 \be
S^{(1)}_1(\bt({\bf x}^{(2n)}),2n,U)\,=
\,\frac{1}{\Delta_{2n}(x)}\Pf \left[ \sum_{m>n\ge 0 }e^{-U_n-U_m}(x_j^mx_i^n-x_i^mx_j^n) \right]_{i,j=1,\dots,2n}
 \ee
Choosing $U_m=0,\,m\le L+2n-1$ and $U_m=+\infty,\, m > L+2n-1$ we obtain
 \be
\sum_{\lambda\in P\atop \lambda_1\le L}\,s_{\lambda}(\bt({\bf x}^{(2n)}))\,=\,
\frac{1}{\Delta_{2n}(x)}\Pf \left[\frac{x_j-x_i}{(1-x_i)(1-x_j)(1-x_ix_j)}\left(1-(x_ix_j)^{L+N}+
\frac{x_j^{L+N}-x_i^{L+N}}{x_j-x_i} \right)\right]_{i,j=1,\dots,2n}
 \ee

\ep

 \bp \em

 \be
S^{(1)}_4(\bt({\bf x}^{(2n)}),2n,U)\,=\,\frac{1}{\Delta_{2n}(x)}\Pf \left[(x_j-x_i)f(x_ix_j,U)\right]_{i,j=1,\dots,2n}
 \ee
where
 \[
  f(z,U)=\sum_{m=0}^\infty e^{-U_m-U_{m+1}}z^m\,=\,S^{(1)}_4(\bt(x_i,x_j),U)
 \]
Choosing $U_m=0,\,m\le L+2n-1$ and $U_m=+\infty,\, m > L+2n-1$ we obtain
 \be
\sum_{\lambda\in P\atop \lambda_1\le L}\,s_{\lambda\cup\lambda}(\bt({\bf x}^{(2n)}))\,=\,
\frac{1}{\Delta_{2n}(x)}\Pf \left[(x_j-x_i)\frac{1-(x_ix_j)^{L+2n-1}}{1-x_ix_j}\right]_{i,j=1,\dots,2n}
 \ee

\ep

Next, as a corollary of Proposition \ref{Pfaff-Miwa} we obtain

 \bp \em

 \be\label{S^{(1)}2-U=0-Pf}
\sum_{\lambda\in\SCP(c)} s_\lambda(\bt({\bf x}^{(2n)}))\,=\,
\frac{1}{\Delta_{2n}(x)}
\Pf \left(\frac{\left( x_j^{[c+\frac 12]}-x_i^{[c+\frac 12]} \right)^2}{x_j - x_i}\right)_{i,j=1,\dots,2n} 
 \ee
 \be
\sum_{\lambda\in\SCP(c)}(-1)^{\sum_{i=1}^n(\lambda_i-i+2n)} s_\lambda(\bt({\bf x}^{(2n)}))\,=\,
\frac{1}{\Delta_{2n}(x)}
\Pf \left(\left( x_j^{[c+\frac 12]}-x_i^{[c+\frac 12]}\right)
\frac{x_j^{[c+\frac 12]}-(- x_i)^{[c+\frac 12]} }{x_j + x_i}\right)_{i,j=1,\dots,2n} 
 \ee
where $[a]$ is equal to the integer part of $a$. Notice that in case $c=n$ we have only one term related to $\lambda=0$
and thus the both sides of identity \eqref{S^{(1)}2-U=0-Pf} are equal to 1 (compare to Lemma 5.7 in \cite{Ishikawa}).

 \ep

\subsection{Specializations and Examples}

\paragraph{Links with group characters}
 There is a known relation (see \cite{Kr}) between the Schur
functions and the odd orthogonal character $so_\lambda$ of
rectangular shape as follows
 \be
\sum_{\lambda_1 \le p}\,s_\lambda(x_1,\dots,x_m)=(x_1\dots
x_m)^{\frac 12 p}\,so_{\left(\left(\frac
p2\right)^m\right)}(x_1^{\pm 1},\dots,x_m,x_m^{\pm 1},1)
 \ee
 The odd orthogonal characters $so_\lambda(x^{\pm 1}_1 , x^{\pm 1}_2 , \dots , x^{\pm 1}_m , 1)$,
  where $x^{\pm 1}_1$ is a shorthand
notation for $x_1, x^{-1}_1 , \dots$, and where $\lambda$ is an
$m$-tuple $(\lambda_1, \lambda_2, \dots , \lambda_m)$ of integers,
or of half-integers, is defined by
 \be\label{char-so-odd}
so_{\lambda}(x_1^{\pm 1},\dots,x_m^{\pm 1},1)\,:=\,
\frac{\det\left(x_j^{\lambda_i-i+m+\frac
12}-x_j^{-(\lambda_i-i+m+\frac
12)}\right)}{\det\left(x_j^{-i+m+\frac 12}-x_j^{-(-i+m+\frac
12)}\right)}
 \ee
  (see, say, (3.3) in \cite{Kr}).
Thus, $S^{(1)}_1$ may be equated to a special character of the orthogonal group.

There is the similar link between $S^{(1)}_4$ and a character of the symplectic group.

\paragraph{Links between sums and matrix models I. Sums as perturbation series for matrix models.}
This topic will be considered separately in \cite{OST-II}.

\paragraph{Links between sums and matrix models II. Discrete analogs of matrix models.} We know \cite{OS} few (basically three) ways to choose parameters $\bt$ 
in order to convert series in the Schur function $s_\lambda(\bt)$ to discrete analogues of matrix
integrals where integrals over eigenvalues are replaced by sums over integers. These are
 \bea
\label{t-infty}
&(A1)&\qquad\qquad \bt=\bt_\infty\, :=\,(1,0,0,\dots)
  \\ 
\label{t(a,x)}
&(A2)&\qquad\qquad \bt=\bt(a,x)\, :=\, a[x]
  \\ 
\label{t(q)}
&(B1)&\qquad\bt=\bt(q)\,:=\,\left(t_1(q), t_2(q), t_3(q),\dots\right)\,,\quad t_m(q)\,:=\,\frac 1m\frac{1}{1-q^m}
  \\
 \label{t(a;q)}
&(B2)&\qquad\bt=\bt(a;q)\,:=\,\left(t_1(a;q), t_2(a;q), t_3(a;q),\dots\right)\,,
\quad t_m(a;q)\,:=\,\frac 1m\frac{1-q^{am}}{1-q^m}
  \\
  \label{t(x)}
&(C)&\,\qquad\qquad\bt =\bt({\bf{x}}^N)\,:=\,\sum_{i=1}^N\,[x_i]
  \eea
 The notations $[x]$ and $a[x]$ are standard in soliton theory and denotes Miwa variables
 \be\label{[x]}
[x]=\left(x,\frac{x^2}{2},\frac{x^3}{3},\dots\right)\,,\quad
a[x]=\left(ax,a\frac{x^2}{2},a\frac{x^3}{3},\dots\right)
 \ee

These specializations of $\bt$ variables will be refereed as respectively the cases (A),(B) and (C) below.
See Appendix \ref{evaluated-section} for cases \eqref{t-infty}-\eqref{t(a;q)}.

We have the following observation

\begin{Proposition} \em\label{Prop-spec-A}

Let us choose specializations of $\bt$ variables according to either \eqref{t-infty} or \eqref{t(a,x)} in sums \eqref{S^{(1)}}
and put
 \be\label{U_n(bt^*,{bar bt^*})}
U_n=U_n(\bt^*,{\bar \bt^*})=U_n^{(0)}+\sum_{m=1}^\infty\,\left(\frac{an+b}{cn+d}\right)^mt^*_m+
\, t_0^*\,\ln\,\left(\frac{an+b}{cn+d}\right)\,-\,\sum_{m=1}^\infty\, \left(\frac{an+b}{cn+d}\right)^{-m} {\bar t}^*_{m}
 \ee
where $U_n^{(0)}$ and $a,b,c,d,$ are arbitrary parameters conditioned by $ad-bc\neq 0$.
Then sums \eqref{S^{(1)}} are tau functions of the ``large'' BKP, introduced in \cite{KvdLbispec} with respect to
the time variables $\bt^*=(t^*_1,t^*_2,\dots)$.

\end{Proposition}
\begin{Proposition} \em\label{Prop-spec-B}

Let us choose specializations of $\bt$ variables according to either \eqref{t(q)} or \eqref{t(a;q)} in sums \eqref{S^{(1)}}
and put
 \be\label{U_n(bt^*,{bar bt^*},q)}
U_n=U_n(\bt^*,{\bar \bt^*},q)=U_n^{(0)}+\sum_{m=1}^\infty\left(\frac{aq^n+b}{cq^n+d}\right)^mt^*_m +
\, t_0^*\,\ln\,\left(\frac{aq^n+b}{cq^n+d}\right)\,-\,\sum_{m=1}^\infty\, \left(\frac{aq^n+b}{cq^n+d}\right)^{-m} {\bar t}^*_{m}
 \ee
where $U_n^{(0)}$ and $a,b,c,d,$ are arbitrary parameters.
Then sums \eqref{S^{(1)}} are tau functions of the ``large'' BKP, introduced in \cite{KvdLbispec} with respect to
the time variables $\bt^*=(t^*_1,t^*_2,\dots)$.

\end{Proposition}

We have
 \be
s_\lambda(\bt_\infty)={\Delta_N(h)}\prod_{i=1}^N\,\frac{1}{h_i!}\,,\quad h=(h_1,\dots,h_N)
 \ee
where $h_i:=\lambda_i-i+N$. 

In case $\lambda=(\lambda_1,\dots,\lambda_{2n})\in\SCP$ (see \eqref{SCP}) then in the variables $y_i$ 
introduced in \eqref{y-h} thanks to \eqref{y+y=0} we can write
 \be
s_\lambda(\bt_\infty)=\left(\Delta_n(y^2)\right)^2\prod_{i=1}^n\,\frac{1}{(c-y_i)!(c+y_i)!}\,,\quad y=(y_1,\dots,y_n)
 \ee
where $y$ are related to $\lambda$ as
 \be
y_i=\lambda_i-i+2n-c\,,\quad i=1,\dots,n
 \ee

If $\lambda$ is rewritten in the Frobenius notations, 
$\lambda=(\alpha_i,\dots,\alpha_k|\beta_1,\dots,\beta_k)$, then the last relation may be written 
as
\be
s_{(\alpha|\beta)}(\bt_\infty)=
\frac{\Delta_k(\alpha)\Delta_k(\beta)}{\prod_{i,j=1}^k(\alpha_i+\beta_j+1)}\,
\prod_{i=1}^k\,\frac{1}{\alpha_i!\beta_i!}
 \ee

We also have (see Remark \ref{Q(t-infty)})
  \be
Q_\alpha(\tfrac
12\bt_\infty)=\Delta^{(3)}_k(\alpha)\,\prod_{i=1}^k\,\frac{1}{\alpha_i!}\,,
\quad \alpha=(\alpha_1,\dots,\alpha_k)
  \ee
 where
  \be\label{Delta}
\Delta_N(h):=
 \prod_{0<i<j\le N}\,(h_i-h_j)
\,,\qquad
 \Delta^{(3)}_k(\alpha):=
 \prod_{0<i<j\le k}\,\frac{\alpha_i-\alpha_j}{\alpha_i+\alpha_j}
 \ee

(A) First we choose the specialization \eqref{t-infty}. Then the parametrization \eqref{U_n(bt^*,{bar bt^*})} is available.
Putting $\bt=\bt_\infty$ we obtain from \eqref{S^{(1)}}
 \be
 \label{S^{(1)}-discrete-1}
S^{(1)}(\bt_\infty,N,U,{\bar A}):=\frac{1}{N!}\,\sum_{h_1,\dots,h_N=0}^M\,{\bar A}_{h(\lambda)}\,e^{-U_{\{h\}}(\bt^*,{\bar \bt^*})}
\,\Delta_N(h)
 \ee
in particular
 \bea
 \label{S^{(1)}1-discrete-1}
S^{(1)}_1\left(\bt_\infty,N,U(\bt^*,{\bar \bt}^*)\right)&:=&\frac{1}{N!}\,\sum_{h_1,\dots,h_N=0}^M\, |\Delta_N(h)|\,\prod_{i=1}^N\, 
\mu_1(h_i,\bt^*,{\bar \bt}^*)
\\
 \label{S^{(1)}2-discrete-1}
S^{(1)}_2\left(\bt_\infty,N=2n,U(\bt^*,{\bar \bt}^*)\right)&:=&\frac{1}{n!}\,\sum_{y_1,\dots,y_n=}^{}\, 
\left(\Delta_N(y^2)\right)^2\,\prod_{i=1}^n\, 
\mu_2(y_i,\bt^*,{\bar \bt}^*)
\\
 \label{S^{(1)}3-discrete-1}
S^{(1)}_{3}\left(\bt_\infty,N,\bt_\infty,U(\bt^*,{\bar \bt}^*)\right)&:=&
\sum_{h_1,\dots,h_N=0}^M\,  {\Delta_N(h)}{\Delta^{(5)}_N(h)}
\,\prod_{i=1}^N\, 
\mu_3(h_i,\bt^*,{\bar \bt}^*)
\\
 \label{S^{(1)}4-discrete-1}
S^{(1)}_{4}\left(\bt_\infty,N=2n,U(\bt^*,{\bar \bt}^*)\right)&:=&\frac{1}{(2n)!}\,\sum_{h_1,\dots,h_n=0}^M\,
 {\tilde{\Delta}^4_n(h)}\,\prod_{i=1}^N\, 
e^{-2{\tilde V}(h_i,\bt^*)}\mu_4(h_i,{\bt}^*,{\bar \bt}^*)\,, 
 \eea
where
 \be
\mu_i(n,\bt^*,{\bar \bt}^*)=e^{-V(n,\bt^*,{\bar \bt}^*)}\mu_i(n)\,, \quad i=1,3
 \ee

 \be\label{tildeDelta^4}
{\tilde{\Delta}}(h)^4\,:=\,\prod_{i<j\le N}
{(h_i-h_j)^2\left((h_i-h_j)^2-1\right)}.
 \ee
(compare with \cite{BorStrahov} where the same expression as the right-hand side was considered in the 
context of random partitions). Here
 \be\label{t^*}
V(n,\bt^*,{\bar \bt}^*)=V^{(0)}(n)+\,\sum_{m=1}^\infty\,\left(\frac{an+b}{cn+d}\right)^mt^*_m \,
 +\, t_0^*\,\ln\,\left(\frac{an+b}{cn+d}\right)\,-\,\sum_{m=1}^\infty\, \left(\frac{an+b}{cn+d}\right)^{-m} {\bar t}^*_{m}
 \ee
where $a,b,c,d$ are arbitrary constants, conditioned by $ad-bc \neq 0$.

Sums
\eqref{S^{(1)}-discrete-1}-\eqref{S^{(1)}4-discrete-1},  \eqref{S^{(2)}11-discrete-1}-\eqref{S^{(2)}44-discrete-1} 
and \eqref{S1-DI}-\eqref{S5-DI}  may be viewed as discrete analogues of random matrix ensembles
 (compare with \cite{OS}), where eigenvalues of matrices are real non-negative numbers. Then
\eqref{S^{(1)}4-discrete-1} is a discrete analogue of the symplectic ensemble, \eqref{S^{(1)}1-discrete-1}
is a discrete analogue of orthogonal ensemble, \eqref{S^{(1)}2-discrete-1}
is a discrete analogue of ensemble of anti-symmetric Hermitian matrices (see Section 3.4 in \cite{Mehta}), and
\eqref{S^{(1)}3-discrete-1} is the so-called Bures ensemble, which describes random density matrices
in quantum chaos problems, see \cite{OsipovSommers} for the details.

From double series \eqref{S^{(2)}} over Frobenius coordinates of partitions we have
 \be
\label{S^{(2)}-discrete-1}
S^{(2)}(\bt_\infty, {\bar A}, B^c):=
1+\sum_{k=1}\,\frac{1}{(k!)^2}\,
\sum_{\alpha_1,\dots,\alpha_k=0}^{M+1}
\,\sum_{\beta_1,\dots,\beta_k=0}^{N+1}\,e^{V(\beta ,\bt^*)-V(\alpha,\bt^*)} 
\frac{{\bar A}_{\alpha}\,\Delta_k(\alpha)\Delta_k(\beta)\,B^c_{\beta}\,}{\prod_{i,j=1}^k(\alpha_i+\beta_j+1)}
 \ee
in particular
 \be
 \label{S^{(2)}11-discrete-1}
S^{(2)}_{11}(\bt_\infty,U ):=\,1\,+\,\sum_{k=1}\,\frac{1}{(k!)^2}\,
\sum_{\alpha_1,\dots,\alpha_k=0}^{M+1}
\,\sum_{\beta_1,\dots,\beta_k=0}^{N+1}\,
e^{V(\beta ,\bt^*)-V(\alpha,\bt^*)} 
\frac{|\Delta_k(\alpha)\Delta_k(\beta)|}{\prod_{i,j=1}^k(\alpha_i+\beta_j+1)},
 \ee
 \be
\label{S^{(2)}33-discrete-1}
S^{(2)}_{33}(\bt_\infty,\bt_\infty,\bt_\infty,\bt^*):=\,1\,+\,\sum_{k=1}\,\frac{1}{(k!)^2}\,
\sum_{\alpha_1,\dots,\alpha_k=0}^{M+1}
\,\sum_{\beta_1,\dots,\beta_k=0}^{N+1}\,\,e^{V(\beta ,\bt^*)-V(\alpha,\bt^*)} 
\frac{\Delta_k(\alpha)\Delta_k^*(\alpha)\Delta_k(\beta)\,\Delta_k^*(\beta)}{\prod_{i,j=1}^k(\alpha_i+\beta_j+1)},
 \ee
 \be
 \label{S^{(2)}41-discrete-1}
S^{(2)}_{41}(\bt_\infty,\bt^*):=\,1\,+
 \ee
 \[
 \sum_{k=1}\,\frac{1}{((2k)!)^2}\,
\sum_{\alpha_1,\dots,\alpha_k=0}^{M+1}
\,\sum_{\beta_1,\dots,\beta_{2k}=0}^{N+1}\,
e^{V(\beta ,\bt^*)-2{\tilde V}(\alpha,\bt^*)} \,
\frac{{\tilde\Delta}^4_k(\alpha)\,{\Delta}_{2k}(\beta)}
{\prod_{i=1}^k \prod_{j=1}^{2k}(\alpha_i+\beta_j+1)(\alpha_i+\beta_j+2)}
 \]
 \be
 \label{S^{(2)}44-discrete-1}
S^{(2)}_{44}(\bt_\infty,\bt^*):=\,1\,+
 \ee
 \[
 \sum_{k=1}\,\frac{1}{((2k)!)^2}\,
\sum_{\alpha_1,\dots,\alpha_k=0}^{M+1}
\,\sum_{\beta_1,\dots,\beta_k=0}^{N+1}\,
e^{2{\tilde V}(\beta ,\bt^*)-2{\tilde V}(\alpha,\bt^*)} \,
\frac{{\tilde\Delta}^4_k(\alpha)\,{\tilde\Delta}^4_k(\beta)}
{\prod_{i,j=1}^k(\alpha_i+\beta_j+1)(\alpha_i+\beta_j+2)^2(\alpha_i+\beta_j+3)}
 \]

\br \em
Compare with sums obtained in \cite{HLO} 
 \bea
 \label{S1-DI}
& \sum_{\alpha\in\DP}\,\Delta^*(\alpha)
\,\prod_{i=1}^k\,\frac{e^{-U_{\alpha_i}(\bt^*)}}{\alpha_i!}\,,
\\
 \label{S2-DI}
&\sum_{\alpha\in\DP}\,\Delta^*(\alpha)^2
\,\prod_{i=1}^k\,\frac{e^{-U_{\alpha_i}(\bt^*)}}{(\alpha_i!)^2}\,,
\\
 \label{S4-DI}
&
\sum_{\alpha\in\DP'}\,{\tilde{\Delta}}^*(\alpha)^4
\,\prod_{i=1}^k\,\frac{e^{-U_{\alpha_i}(\bt^*)-U_{\alpha_i+1}(\bt^*)}}
{\alpha_i!(\alpha_i+1)!}\,,
\\
 \label{S5-DI}
& \sum_{k=0}^\infty \,\,
\frac {1}{k!}\sum_{\alpha,\beta\in\DP\atop
\ell(\alpha)=\ell(\beta)=k}\,
\Delta^*(\alpha)\Delta^*(\beta)\,\prod_{i=1}^k \, \frac
{D_{\alpha_i,\beta_i}}{\alpha_i!\beta_i!}
 \eea
where we remind that $\DP'$ is the set of all strict partitions
$(\alpha_1,\alpha_2,\dots,\alpha_N>0)$ with the property
$\alpha_{i}>\alpha_{i+1} +1,\,i=1,\dots,N-1$, and
 \be
{\tilde{\Delta}}^*(\alpha)^4\,:=\,\prod_{i<j\le N}
\frac{(\alpha_i-\alpha_j)^2\left((\alpha_i-\alpha_j)^2-1\right)}
{(\alpha_i+\alpha_j)^2\left((\alpha_i+\alpha_j)^2-1\right)} .
 \ee

 \er

Interpretation of series \eqref{S^{(2)}11-discrete-1}-\eqref{S^{(2)}44-discrete-1} as discrete version
of ensembles of random matrices stays unclear for us.

(B) In the same way the specialization \eqref{t(q)} yields discrete analogues of circular ensembles 
in case $q$ lies on the unit circle of the complex plane, 
$q=e^{\sqrt{-1}\phi}$, $\phi\neq \pi n, \,,n\in\mathbb{Z}$. 
Here the parameterization \eqref{U_n(bt^*,{bar bt^*},q)}

For instance 
 \bea
 \label{S^{(1)}1-discrete-2}
S^{(1)}_1(\bt(q),N,\bt^*,{\bar \bt}^*)&:=&\frac{1}{N!}\,
\sum_{h_1,\dots,h_N=0}^M\,e^{-V(q^h,\bt^*,{\bar \bt}^*)} \,|\Delta_N(q^h)|
\\
 \label{S^{(1)}4-discrete-2} 
S^{(1)}_{4}(\bt(q),N,\bt^*,{\bar \bt}^*)&:=&\frac{1}{(2n)!}\,\sum_{h_1,\dots,h_n=0}^M\,
e^{-2{\tilde V}(q^h,\bt^*,{\bar \bt}^*)} \,{\tilde{\Delta}^4_n(q^h)}\,, \quad N=2n
 \eea
where now
 \be\label{t^*-q}
V(q^n,\bt^*,{\bar \bt}^*)=
V^{(0)}(q^n)+\,\sum_{m=1}^\infty\,\left(\frac{aq^n+b}{cq^n+d}\right)^m t^*_m \, +\, 
t_0^*\,\ln\,\left(\frac{aq^n+b}{cq^n+d}\right)\,-\,\sum_{m=1}^\infty\,\left(\frac{aq^n+b}{cq^n+d}\right)^{-m}{\bar t}^*_m
 \ee

\br \em For $q$ real the correspondent sums may be identified with the so-called Jackson integrals \cite{KV}
\er

(C) The specialization \eqref{t(x)} where put $x_i=e^{y_i}$ allows to rewrite \eqref{S^{(1)}1} as
  \be
S^{(1)}_1=\,\frac{1}{\Delta_N(x)} \sum_{h_1,\dots,h_N=1}^M\, 
e^{V(h,\bt^*)}\,\det\left(e^{y_jh_i}\right)\,\sgn \Delta_N(h)
  \ee
which is a discrete analogue of the following two-matrix integral
  \be
\int \, dU \int \, dR\,\det\,R^n\,\exp\,\left(\Tr\,\left(UYU^\dag R+ \,\sum_{m\neq 0}\,t^*_mR^m\right) \right)
  \ee
where the first integral is the integral over unitary matrices and the second is the integral over real 
symmetric ones, $dU$ and $dR$ denote the correspondent Haar measures. $Y$ is any diagonal matrix (a source). 
The matrices are $N$ by $N$ ones.
This integral may be viewed as an analogue of the Kontsevich integral.

Then with the same specialization of higher times $\bt$ we rewrite \eqref{S^{(2)}11} as
 \be
S^{(2)}_{11}=1+\sum_{k=1}^\infty \sum_{\alpha_1,\dots,\alpha_k=0}^M 
\sum_{\beta_1,\dots,\beta_k=0}^N\, \left(\,\sgn \Delta_k(\alpha)\,
\left(\det\,\frac{1}{\alpha_i+\beta_j+1}\right)\,\sgn \Delta_k(\beta)\,\right)\,
e^{V(\beta,\bt^*)-V(\alpha,\bt^*)}
 \ee
Each partial sum related to a given $k$ may be considered as a discrete version of the 3-ple integral
 \be
\int  dU \int  dA \int dB\,\det\,A^n\,\det \,B^n \,\det\left(B+UAU^\dag \right)^{-k}\,
\exp\,\Tr\left( \,\sum_{m\neq 0}\,t^*_mA^m -\,\sum_{m\neq 0}\,t^*_{m}B^m \right)
  \ee
where $U$ is an unitary and both $A$ and $B$ are real symmetric matrices of size $k\times k$ and where
 $dU$, $dA$ and $dB$ are related Haar measure.

\paragraph{New hypergeometric functions.} Now we specify factors $e^{-U_\lambda}$ in series 
\eqref{S^{(1)}1}-\eqref{S^{(1)}4} in order to get certain generalizations of hypergeometric functions and basic
hypergeometric functions.

\,

First, we introduce the following hypergeometric series
\be\label{lBKP-hyp-series}
{{_p}\Phi}_r^{(\beta,N)}({\bf a}+n;{\bf
b}+n;\bt)\,:=\,\sum_{\lambda\in \Pa_\beta\atop \ell(\lambda)\le
N}\,\,\, \frac{\prod_{i=1}^p \,(a_i+n)_{\lambda}}{\prod_{i=1}^{r}\,
(b_i+n)_{\lambda}}\,
\,s_\lambda(\bt)\,,\quad \beta=1,2,4
 \ee
 and their $q$-deformed version
\be\label{lBKP-hyp-series-q}
{{_p}{\tilde\Phi}}_r^{(\beta,N)}({\bf a}+n;{\bf
b}+n;q,\bt)\,:=\,\sum_{\lambda\in
\Pa_\beta\atop \ell(\lambda)\le N}\,\, \frac{\prod_{i=1}^p
(q^{a_i+n};q)_{\lambda}}{\prod_{i=1}^r (q^{b_i+n};q)_{\lambda}}\,
s_\lambda(\bt)\,,\quad \beta=1,2,4
 \ee
where $\Pa_1=\Pa,\,\Pa_2=\SCP,\,\Pa_4=\FP$.

Here by ${\bf a}+n$ and by ${\bf b}+n$ we denote a set of parameters (``indices'') $(a_1+n,\dots,a_p+n)$
and $(b_1+n,\dots,b_r+n)$ where $n$ and $N$ are integer parameters and where the sum ranges over all partitions whose length 
(i.e. the number of non-vanishing parts) do not exceed $\,N\,$.
 The notation $\,(a)_\lambda\,$ where $\,\lambda\,$ has
 $\,n\,$ non-vanishing parts serves for
\be\label{Poch-lambda}
(a)_{\lambda} :=(a)_{\lambda_1}(a- 1 )_{\lambda_2}\cdots
(a-n+1)_{\lambda_n},\quad (a)_0=1
  \ee
where $\,(a)_m:=\frac{\Gamma(a+m)}{\Gamma(a)}\,$ is the so-called
 Pohgammer symbol. Then $\,(q^a;q)_\lambda\,$ is the $q$-deformed version of $\,(a)_\lambda\,$ 
  \be\label{Poch-q-lambda}
(q^a;q)_\lambda:\,=(q^a;q)_{\lambda_1}\cdots
(q^{a-n+1};q)_{\lambda_n}
 \ee
defined via the  $q$-deformed Pochhammer's symbols:
 \be\label{}
(q^a;q)_0:\,=1,\,(q^a;q)_n:\,=(1-q^a)\cdots(1-q^{a+n-1})
\ee 

\br \em
In case $\,t_m=\frac 1m \sum_{i=1}^L\,x_i^m\,$ the summation range
is restricted by the condition $\ell(\lambda)\le L$ because the
Schur functions vanish on the partitions whose length exceed $L$.
\er

\br \em
As one may notice that by specialization of $\bt =(t,0,0,\dots)$ and $N=1$ we obtain the generalized
hypergeometric function \cite{GasparRa}
 \[
  {{_p}\Phi}_r^{(1,1)}({\bf a}+n;{\bf
b}+n;\bt)=\sum_{k=0}^\infty\,\,\, \frac{\prod_{i=1}^p \,(a_i+n)_k}{\prod_{i=1}^{r}\,
(b_i+n)_k}\,
\,\frac{t^k}{k!}
 \]
Taking $\bt =\left(\frac x1,\frac{x^2}{2},\frac{x^3}{3},\dots \right)$ we obtain
 \[
  {{_p}\Phi}_r^{(1,1)}({\bf a}+n;{\bf
b}+n;\bt)=\sum_{k=0}^\infty\,\,\, \frac{\prod_{i=1}^p \,(a_i+n)_k}{\prod_{i=1}^{r}\,
(b_i+n)_k}\,
\,{x^k}
 \]
 \[
  {{_p}{\tilde\Phi}}_r^{(1,1)}({\bf a}+n;{\bf
b}+n;q,\bt)\,:=\,\sum_{k=0}^\infty\,\, \frac{\prod_{i=1}^p
(q^{a_i+n};q)_k}{\prod_{i=1}^r (q^{b_i+n};q)_k}\,
{x^k}
 \]
which, say, for $b_1+n=1$  yields respectively the known generalized and the basic 
hypergeometric functions \cite{GasparRa}.

\er

The hypergeometric function \eqref{lBKP-hyp-series} is obtained as a specific case of
\eqref{S^{(1)}1} if we choose $U$ variables as follows
 \be\label{hypU} U_m \,=
\,\sum_{i=1}^q\,\log (b_i)_m\, -\,\sum_{i=1}^p\,\log (a_i)_m
 \ee
The choice
 \be\label{hypU-q}
U_m =\sum_{i=1}^r\,\log (q^{b_i};q)_m\, - \,\sum_{i=1}^p\,\log
(q^{a_i};q)_m
 \ee
gives rise to \eqref{lBKP-hyp-series}.

\paragraph{Hypergeometric functions II}

We also introduce
\be\label{lBKP-hyp-series-Psi4}
{{_p}\Psi}_r^{(4,N)}({\bf a}+n;{\bf
b}+n;\bt)\,:=\,\sum_{\lambda\in \Pa\atop \ell(\lambda)\le
N}\,\,\, \frac{\prod_{i=1}^p \,(a_i+n)_{\lambda,4}}{\prod_{i=1}^{r}\,
(b_i+n)_{\lambda,4}}\,
\,s_{\lambda\cup\lambda}(\bt)
 \ee
\be\label{lBKP-hyp-series-Psi1}
{{_p}\Psi}_r^{(1,N)}({\bf a}+n;{\bf
b}+n;\bt)\,:=\,\sum_{\lambda\in \Pa_{ev}\atop \ell(\lambda)\le
N}\,\,\, \frac{\prod_{i=1}^p \,(a_i+n)_{\lambda,1}}{\prod_{i=1}^{r}\,
(b_i+n)_{\lambda,1}}\,
\,s_{\lambda}(\bt)
 \ee
 and their $q$-deformed version
\be\label{lBKP-hyp-series-q-Psi4}
{{_p}{\tilde\Psi}}_r^{(4,N)}({\bf a}+n;{\bf
b}+n;q,\bt)\,:=\,\sum_{\lambda\in
\Pa\atop \ell(\lambda)\le N}\,\, \frac{\prod_{i=1}^p
(q^{a_i+n};q)_{\lambda,4}}{\prod_{i=1}^r (q^{b_i+n};q)_{\lambda,4}}\,
s_{\lambda\cup\lambda}(\bt)
 \ee
\be\label{lBKP-hyp-series-q-Psi1}
{{_p}{\tilde\Psi}}_r^{(1,N)}({\bf a}+n;{\bf
b}+n;q,\bt)\,:=\,\sum_{\lambda\in
\Pa_{ev}\atop \ell(\lambda)\le N}\,\, \frac{\prod_{i=1}^p
(q^{a_i+n};q)_{\lambda,1}}{\prod_{i=1}^r (q^{b_i+n};q)_{\lambda,1}}\,
s_{\lambda}(\bt)
 \ee
where
  \be\label{Poch-lambda-beta}
(a)_{\lambda,\beta} :=(a)_{\lambda_1}(a-\frac 1 2\beta)_{\lambda_2}\cdots
(a-\frac 12(n-1)\beta)_{\lambda_n},\quad (a)_{0,4}=1
  \ee
Thanks to the formula $s_\lambda(-\bt)=(-1)^{|\lambda|}s_{\lambda^{tr}}(\bt)$ we have
\be\label{beta-duality}
{{_p}{\Psi}}_r^{(1,N)}({\bf a}+n;{\bf
b}+n;q,\bt)={{_p}{\Psi}}_r^{(4,N)}({\bf a}+n;{\bf
b}+n;q,-\bt)
 \ee
 \be\label{beta-duality-q}
{{_p}{\tilde\Psi}}_r^{(1,N)}({\bf a}+n;{\bf
b}+n;q,\bt)={{_p}{\tilde\Psi}}_r^{(4,N)}({\bf a}+n;{\bf
b}+n;q,-\bt)
 \ee

\br \em

Having in mind the known relation (see \cite{Mac},\cite{V})
 \[
  \int_{U\in\mathbb{U}(N,\mathbb{F}_\beta)}\,s_{2\lambda}(XU)dU\,=
\,\frac{J^{\left(\frac2\beta\right)}_\lambda(XX^\dag)}{J^{\left(\frac2\beta\right)}_\lambda({\bf{1}}_N)}
 \]
we obtain
\be\label{lBKP-hyp-series-Jack}
\int_{U\in\mathbb{U}(2k,\mathbb{F}_\beta)}\,   {{_p}\Psi}_r^{(\beta,N)}({\bf a}+n;{\bf
b}+n;XU)dU\,:=\,\sum_{\lambda\in \Pa\atop \ell(\lambda)\le k}\,\,\, \frac
{\prod_{i=1}^p \,(a_i+n)_{\lambda,\beta}}
{\prod_{i=1}^{r}\,
(b_i+n)_{\lambda,\beta}}\,
\,\frac{J^{\left(\frac2\beta\right)}_\lambda(XX^\dag)}{J^{\left(\frac2\beta\right)}_\lambda({\bf{1}}_N)}
\,,\quad \beta=1,2,4
 \ee
 and their $q$-deformed version
\be\label{lBKP-hyp-series-q-Jack}
\int_{U\in\mathbb{U}(2k,\mathbb{F}_\beta)}\,{{_p}{\tilde\Psi}}_r^{(\beta,N)}({\bf a}+n;{\bf
b}+n;q,XU)dU\,:=\,\sum_{\lambda\in
\Pa\atop \ell(\lambda)\le k}\,\, \frac{\prod_{i=1}^p
(q^{a_i+n};q)_{\lambda,\beta}}{\prod_{i=1}^r (q^{b_i+n};q)_{\lambda,\beta}}\,
\frac{J^{\left(\frac2\beta\right)}_\lambda(XX^\dag)}{J^{\left(\frac2\beta\right)}_\lambda({\bf{1}}_N)}
\,,\quad \beta=1,2,4
 \ee

\er

 Thanks to results of Subsection \ref{Pfaffian representations-1} we have Pfaffian representation for each
of the introduced hypergeometric functions. In particular
 \be
{{_p}\Phi}_r^{(4,N)}({\bf a}+n;{\bf
b}+n;\bt)\,=\,\frac{1}{\Delta_{N}(x)}\Pf\,(x_j-x_i){{_p}\Phi}_r^{(4,N)}({\bf a}+n;{\bf
b}+n; x_ix_j)
 \ee
\be
{{_p}\Psi}_r^{(4,N)}({\bf a}+n;{\bf
b}+n;\bt)\,=\,\frac{1}{\Delta_{N}(x)}\Pf\,(x_j-x_i){{_p}\Phi}_r^{(4,N)}({\bf a}+n;{\bf
b}+n; x_ix_j)
 \ee

 \br \em

 Let us note that the hypergeometric series \eqref{lBKP-hyp-series},\eqref{lBKP-hyp-series-Psi4},\eqref{lBKP-hyp-series-Psi4}
  are different from the so-called (case $\mathbb{C}$) hypergeometric function of matrix argument \cite{GR},\cite{KV}
 \[
\,\sum_{\lambda\in \Pa \atop \ell(\lambda)\le N}\,\,
\frac{\prod_{i=1}^p (q^{a_i};q)_\lambda}{\prod_{i=1}^r
(q^{b_i};q)_\lambda}\,\frac{s_\lambda(x^{(N)})}{H_\lambda}
  \]
and hypergeometric series \eqref{lBKP-hyp-series-q}  are different from
 Milne's hypergeometric series \cite{Milne},\cite{KV}
  \[
\,\sum_{\lambda\in \Pa \atop \ell(\lambda)\le N}\,\,
\frac{\prod_{i=1}^p (q^{a_i};q)_\lambda}{\prod_{i=1}^r
(q^{b_i};q)_\lambda}\, \frac{s_\lambda(x^{(N)})}{H_\lambda(q)}
  \]
 which are examples of KP tau functions \cite{OSch1},\cite{OSch2}. In these formulas $H_\lambda$
 and $H_\lambda(q)$ are the hook-product and the $q$-deformed hook product respectively.

 \er

\section{Fermionic representation}

We suppose that the reader is familiar with the definition of the Fermi operators and the vacuum expectation value,
for notations see Appendix \ref{fermions-section}.

One may prove the following relations
 \be\label{S^{(1)}-f}
S^{(1)}(\bt,N,U,{\bar A})=\l N|\,\g (\bt)\, \mathbb{T}(U)\, g^{--}({\bar A})\,|0\r
 \ee
where
 \be \label{g^{--}(A)}
g^{--}({\bar A})=g^{--}(A,a)=e^{\frac 12 \sum_{n,m\in\mathbb{Z}}\,A_{nm}\psi_n\psi_m +\sum_{n\in\mathbb{Z}}\,a_n\psi_n\phi_0}
 \ee

In particular we have
 \bea
 \label{S^{(1)}0-f}
S^{(1)}_{0}(\bt,N;M,U)&:=&\l N|\,\g (\bt)\, 
e^{\sum_{ M \ge m > n}\,\psi_m\psi_n +\sum_{m\le M}\,\psi_m\phi_0} \,|0\r
\\
 \label{S^{(1)}1-f}
S^{(1)}_{1}(\bt;N,U)&:=&
\l N|\,\g (\bt)\, \mathbb{T}(U)\,
e^{\sum_{m>n}\,\psi_m\psi_n +\sum_{m\in\mathbb{Z}}\,\psi_m\phi_0} \,|0\r
\\
 \label{S^{(1)}2-f}
S^{(1)}_{2}(\bt,N=2n;U,c)&:=&
\l N|\,\g (\bt)\, \mathbb{T}(U)\,
e^{\sum_{m<n}\,\psi_{2c-m+1}\psi_{m} } \,|0\r
\\
 \label{S^{(1)}3-f}
S^{(1)}_{3}(\bt,N,\bt';U)&:=&
\l N|\,\g (\bt)\, \mathbb{T}(U)\,
e^{\sum_{m>n}\,Q_{(\alpha(n),\alpha(m))}(\tfrac 12\bt')\psi_m\psi_n +
\sum_{m\in\mathbb{Z}}\,Q_{(\alpha(m))}\psi_m\phi_0} \,|0\r
\\
 \label{S^{(1)}4-f}
S^{(1)}_{4}(\bt,N=2n;U)&:=&
\l N|\,\g (\bt)\, \mathbb{T}(U)\,
e^{\sum_{m \in\mathbb{Z}}\psi_m\psi_{m-1} } \,|0\r
\\
 \label{S^{(1)}5-f}
S^{(1)}_{5}(\bt,N;U,f)&:=&
\l N|\,\g (\bt)\, \mathbb{T}(U)\,
e^{\sum_{m>n}\,\frac{f(m)-f(n)}{f(m)+f(n)}\psi_m\psi_n +
\sum_{m\in\mathbb{Z}}\,\psi_m\phi_0} \,|0\r
\\
 \label{S^{(1)}6-f}
S^{(1)}_{6}(\bt,N;U,f)&:=&
\l N|\,\g (\bt)\, \mathbb{T}(U)\,
e^{\sum_{m>n}\,\frac{f(m)-f(n)}{1-f(m)f(n)}\psi_m\psi_n +
\sum_{m\in\mathbb{Z}}\,\psi_m\phi_0} \,|0\r
 \\
 \label{S^{(1)}7-f}
S^{(1)}_{7}(\bt,N;U,f)&:=&
\l N|\,\g (\bt)\, \mathbb{T}(U)\,
e^{\sum_{m>n}\,\frac{f(m)-f(n)}{(f(m)+f(n))^2}\psi_m\psi_n +
\sum_{m\in\mathbb{Z}}\,\psi_m\phi_0} \,|0\r
\eea
where
  \be\label{gamma(t)}
\g(\bt)=e^{J(\bt)},\quad {\bar \g}(\bt)=e^{{\bar J}(\bt)}
  \ee
 \be\label{currentJ}
J(\bt):=\sum_{n=1}^\infty J_n t_n,\quad {\bar J}({\bar
\bt}):=\sum_{n=1}^\infty J_{-n}{\bar t}_n,\quad
J_n:=\sum_{m\in\mathbb{Z}}\psi_{m}\psi^\dag_{m+n}
 \ee
 and
\be\label{mathbb T'}
\mathbb{T}(U):=\exp \, \, \sum_{i < 0} U_i\psi_i\psi_i^\dag
 -\sum_{i\ge 0}U_i\psi_i\psi_i^\dag
 \ee
where the fermionic operators are defined as in \cite{JM}, see Appendix \ref{fermions-section}

\br \em 
Pfaffian representation presented in Subsection \ref{Pfaffian representations-1} considered above may be 
obtained from the Wick's rule.

\er

\br \em

One can write
\bea
 \label{S^{(1)}1-f-int}
g^{--}({\bar A}_1)&:=&
e^{\sum_{m>n}\,\psi_m\psi_n +\sum_{m\in\mathbb{Z}}\,\psi_m\phi_0} =
\,e^{\oint\,\psi(x^{-1})\left(\psi(x)+\phi_0 \right)\frac{1}{1-x} dx}
\\
 \label{S^{(1)}2-f-int}
g^{--}({\bar A}_2)&:=&
e^{\sum_{m<n}\,(-1)^m\psi_{2c-m+1}\psi_{m} } =
e^{\oint\,x^{-2c-2}\psi(x)\psi(-x) dx}
\\
 \label{S^{(1)}4-f-int}
g^{--}({\bar A}_4)&:=&
e^{\sum_{m \in\mathbb{Z}}\psi_m\psi_{m-1} } =
e^{\oint\,\psi(x^{-1})\psi(x) dx}
\eea
The corollary of the right hand side expressions is the fact that sums 
\eqref{S^{(1)}1},\eqref{S^{(1)}2} and \eqref{S^{(1)}4} may be re-written as certain multiply integrals ($\frac 12 N$-ply integrals
for $S^{(1)}_2$, $S^{(1)}_4$, and $N$-ply integrals for $S^{(1)}_1$), this will considered in details in \cite{OST-II}.
Now, we shall mention a general remark.

Imagine that a sum \eqref{S^{(1)}} we can present $A_{nm}$ as moments, or, the same we can solve the following inverse moment problem:
given $A_{nm}=-A_{mn},\,m,n\ge 0$ to find such an integration domain $D$ and an antisymmetric measure 
$dA(x,y)=-dA(y,x)$ such that
 \be\label{IMP}
A_{nm}=\int_D\, x^ny^m d A(x,y),\quad n,m\ge 0
 \ee
Also
 \be
a_n=\int_\gamma\,x^n da(x)
 \ee

If we have \eqref{IMP} then in case $N=2n$ we can write $N$-ply integral
 \be
S^{(1)}(\bt,2n,U,{\bar A})=e^{-\sum_{i=0}^{N-1} U_{i}}\int_{D^n}\,
\left(\prod_{i=1}^{2n}e^{\xi_r(\bt,x_i)}\cdot\Delta_{2n}(x)\right) \Pf \left[ dA(x_i,x_j) \right] 
 \ee
where $\xi_r(\bt,x)$ is the following $\Psi DO$ operator
 \be
\xi_r(\bt,x)=\sum_{m=1}^\infty\,t_m \left(x r(D) \right)^m\,,\quad D=x\partial_x
 \ee
and $r$ is related to $U$ as follows
 \be
r(n)=e^{U_{n}-U_{n+1}}
 \ee
The case $U=0$ causes $
  \xi_r(\bt,x)=\sum_{m=1}^\infty\,t_m x ^m $ and we obtain more familier expression
 \be
S^{(1)}(\bt,2n,U=0,{\bar A})=\int_{D^n}\,
\prod_{i=1}^{2n}e^{\sum_{m=1}^\infty\,t_m x_i^m }\,\Delta_{2n}(x) \Pf \left[ dA(x_i,x_j) \right] 
 \ee

In case $N=2n+1$ we have more involved expressions which will be written down in a more detailed version.

 In case the solution of the inverse problem is not unique we have a set of different
integral representations for the sum \eqref{S^{(1)}}.
 \er

\paragraph{Other representations}:

For certain sums like \eqref{S^{(1)}3-f} we present a different representation as follows
 \be \label{S^{(1)}3-quatric}
S^{(1)}_{3}(\bt,N,\bt';U):=
\l N|\,\g (\bt)\, \g_B (\bt')\,\mathbb{T}(U)\,
e^{\sum_{m>n}\,\phi_{\alpha(n)}\phi_{\alpha(m)}\psi_m\psi_n +
\sum_{m\in\mathbb{Z}}\,Q_{(\alpha(m))}\psi_m\phi_0} \,|0\r
 \ee

\paragraph{Fermionic representation for $S^{(2)}$.}

For $S^{(2)}$ of \eqref{S^{(2)}} we have a similar relations
\be\label{S^{(2)}-f}
S^{(2)}(\bt;U,{\bar A},{\bar B})=\l 0|\,\g (\bt)\, \mathbb{T}(U)\,g^{+}({\bar B}) g^{-}({\bar A})\,|0\r
 \ee
where $\mathbb{T}(U)$ may found in \eqref{mathbb T'}
  \be \label{g^{-}(A)}
g^{-}({\bar A})=g^{-}(A,a)=e^{\frac 12 \sum_{n,m \ge 0}\,A_{nm}\psi_n\psi_m +\sum_{n > 0}\,a_n\psi_n\phi_0}
 \ee
 \be \label{g^{+}(A)}
g^{+}({\bar B})=g^{+}(B,b)=e^{\frac 12 \sum_{n,m\ge 0}\,B_{nm}(-1)^{n+m}\psi^\dag_{-n-1}\psi^\dag_{-m-1} +
\sum_{n > 0}\,(-1)^n b_n\phi_0\psi_{-n-1}^\dag}
 \ee

For those who are familiar with \cite{KvdLbispec} these fermionic relations yields a direct proof of 
Propositions \ref{Prop1} and \ref{Prop3}. 
In the next section we will consider it in more details.

\section{Small and Large BKP tau functions}

\subsection{A  class of "small" BKP (sBKP) tau functions}

We start with the "small" BKP case because it is more simple and
illustrative. Concerning this case see \cite{HLO}.

A general tau function of the sBKP hierarchy may be written as
 \be\label{sBKPgeneral}
 \tau^{sBKP}(\bt')=\l
0|\,\Gamma_B(\bt')\,e^{\sum_{n,m \in\mathbb{Z}
}\,A_{nm}\phi_n\phi_m }\,|0\r
 \ee
 and a tau function of 2-sBKP hierarchy as
 \be\label{2- sBKP}
 \tau^{sBKP}(\bt',{\bar\bt}')=\l
0|\,\Gamma_B(\bt')\,e^{\sum_{n,m \in\mathbb{Z}
}\,A_{nm}\phi_n\phi_m }\,\Gamma_B({\bar\bt}')\,|0\r
 \ee

The sBKP hierarchy  is obtained from the KP hierarchy by a
reduction. In the sBKP reduction, even times are set equal to zero
and we shall mark it by " ' " : $\bt'
=(t_1',0,t_3',0,t_5',\dots)$, then
\be
 \Gamma_B(\bt')=\exp\, \sum_{n\ge 1,\;\text{odd}}H^B_nt_n'\,
 ,\quad {\bar\Gamma}_B({\bar \bt}')=\exp\, \sum_{n\ge
 1,\;\text{odd}}H^B_{-n}{\bar t}_n'
 \ee
where
 \be
    H^B_n=\frac12\sum_{i\in\mathbb{Z}}(-1)^{i+1}\phi_i\phi_{-i-n}.
 \ee
The sets of parameters $\bt'$ and ${\bar\bt}'$ are called sBKP
higher times.

\paragraph{``Easy'' tau functions} We shall consider a simple case, where all terms in the sum of the
exponent \eqref{sBKPgeneral} commute,  namely, tau functions
 \be\label{sBKPour}
\tau^{sBKP}(\bt',U,A)=\l
0|\,\Gamma_B(\bt')\,\mathbb{T}_B(U)\,e^{\sum_{n>m
> 0}\,A_{nm}\phi_n\phi_m\,
+\,\sum_{n>0}\,a_n\phi_n\phi_0\sqrt{2}}\,|0\r
 \ee
where
 \be
\mathbb{T}_B(U)=\exp\,\left(-\sum_{n>0}\,(-1)^{n+1}U_n\phi_n\phi_{-n}\right)
 \ee
We want to single out the $U$-dependence though it may be included
into the redefinition of $A$ as $A_{nm}\to e^{-U_m-U_n}A_{nm}$.

We shall refer tau functions  \eqref{sBKPour} as easy sBKP tau functions. 

 We have
 \bp \em
 \be\label{sBKP-tau-simple}
\tau^{sBKP}(\bt',U,A)=\sum_{\alpha\in
\DP}\,e^{-U_{\{\alpha\}}}A_{\alpha}Q_\alpha(\bt')
 \ee
 where the sum ranges over all strict partitions
 $\alpha=(\alpha_1,\dots,\alpha_k ),\, k=0,1,2,\dots$,
 where
 \be
U_{\{\alpha\}}=\sum_{i=1}^k\, U_{\alpha_i}
 \ee
  and where
 ${ A}_{\{\alpha\}}$  is the Pfaffian of the $k \times k$  antisymmetric
 matrix ${\tilde A}$  defined as follows:

  for even $k$ its entries are
  \be
 {\tilde A}_{nm}=A_{\alpha_n\alpha_m},\quad
 {n,m=1,\dots,k};
  \ee

for $k$ odd we take
  \be
 {\tilde A}_{nm}=A_{\alpha_n\alpha_m},\quad
 {n,m=1,\dots,k};\qquad {\tilde A}_{n,k+1}=-{\tilde
 A}_{k+1,n}=a_n,\quad n=1,\dots,k+1
  \ee
It is assumed that for $\alpha=0$
$A_{\{0\}}=Q_{0}=e^{-U_{\{0\}}}=1$.

 \ep

 For proof we notice that
 $e^{\sum_{n>0}\,a_n\phi_n\phi_0\sqrt{2}}=1+\sum_{n>0}\,a_n\phi_n\phi_0\sqrt{2}$
 , and take into account that
   \[
\mathbb{T}_B(U)\phi_{\alpha_1}\cdots
\phi_{\alpha_k}|0\r=e^{-U_{\{\alpha\}}}\phi_{\alpha_1}\cdots
\phi_{\alpha_k}|0\r
   \]
  Then we obtain \eqref{sBKP-tau-simple} thanks to Lemma \ref{You}
 in Appendix \ref{projective-section}.

\paragraph{Example 1.} Choosing
 \[
A_{nm}=Q_{(n,m)}({\tfrac 12{\bar\bt}}'),\quad n>m>0,\quad
a_n=Q_{(n)}({\tfrac 12{\bar\bt}}')
 \]
where $Q_{(n,m)}$ is the projective Schur function related to a
partition $(n,m)$ and ${\bar\bt}'$ are parameters we obtain
 \be\label{2-sBKP-hyptau}
\tau^{sBKP}(\bt',U,A)=\sum_{\alpha\in
\DP}\,e^{-U_{\{\alpha\}}}Q_\alpha(\bt')Q_\alpha({\bar\bt}')
 \ee
which is actually an example of a tau function \eqref{2- sBKP},
see \cite{Q}.

\paragraph{Example 2.} Choosing
 \[
A_{nm}=1,\quad n>m,\qquad a_n=1
 \]
  we obtain
 \be\label{sBKP-hyptau}
\tau^{sBKP}(\bt',U)=\sum_{\alpha\in
\DP}\,e^{-U_{\{\alpha\}}}Q_\alpha(\bt')
 \ee

The right-hand side of \eqref{sBKP-hyptau} appeared in \cite{LO}
as a generating function for partition functions related to
oscillating strict partitions.

\br \em\label{exclusion} Summation range $\infty>\alpha_1>\cdots >
\alpha_k>0,\, k=0,1,2$ in sums over all strict partitions may be
replaces by sets of strict partitions whose parts may take values
in a given set of natural numbers which we can write as a strict
partition, say, $\gamma=(\gamma_1,\dots,\gamma_N)$,
$\gamma_1>\cdots>\gamma_N$, $N$ may be infinite. This may be
obtained by equating $e^{-U_n}$ to zero in case $n$ is not equal
to any of $\gamma_i$. We obtain the following sBKP tau function
 \be\label{tau-gamma-sBKP}
\tau^{sBKP}_\gamma(\bt',U)\,:=\,\sum_{\alpha\in \DP \atop \alpha
\subseteq \gamma}\,e^{-U_{\{\alpha\}}}Q_\alpha(\bt')
 \ee
 \er

\paragraph{Relation to solitons.}
Let us note that for $\bt'=(1,0,0,0,\dots)=:\bt_\infty$ and
 \[
 U_n=U_n(\bt^*,{\bar\bt}^*))=
 U_n^{(0)}\,-\,\log n!\,-\,\sum_{m=1,3,\dots}\,n^m t_m^*
 +\,\sum_{m=1,3,\dots}\,n^{-m} {\bar t}_m^*,\quad n>0
 \]
thanks to Lemma \ref{Q-t-infty} in Appendix
\ref{evaluated-section} the right-hand side of \eqref{sBKP-hyptau}
gives rise to a multisoliton  2-sBKP tau function where $\bt^*$
and ${\bar\bt}^*$ play the role of higher times. Indeed, with the
help of \eqref{phi(t)} the right-hand side of \eqref{sBKP-hyptau}
reads as
\bea
\lefteqn{
\l
0|\,\Gamma_B(\bt_\infty)\,\mathbb{T}_B(U(\bt^*,{\bar\bt}^*))\,e^{\sum_{n>m
> 0}\,\phi_n\phi_m\,
+\,\sum_{n>0}\,\phi_n\phi_0\sqrt{2}}\,|0\r\,=
}\nonumber\\
&=&1+\sum_{k=1}^\infty\,\sum_{\alpha_1>\cdots
>\alpha_k>0}\,\prod_{i=1}^k
e^{\sum_{m=1,3,\dots}  (\alpha_i^m t_m^*-\alpha_i^{-m} {\bar
t}_m^*-U_{\alpha_i}^{(0)})}\,
\prod_{i<j}\frac{\alpha_i-\alpha_j}{\alpha_i+\alpha_j}\,=
\nonumber\\
&=&\l 0|\,\Gamma_B(\bt^*)\,e^{\sum_{n>m>0
}\,e^{-U_m^{(0)}-U_n^{(0)}}\phi(n)\phi(m)} \, e^{\sum_{n>0}
e^{-U_n^{(0)}}\phi(n)\phi_0\sqrt{2} }\,\Gamma_B({\bar\bt}^*)\,|0\r
 \eea
 where
 \[
\phi(z):=\,\sum_{n\in\mathbb{Z}}\,z^n\phi_n
 \]
 \br \em On soliton solutions of integrable equations see \cite{TeorSol},
 on multisolitons of sBKP hierarchy see \cite{JM}.
The constant $U_n^{(0)}$ plays the role of the initial phase which
defines the initial location of the soliton  marked by $n$. One
can "remove" any soliton by 'sending it to infinity', i.e. via
$e^{-U_n^{(0)}}\to 0$. See Remark \ref{exclusion}, where $\gamma$
may be interpreted as the soliton  momentums  in the
$N$-soliton solution to sBKP hierarchy. \er

\section{The "large" BKP (lBKP) and "large" 2-BKP tau functions
\label{large-BKP-sDKP}}

A general large BKP and large 2-BKP tau functions may be
expressed as the following fermionic expectation value,
respectively :
 \be\label{lBKPtau}
\tau_N(l,\bt)=\langle N+l|\,\g(\bt) g \,|l\rangle
 \ee
and
 \be\label{2-lBKPtau} \tau_N(l,\bt,{\bar \bt})=\langle N+l|\,\g(\bt)
g {\bar \g}({\bar\bt})\,|l\rangle
 \ee
 where $N$ is an integer where $\g(\bt)$ is the same as in \eqref{gamma(t)} and where
 \be\label{lBKPg}
g=\exp \, \sum_{n,m}\,
A_{nm}\psi_n\psi_m+B_{nm}\psi^\dag_n\psi^\dag_m
+D_{nm}\psi_n\psi^\dag_m + \sum_{n}\,(a_n\,\psi_n +
b_n\psi^\dag_n)\phi_0
 \ee
Here $A_{nm}=-A_{mn}$ and $B_{nm}=-B_{mn}$. 
It is due to the presence of $\phi_0$ and thanks to equations
\eqref{zero-phi-on-vacuum}, \eqref{phi-psi} the number $N$ may be
odd as well as even.

  Parameters $\bt=(t_1,t_2,\dots)$ and
 ${\bar\bt}=({\bar t}_1,{\bar t}_2,\dots)$ are called higher times
 of the 2-BKP hierarchy.

 The large BKP tau function \eqref{lBKPtau} was introduced in \cite{KvdLbispec}, it
 solves large BKP Hirota equation written down in \cite{KvdLbispec}. Hirota equations
  for 2-lBKP are written down in the Appendix.

 At the present paper  lBKP tau functions \eqref{lBKPtau} are
  mainly used to study multiple sums, while 2-lBKP tau functions \eqref{2-lBKPtau}
 will be used to study multiple integrals in \cite{OST-II}.

\paragraph{Easy tau functions.} If for $g$ we choose any product of three special $g$ which are

 \bea\label{simplification-g-}
 g=g^{-}({\bar A}):=\left(\exp \,
\sum_{n,m \ge L} \,A_{nm}\psi_n\psi_m \,+\,\sum_{n\ge L}\,a_n\psi_n\phi_0\right),\\
\label{simplification-g+} g=g^{+}({\bar B}):=\left( \exp \, \sum_{n,m
\ge L} \ B_{nm}\psi^\dag_{-n-1}\psi^\dag_{-m-1} \,+\,\sum_{n\ge L}\,b_n\phi_0\psi^\dag_{-n-1}\right)
 \eea
where $l$ is the right hand charge. We will also use
 \be\label{simplification-g-+}
g=g^{-+}(D):=\left(\exp \, \sum_{n,m\ge L}
\,D_{nm}\psi_{n}\psi_{-m-1}^\dag \right)
 \ee
we obtain simply expressions for \eqref{lBKPtau} and \eqref{2-lBKPtau}. This is because of the fact that all 
fermionic operators in the exponents in \eqref{simplification-g-},\eqref{simplification-g+},\eqref{simplification-g-+}    
anticommute. (In this sense we treat fermions as Grasmannian variables).

The integer $L$ indicating the summation range is arbitrary. Next we consider basic examples.

First, let us consider
 \be
\tau = \l l|\,\g(\bt)\,\mathbb{T}(U)g^{-+}(D)\,|l\r\,=\,\sum_{}
\,\det[D_{h_i,h_j}]\,e^{-U_{\{h\}}}s_{\{h\}}(\bt)
 \ee
then choosing $D_{nm}=s_{(n|m)}({\bar\bt})$ we obtain that it is equal to
 \be\label{KP-hyp-tau}
c_l\sum_{\lambda\in \Pa}\, e^{-U_\lambda(l)}s_\lambda(\bt)s_\lambda({\bar\bt})
 \ee
where $U_\lambda(l)$ and $c_l$ are the same as written down below in Proposition \ref{lBKP-Schut-prop}.
This is the well-known solution of TL hierarchy where the set $l,\bt,{\bar\bt}$ plays the role of higher times. These series were called hypergeometric tau functions in \cite{OSch1} because
they keep many properties of ordinary generalized hypergeometric functions 
(where the role of Gauss equation takes the so-called string equation). Various specifications of this tau function were widely 
used in various problems: 
 in analyze of generalized Kontsevich model \cite{Mironov}, in 2D chromodynamics \cite{Mig},\cite{SK},\cite{MirMorSem}, 
some $c=1$ string theory calculations \cite{NTT}, evaluation of Hurwitz numbers \cite{HurwitzOk}, generalized hypergeometric 
functions \cite{OS} (where the general form \eqref{KP-hyp-tau} was studied),for models of random partitions 
\cite{Ok-SchurMeasure}, for perturbation series in coupling constants for 
two-matrix and for normal matrix models \cite{HO}, \cite{OS}, for construction of solvable matrix integrals \cite{O2004}, 
for some calculus in Seiberg-Witten theory \cite{NOk}, Gromov-Witten theory \cite{OP},\cite{LQW}, physics of electronic 
liquid \cite{ABW}, models of random turn motion by M.Fisher \cite{HO}, so-called melting crystals problem 
\cite{NakTak},\cite{Foda} 6-vertex model \cite{FodaWZ-XXZ},\cite{ZJ6vert},\cite{Takasaki-Bethe},\cite{Zabr-Bethe},
 and in many others problems.

We hope that the relatives of this series which will be presented below will also find wide applications.

\subsection{lBKP tau functions $\tau_N(l,\bt,U,{\bar A})$ \label{B-tauUA-section} }

We will consider
 \be\label{lBKP-tauUA}
\tau_{N}(l,\bt,U,{\bar A}):\,= \,\langle
N+l|\,\g(\bt)\,\mathbb{T}(U)\,g^{--}(A,a)\, | l\rangle,
 \ee
where
  \be\label{g^{--}lBKP}
g^{--}(A,a):=e^{ \frac 12 \sum_{n,m}\,A_{nm} \psi_m \psi_n + \sum_{n\in\mathbb{Z}}\,a_n\psi_n\phi_0\sqrt{2}}
 \ee
and $\mathbb{T}(U)$ is as in \eqref{mathbb T'}. Tau function
\eqref{lBKP-tauUA} vanishes if $N<0$. We remind that we deal with a pair ${\bar A}=(A,a)$ which consists of an infinite
antisymmetric matrix $A$ and an infinite vector $a$.

 We have the following (compare with Proposition \ref{Prop1})

\bp \label{lBKP-Schut-prop} \em
 \be\label{lBKP-tau-Schur}
\tau_{N}(l,\bt,U,{\bar A})\,= \,c_{l} \,\sum_{\lambda\atop
\ell(\lambda)\le N} \, e^{-U_\lambda(l)}\,
{\bar A}_{\{h\}}(l)\,s_\lambda(\bt)
 \ee
where
 \be\label{U(l)}
 U_\lambda(l):=\sum_{i=1}^N\, U_{\lambda_i-i+N+l}
 \ee
 and ${\bar A}_{\{h\}}(l)$ is the Pfaffian of a matrix ${\tilde A}$ defined as the Pfaffian of an antisymmetric $2n
\times 2n$ matrix ${\tilde A}$ as follows:
  \be
  \label{A-bar}
{\bar A}_{h}(l):=\,\Pf[{\tilde A}]
  \ee
where for $N=2n$ even
  \be
  \label{A-h-even-n}
{\tilde A}_{ij}=-{\tilde A}_{ji}:=A_{h_i+l,h_j+l},\quad 1\le
i<j \le 2n
  \ee
and for $N=2n-1$ odd
 \be \label{A-h-odd-n} {\tilde
A}_{ij}=-{\tilde A}_{ji}:=
\begin{cases}
A_{h_i+l,h_j+l} &\mbox{ if }\quad 1\le i<j \le 2n-1 \\
a_{h_i+l} &\mbox{ if }\quad 1\le i < j=2n  .
 \end{cases}
  \ee
We set ${\bar A}_0(l) =1$. 

The constant $c_{l}$ is defined by 
\be\label{c_n}
 c_{l}  =\begin{cases}
    \displaystyle  e^{-U_{l-1}-\dots -U_{0}} & l > 0\\
         1 & l=0        \\
    e^{U_{l} +\dots +U_{-1}}  & l < 0
  \end{cases}.
\ee
\ep

As we see ${\bar A}_h(0)$ and $U_\lambda(0)$ coincides respectively with ${\bar A}_h$ and $U_\lambda$  defined in 
Section \ref{Sums of Schur functions}. This proves Proposition \ref{Prop1}.

 \br \em

The right-hand side of \eqref{lBKP-tau-Schur} may be also obtained
as a lDKP tau function.

 \er

 \br \em { The right-hand side of \eqref{lBKP-tau-Schur} may be also obtained as a special
 limit of a tau function (\ref{tau-UAB}) below. This is a case where $e^{-U_\lambda}$ vanishes
 if the length of partition $\lambda$ exceeds $N$,  to ensure this we
 put $e^{-U_{-N-1}}=0$, or, the same, we put $U_{-N-1}=+\infty$.}
 \er

As examples we have
 \be\label{restricted-tau+NBKP U}
  \tau_N^{lBKP-sBKP}(l,\bt,\bt',U): =
  \sum_{\lambda\atop \ell(\lambda)\le N} \, e^{-U_\lambda(l)}\,\,
Q{_{l+\lambda^-}\left({\tfrac 12 \bt}'\right)}\,s_\lambda(\bt)
 \ee
 \be\label{restricted-tau+NBKP U=0}
  \tau_N^{lBKP-sBKP}(l,\bt,\bt',U=0): =
  \sum_{\lambda\atop \ell(\lambda)\le N} \,
Q{_{l+\lambda^-}\left({\tfrac 12  \bt}'\right)}\,s_\lambda(\bt)
 \ee
 \be\label{restricted-tau+N U}
  \tau_N(l,\bt,\bt',U): =
  \sum_{\lambda\atop \ell(\lambda)\le N} \, e^{-U_\lambda(l)}\,s_\lambda(\bt)
 \ee
 \be\label{restricted-tau-N}
  \tau_N(\bt): =
  \sum_{\lambda\atop \ell(\lambda)\le N} \,s_\lambda(\bt)
 \ee
where $\,l+\lambda^-\,$ denotes the strict partition whose $i$-th
 part is equal to $
\lambda_i-i+N+l\,$. Here $\,\bt'\,$ is the set of variables
denoted $\,\left( t_1',t_3',t_5',\dots\right)\,$ and
$\,Q_{\alpha}(\frac 12 \bt')\,$ is the projective Schur functions
\cite{Mac} related to a strict partition $\alpha$. As we shall
show later lBKP tau functions \eqref{restricted-tau+NBKP U} and
\eqref{restricted-tau+NBKP U=0}
 are also  sBKP tau function whose
higher times are $\bt'$.

\subsection{lBKP tau functions $\tau_N(l,\bt,U,{\bar A},{\bar B})$ \label{B-tauUAB-section} }

We begin with a rather special $O(2\infty+1)$ element as follows
 \be
 g_o=g_o^{+}\,g_o^{-},
  \ee
  where 
  \be
g_o^{-}= e^{ \sum_{n>m\ge 0} \,\psi_n\psi_m\,+\,\sum_{n\ge 0} \psi_n\phi_0},\quad
g_o^{+}=e^{\sum_{n>m\ge 0}\,
(-)^{n+m}\psi^\dag_{-m-1}\psi^\dag_{-n-1}\,+\,\sum_{n\ge 0} \phi_0\psi^\dag_{-n-1}}
 \ee
 which are exponentials of nilpotents and mutually commuting
 operators in the Fock space.

Let us recall that the Fock space $\FF$  admits a decomposition as
an orthogonal direct sum of the subspaces $\FF_N$  of states with
charge $N$
 \be\label{charge-splitting}
\FF \,= \,\oplus_{N\in \mathbb{Z}} \FF_N.
 \ee
 We have
 \be\label{Omega}
g_o|0\rangle\, =\,|\Omega\rangle, \qquad |\Omega\rangle\,
=\,\sum_{N\in\mathbb{Z}}\, |\Omega_{N}\rangle
 \ee
 where each vector $|\Omega_N\rangle$ belongs to the subspace
 $\FF_N$. The result we need is

  \bl \label{Lemma1} \em The vector $|\Omega_0\rangle$ is the sum of all basis Fock
  vectors in $\FF_0$:
  \be
|\Omega_0\rangle =\sum_{\lambda\in \Pa}\,|\lambda\rangle
  \ee
where $\lambda$ runs over the totality $\Pa$ of all partitions
 $\lambda=(\lambda_1,\lambda_{2},\dots)$,  and
 $|\lambda\rangle$ is defined as
  \be
 |\lambda\rangle =(-)^{\beta_1+\cdots +\beta_k} \,\psi_{\alpha_1}\cdots
 \psi_{\alpha_{k}}\psi^\dag_{-\beta_k-1}\cdots
 \psi^\dag_{-\beta_1 -1} \,|0\rangle
  \ee
  where we use the Frobenius notation for partitions \cite{Mac}:
  $\lambda = (\alpha_1,\dots,\alpha_k|\beta_1,\dots,\beta_k)$  $= (\alpha|\beta)$ where
$\alpha_1>\cdots>\alpha_k\ge 0$, $\beta_1>\cdots>\beta_k\ge 0$,
$k=0,1,2,\dots$\footnote{Here and below $k=0$ will be related to
$\lambda =0$. \label{k=0}}.
 \el

  Then, we obtain the following simplest nontrivial DKP tau
  function (a version of a ``vacuum tau function''):

  \bp {\em We have the following DKP tau function
 \be\label{the simplest}
  \tau_o(\bt):=\,\langle 0|\,\g(\bt)  \,|\Omega\rangle\,
  =\,\sum_{\lambda\in \Pa}\,s_\lambda(\bt)=e^{\sum_{m=1}^\infty
  \,\frac 12 mt_m^2+\sum_{m=1}^\infty\, t_{2m-1}}
 \ee
 where $s_\lambda(\bt)$ are Schur functions. }
  \ep

 The
 second equality follows from the
 well-known formula \cite{JM}
 \be\label{KP-schur}
\langle 0|\,\g(\bt)\,|\lambda\rangle =s_\lambda(\bt),
 \ee
which is an example of the KP tau function.
 The third equality in \eqref{the simplest} follows from the Exercise
 I-5-4 in \cite{Mac}  which should be re-written
 in terms of power sums.

{\bf Corollary}. From consideration similar to \cite{TauFuncMI} we obtain
 \be
e^{
\frac 12\sum_{m=1}^\infty\frac{1}{m}J_{-m}^2\,+\,\sum_{m=1,3,\dots}^\infty \frac1m J_{-m}}\,|l\r\,=
\,\sum_{\lambda\in\Pa}\,|\lambda,l\r
 \ee
Similar relations follows from each equation of Propositions \ref{exponentials} and \ref{polynomials}.

Now, let us introduce 
 \be\label{tau-UAB}
\tau_N(l,\bt,U,{\bar A},{\bar B})\,:=\,\,\langle N+l|\,\g(\bt)\,\mathbb{T}(U)\,g^{+}({\bar B})\,g^{-}({\bar A})  \,|l\rangle\
 \ee
where $g^{-}({\bar A})$ and $g^{+}({\bar B})$ are given by \eqref{simplification-g-},\eqref{simplification-g+}. 
Then by a direct evaluation of vacuum expectation value we obtain 
 \be
S^{(2)}(\bt;U,{\bar A},{\bar B}) = \tau_0(0,\bt,U,{\bar A},{\bar B})
 \ee
which is the content of Preposition \ref{Prop3}.

\subsection{Mixed $lBKP$ and  $sDKP$ tau function}.

 \be
\tau:=\l 0|\Gamma(\bt,\bt',\bt'')e^{\sum_{n,m\ge 0}\,
\left(D_{nm}\p_{n}\psi_m+D^*_{nm}\hp_{n}\pd_{-m-1}\right)}|0\r
 \ee
 and
 \be
\tau:=\l N|\Gamma(\bt,\bt')\, e^{\sum_{n,m\ge 0}\,
D_{nm}\p_{n}\psi_m}|0\r
 \ee
 \be\label{AQs}
=\sum_{h_1>\cdots > h_N\ge 0}\,\sum_{\alpha\in \DP\atop
\ell(\alpha)\le N}\,2^{-\frac{\ell(\lambda)}{2}}\,
D_{\{\alpha,h\}}\,Q_{\alpha}(\tfrac12\bt')\,s_{\{h\}}(\bt)
 \ee
where
 \be
\Gamma(\bt,\bt',\bt''):=\Gamma(\bt)\Gamma_B(\bt')\Gamma_B(\bt''),\quad
\Gamma(\bt,\bt'):=\Gamma(\bt)\Gamma_B(\bt')
 \ee
and where $D_{\{\alpha,h\}}$ is given by \eqref{D-alpha-beta}.

 The particular cases are ($D_{nm}=e^{-U_n}\delta_{nm}$)
 \be\label{tau-t-t'-U}
\tau(\bt,\bt',U)=\sum_{\lambda\in \Pa\atop \ell(\alpha)\le
N}\,\,2^{-\frac{\ell(\lambda)}{2}}\,
e^{-U_\lambda}\,s_\lambda(\bt)\,Q_{\lambda^-}(\tfrac12 \bt'),\quad
 \ee
\be\label{tau-t-t'} \tau(\bt,\bt',U)=\sum_{\lambda\in \Pa\atop
\ell(\alpha)\le N}\,\,2^{-\frac{\ell(\lambda)}{2}}\,
\,s_\lambda(\bt)\,Q_{\lambda^-}(\tfrac12\bt'),\quad
 \ee
\be\label{tau-t-UI} \tau(\bt,U)=\sum_{\lambda\in \Pa\atop
\ell(\alpha)\le N}\, e^{-U_\lambda}\,s_\lambda(\bt)
 \ee
 \be\label{tau-t-0I}
\tau_0(\bt):=\sum_{\lambda\in \Pa\atop \ell(\alpha)\le
N}\,s_\lambda(\bt)
  \ee
 where $\lambda^-$ is the partition with shifted parts: $\lambda^-_i \,:=
 \lambda_i-i+N,\, i=1,\dots,N$, and where
  \be
U_\lambda:=\sum_{i=1}^N\, U_{\lambda_i-i}
  \ee

\subsection{Sums. Modifications of the Schur measure and tau functions. Discrete analogs
of matrix ensembles \label{sums-section}}

Here we plan to get use of the considered series in partitions
 \be\label{hyp-tau-collection1}
\sum_{\lambda\in\Pa}\, e^{-U_\lambda}s_\lambda(\bt),\quad
\sum_{\lambda\in\Pa}\,
\,2^{-\frac{\ell(\lambda)}{2}}\,e^{-U_\lambda}Q_{\lambda^-}(\tfrac
12 \bt'),\quad
\sum_{\lambda\in\Pa}\,\,2^{-\frac{\ell(\lambda)}{2}}\,e^{-U_\lambda}s_\lambda(\bt)
Q_{\lambda^-}(\tfrac 12 \bt'),\quad
 \ee
and also of the series
\be\label{hyp-tau-collection2} \sum_{\lambda\in\Pa}\,
e^{-U_\lambda}s_\lambda(\bt)s_\lambda({\bar \bt}),\quad
\sum_{\lambda\in\Pa}\,\,2^{-\ell(\lambda)}\,
e^{-U_\lambda}Q_{\lambda^-}(\tfrac 12 \bt')Q_{\lambda^-}(\tfrac 12
{\bar \bt}')
 \ee
earlier studied respectively in \cite{OSch2} and \cite{Q}.

\br  Let us notice that sums over partitions  are studied in the
context of random partitions. Representation theory Random
partitions were started by the school of A. Vershik starting late
60-es and are under intensive studies nowadays (see series of
papers by A. Vershik, S. Kerov, G. Olshansky, A. Okounkov, A.
Borodin on this topic). The elegant fermionic approach to this
subject was developed by A. Okounkov.

Let us write down most studied probability measures on the sets of
partitions. By ideology developed by A. Vershik these are parts of
representation theory of linear and symmetric groups.

 The Plancheral measure on the set of partitions of $n$ is defined as
 \[
n!\prod_{i=1}^N\frac{1}{(h_i!)^2}\prod_{1\le i<j\le
N}(h_i-h_j)^2\, ,\quad n=|\lambda|,\quad h_i:=\lambda_i-i+N
 \]
where $N\ge\ell(\lambda)$ \footnote{As one can check
\eqref{poiss-Plancheral} does not depend on choice of $N$ if $N$
no less than the partition length $\ell(\lambda)$.}. The
$z$-measure is defined as
 \[
\frac{n!}{(zz')_n}\prod_{i=1}^{N}
\frac{(z-i+1)_{\lambda_i}(z'-i+1)_{\lambda_i}}{(h_i!)^2}
\prod_{1\le i<j\le N}(h_i-h_j)^2\, ,\quad n=|\lambda|,\quad
h_i:=\lambda_i-i+N
 \]
where $z$ and $z'$ are parameters, and
$(z)_k:=\frac{\Gamma(z+k)}{\Gamma(z)}$ is the Pochhammer symbol.

The Schur measure on the set of all partitions was introduced by
Okounkov in \cite{Ok-SchurMeasure}. The weight of a partition
$\lambda$ is
\[
W_\lambda(\bt,{\bar \bt})= s_\lambda(\bt)s_\lambda({\bar\bt})
 \]
where $\bt$ and ${\bar\bt}$ are parameters of the measure. The
Schur measure generalizes the (poissonized with a parameter $p$)
Plancheral
 and $z,z'$-measures which may be basically obtained as
evaluations of the Schur measure respectively at the points
$\bt={\bar \bt}=e^{- p/2}\bt_\infty$ and
$\bt=e^{-p/2}\bt(z),{\bar\bt}=e^{-p/2}\bt(z')$ in notations
\eqref{choicetinfty'}, \eqref{choicet(a)'}. The poissonized
Plancharel measure on the set of all partitions assign the weight
 \be\label{poiss-Plancheral}
e^{-p|\lambda|}\prod_{i=1}^N\frac{1}{(h_i!)^2}\prod_{1\le i<j\le
N}(h_i-h_j)^2\, ,\quad h_i:=\lambda_i-i+N
 \ee

 The similarity of \eqref{poiss-Plancheral} to ensembles of random
 Hermitian matrices was observed and intensively worked out to solve combinatorial
 problems in late 90-es in papers by Okounkov, Borodin, Johansson and Baik,
 see \cite{Ok-SchurMeasure},\cite{Joh},\cite{Baik}
 Earlier it was used in physics in \cite{DouglasKazakov},\cite{KazakovW},\cite{KazakovSW}.
  Let us note that evaluation of the Schur measure at other points
\eqref{choicetinftyq'} and \eqref{choicet(a)q'} yields links with
 different matrix models, see \cite{OS}.

\paragraph*{}

Now turn out to the studied series in partitions. They bring us
to consider the following weights on the set of partitions
 \be\label{measure-collection1}
 W_\lambda(\bt,U)=e^{-U_\lambda}s_\lambda(\bt)\, ,
 \qquad W_\lambda(\bt,{\bar\bt},U)=e^{-U_\lambda}s_\lambda(\bt)s_\lambda({\bar
 \bt})\, ,
 \ee
 if  $\lambda=(\alpha|\beta)$
\be\label{measure-collection2}
W_{\lambda}(\bt,\bt',{\bar\bt}',U)=2^{-\ell(\lambda)}\,e^{-U_\lambda}\,
s_\lambda(\bt)\, Q_\alpha\left(\tfrac
12{\bt'}\right)Q_\beta\left(\tfrac 12{{\bar\bt}'}\right)\, ,
 \ee
 and, in case partitions are
restricted to have at most $N$ parts:
 \be\label{measure-collection3}
W_\lambda(\bt,{\bt}',U)=2^{-\frac{\ell(\lambda)}{2}}\,e^{-U_\lambda}s_\lambda(\bt)
{Q_{\lambda^-}\left(\tfrac 12 \bt'\right)}\,,
 \ee
 where $\lambda^-$ is a strict partition defined by
  $\lambda=(\lambda_1,\dots,\lambda_N)$
  as $\lambda^-:=(\lambda_1-1+N,\lambda_2-2+N,\dots,\lambda_N)$.
  On the set of strict partitions $\alpha\in \DP$
  \be\label{measure-collection4}
W_{\alpha}({\bt}',U)=2^{-\frac{\ell(\alpha)}{2}}e^{-U_{\{\alpha\}}}Q_{\alpha}(\tfrac
12 \bt')\, ,\quad
W_{\alpha}(\bt',{\bar\bt}',U)=2^{-\ell(\alpha)}
e^{-U_{\{\alpha\}}}Q_{\alpha}(\tfrac 12 \bt')Q_{\alpha}(\tfrac 12
{\bar \bt}')
 \ee
 At last
  \be\label{weight-beta=4}
W_\lambda(\bt,U)=e^{-U_{\lambda\cup\lambda}}s_{\lambda\cup\lambda}(\bt)
  \ee
restricted on the set of partitions
$\lambda\cup\lambda=(\lambda_1,\lambda_1,\lambda_2,\lambda_2,\dots)\in
\Pa^2$ may be viewed as the weight of $\lambda\in\Pa$.

  Here $e^{-U_\lambda}$ (or $e^{-U_{\{\alpha\}}}$) plays the role of additional
  Gibbs-Boltsmann weight assigned to each configuration $\lambda$ (or $\alpha$) induced
  by external  sources. Let us note that in $U$-dependence of partition functions
  $Z=Z(U)$ shows non-analytic behavior (phase transitions, compare to
  \cite{DouglasKazakov},\cite{KazakovW}, \cite{MirMorSem}, \cite{LO}, or just to, say,
   $_{1}F_0(a)=(1-x)^{-a}$ behavior which is the simplest example of the series under
   consideration).
  Now, $W_\lambda(\bt,{\bar\bt},0)$ is the Schur measure studied by Okounkov in
  \cite{Ok-SchurMeasure} while $W_{\alpha}(\bt',{\bar\bt}',0)$ is the shifted Schur
  measure introduced and studied by Tracy and Widom in \cite{TW-shifted}.

\,

  Then all normalization functions (in physics: "partition functions")
  $Z=\sum_\lambda W_\lambda$ of these ensembles of random partitions are tau
  functions \eqref{hyp-tau-collection1} and \eqref{hyp-tau-collection2}. Provided the
  weights are non-negative the probability of a configuration $\lambda$ is
 \[
p_\lambda=\frac{W_\lambda}{Z}
 \]
 where for each ensemble $Z$ is a tau function.

\,

It is interesting, that if we put all
$\bt,{\bar\bt},\bt',{\bar\bt}'$ to be equal to $(1,0,0,\dots)$ and
deform  $U$ via deformation parameters $\bt^*$ according to
\eqref{choice1} below we obtain that the partition function
$Z=Z(\bt^*)$ is again a tau function, in this case, it is a tau
function with respect to the deformation parameters $\bt^*$.
Moreover, it will be related to discrete versions of ensembles of
random matrices (where in context of considerations in physics
parameters $\bt^*$ are commonly called coupling constants).

\,

The last remark is the following. It is natural to consider
bi-measure on pairs of partitions taking general tau functions as
weight functions. Such models should possess good properties. Tau
functions of 2-KP , 2-lBKP (\eqref{general-coupled-DKPtau}), 2-sBKP
(\eqref{2-sBKP-QQ}) and lBKP coupled to sDKP provide respectively
$\Pa\times\Pa$, $\DP\times \DP$, $\DP\times\Pa$ with bi-measures
according to formulae
\eqref{Takasaki-Schur},\eqref{general-coupled-DKPtau},
\eqref{2-sBKP}, \eqref{AQs}. Objects of these models like
correlation functions should be expressed in terms of
Baker-Akhiezer functions.

\er

Now we shall write down normalization functions $Z$ for ensembles
of random partitions with measures $W_\lambda$ according to
\eqref{measure-collection1}-\eqref{measure-collection4}.
Parameters $\bt,{\bar\bt},\bt',{\bar\bt}'$ will be equated to one
of \eqref{choicetinfty'},
\eqref{choicet(a)'},\eqref{choicetinftyq'},\eqref{choicet(a)q'}
yielding different specifications of measures
\eqref{measure-collection1}-\eqref{measure-collection4}.

\paragraph*{ (i) Ensemble with the measure $W_\lambda(\bt,U(\bt^*))$. Discrete $\beta=1$
ensembles.}
The partition function $Z$ of this ensemble is the tau function
(\ref{restricted-tau+N U}).

\paragraph{}

First we  put $\bt=e^p\bt_\infty$ taking
 \be\label{choice1}
 \bt_\infty =(1,0,0,\dots),\quad U_n=U_n(\bt^*)=U_n^{(0)}-\sum_{m=1}^\infty \,
 n^m{ t}^*_m
 \ee
 Then due to \eqref{schurhook}  the weight function is
  \be\label{W1}
W_\lambda^{(1)}(\bt^*)\, =\,\prod_{n<m\le N}\,(h_n-h_m)\,
\prod_{j=1}^N \,\frac{1}{h_j!}\, e^{ \sum_{n=1}^\infty { t}^*_n
h_j^n\, +ph_j-U^{(0)}_{h_j}}
  \ee
the normalization function $Z=Z^{(1)}$ is equal to
 \be\label{eigenvalue-sum1}
Z^{(1)}(\bt^*,p)=\tau_N(\bt_\infty,U(\bt^*))=
\frac{e^{c_Np}}{N!}\sum_{h_1,\dots,h_N=0}^\infty\, \,
\prod_{n<m\le N}|h_n-h_m|\, \, \prod_{j=1}^N \, e^{
\sum_{n=1}^\infty { t}^*_n h_j^n}\,
\frac{e^{ph_j-U^{(0)}_{h_j}}}{h_j!}
 \ee
where the factor $N!$ appears when we spread the summation over
the cone $h_1>\cdots
>h_N\ge 0$ to the independent summation over each of
$h_j=0,1,2,\dots $ at the same time changing $\prod_{n<m\le
N}(h_n-h_m)$ to $\prod_{n<m\le N}|h_n-h_m|$. $c_N$ is an unrelated
constant. The parameter of Poissonization $p$ can be identified
with $t_1^*$.

If we take the same $U$ as in \eqref{choice1} but take $\bt$ as in
\eqref{choicet(a)'} we obtain an analog of the Poissonized $z$-
measure (with $z=a$):
 \be\label{W2}
W_\lambda^{(2)}(\bt^*,z)\, =\,c_N(a)\prod_{n<m\le N}\,(h_n-h_m)\,
\prod_{j=1}^N \,\frac{(a)_{h_j+1-N}}{h_j!}\, e^{ \sum_{n=1}^\infty
{ t}^*_n h_j^n\, -U^{(0)}_{h_j}}
 \ee
 with the normalization function
\be\label{eigenvalue-sum2} Z^{(2)}(\bt^*)=\tau_N(\bt(a),U(\bt^*))=
\frac{c_N(a)}{N!}\,\sum_{h_1,\dots,h_N=0}^\infty  \, \,
\prod_{n<m\le N}|h_n-h_m|\, \, \prod_{j=1}^N \, e^{
\sum_{n=1}^\infty { t}^*_n h_j^n}\,
\frac{e^{-U^{(0)}_{h_j}}}{h_j!}
 \ee
 where $c_N(a)$ is given by \eqref{c-k-a-q}.

 \br \em Both expressions \eqref{eigenvalue-sum1} and \eqref{eigenvalue-sum1} which are BKP
  tau functions evaluated at points $\bt=\bt_\infty$ and $\bt=\bt(a)$ are also BKP tau
 functions with respect to new parameters $\bt^*$ introduced in \eqref{choice1}.
 See also Section \ref{interlinks-section} about interlinks between different BKP tau
 functions.
 At the same time these tau functions where $\bt^*$ are higher times are examples of the
 $\beta=1$ ensemble \cite{Mehta} where the measure is proportional to a
sum of delta functions and may be treated as a discrete version of
ensemble of random orthogonal matrices with positive eigenvalues,
where the measure deformation parameters are $\bt^*$. \er

\paragraph{}

Next, instead of (\ref{choice1}) we put
  \be\label{choice2}
 t_m=\frac 1m \frac{1-(q^a)^m}{1-q^{m}},
 \quad U_n=U_n(\bt^*;q)=U_n^{(0)}-\sum_{m=-\infty}^\infty \, q^{nm}{ t}^*_m
 \ee
 By \eqref{choicet(a)q'} we obtain a different model with the weight
 \be\label{W3}
W_\lambda^{(3)}(\bt^*;q,a)\, =\,c_N(a,q)\, \prod_{n<m\le
N}\,(q^{h_n}-q^{h_m})\, \, \prod_{j=1}^N \,
\frac{(q^{a};q)_{h_i-n+1}}{(q;q)_{h_j}}\, e^{
\sum_{n=-\infty}^\infty {t}^*_n q^{nh_j}-U_{h_j}^{(0)}}
  \ee
with the following normalization function
 \[
Z^{(3)}(\bt^*;a,q)=\tau_N(\bt(a,q),U(\bt^*))=
 \]
 \be\label{eigenvalue-sum3}
= \frac{c_N(a,q)}{N!}\,\sum_{h_1,\dots,h_N=0}^\infty\, \,
\prod_{n<m\le N}|q^{h_n}-q^{h_m}|\, \, \prod_{j=1}^N
\frac{(q^{a};q)_{h_i-n+1}}{(q;q)_{h_j}}\,e^{
\sum_{n=-\infty}^\infty {t}^*_n q^{nh_j}-U_{h_j}^{(0)}}\,
 \ee
 which, for $q \in {S}^1$, may be considered as a discrete analogue of the
 circulate $\beta=1$ ensemble (\cite{Mehta}). $c_N(a,q)$ is
 defined in \eqref{c-k-a-q}. Let us choose the limit $q^a\to 0$
 (or, the same, $\bt$ is chosen by \eqref{choicetinftyq'})
 then $(q^{a};q)_{h_i-n+1}\to 1$.

 \br \em Expression \eqref{eigenvalue-sum3}  is the BKP tau function evaluated
 at point $\bt=\bt(a,q)$ is also a 2-BKP tau function \eqref{2-lBKPtau} with respect to
 new parameters $\{t^*_n,\,n>0\}$ and $\{t^*_n,\,n<0\}$ introduced in \eqref{choice2}
 and at the same time for $q\in S^1$ is an example of the $\beta=1$ circular ensemble
 \cite{Mehta} where the measure is singular and is equal to a
weighted sum of delta functions where the weight parameters
depends on $\bt^*$. \er

\paragraph*{ (ii) Ensemble with the measure $W_\lambda(\bt,{\bar\bt},U(\bt^*))$ of
\eqref{measure-collection1}.
Discrete $\beta=2$ ensembles.} The partition function $Z$ of this
ensemble is the KP tau function \eqref{KP-hyp-tau}. This case was
considered in the paper \cite{OS}, here we only write down the
most general case where $\bt=\bt(a,q),\,{\bar\bt}=\bt(a',q)$
(which may be treated as a $q$-version of $z,z'$-measure, with
$z=a,z'=a'$) and where $U$ are chosen by \eqref{choice2}:
 \[
W_\lambda^{(4)}(\bt^*;a,a',q)\, =
 \]
\be\label{W4} c_N(a,q)c_N(a',q)\, \prod_{n<m\le
N}\,(q^{h_n}-q^{h_m})^2\, \, \prod_{j=1}^N \,
\frac{(q^{a};q)_{h_i-n+1}(q^{a'};q)_{h_i-n+1}}{\left((q;q)_{h_j}\right)^2}\,
e^{ \sum_{n=-\infty}^\infty {t}^*_n q^{nh_j}-U_{h_j}^{(0)}}
  \ee
with the following normalization function
 \[
Z^{(4)}(\bt^*;a,a',q)=
 \]
\be\label{eigenvalue-sum3'}
\frac{c_N(a,q)c_N(a',q)}{N!}\,\sum_{h_1,\dots,h_N=0}^\infty\,\,
\prod_{n<m\le N}\left(q^{h_n}-q^{h_m}\right)^2\, \, \prod_{j=1}^N
\frac{(q^{a};q)_{h_i-n+1}(q^{a'};q)_{h_i-n+1}}{\left((q;q)_{h_j}\right)^2}\,e^{
\sum_{n=-\infty}^\infty {t}^*_n q^{nh_j}-U_{h_j}^{(0)}}\,
 \ee
 The last expression which is the 2-KP (TL) tau function \eqref{KP-hyp-tau} at the same
 time is 2-KP tau function with respect to the variables $\{t^*_n,\,n>0\}$ and
 $\{t^*_n,\,n<0\}$. For $q\in S^1$ it may be identified with a discrete version of the
 one-matrix model of unitary matrices \cite{Morozov}.

\paragraph*{ (iii) Ensemble \eqref{measure-collection2}:
$W_\lambda(\bt,\bt',{\bar\bt}',U(\bt^*))$}
 where $\bt=\bt'={\bar\bt}'=\bt_\infty $.

This ensemble is defined on the set of all partitions
$\lambda=(\alpha|\beta)\in \Pa$,
 \be
W_\lambda^{(5)}(\bt^*)=\prod_{i>j}^k\,
\frac{(\alpha_i-\alpha_j)^2(\beta_i-\beta_j)^2}{(\alpha_i+\alpha_j)(\beta_i+\beta_j)}
\prod_{i,j=1}^k \frac{1}{\alpha_i+\beta_j+1}\,\prod_{i=1}^k
\frac{e^{U_{-\beta_i
-1}(\bt^*)-U_{\alpha_i}(\bt^*)}}{\left(\alpha_i!\beta_i!
\right)^2}
 \ee
  \be
Z^{(5)}(\bt^*)=1+\sum_{k=1}^\infty\sum_{\alpha}\sum_{\beta}\,\prod_{i=1}^k
\frac{e^{U_{\{\beta\}}(\bt^*)-U_{\{\alpha\}}(\bt^*)}}{\left(\alpha_i!\beta_i!
\right)^2}\prod_{i>j}^k\,
\frac{(\alpha_i-\alpha_j)^2(\beta_i-\beta_j)^2}{(\alpha_i+\alpha_j)(\beta_i+\beta_j)}
\prod_{i,j=1}^k \frac{1}{\alpha_i+\beta_j+1}
  \ee
  where summation ranges over all pairs of strict partitions $\alpha$
  and $\beta$ such that $\ell(\alpha)=\ell(\beta)$.

  One can convert the summations over the cones $\alpha_1>\cdots >\alpha_k\ge
  0$, $\beta_1>\cdots >\beta_k\ge 0$ to the summation over
  independent numbers $\alpha_i$ and $\beta_i$ as follows
   \be
Z^{(5)}(\bt^*)=1+\sum_{k=1}^\infty\frac{1}{k!}\sum_{\alpha_1,\dots,\alpha_k=0}^\infty
\sum_{\beta_1,\dots,\beta_k=0}^\infty\,\prod_{i=1}^k
\frac{e^{U_{\{\beta\}}(\bt^*)-U_{\{\alpha\}}(\bt^*)}}{\left(\alpha_i!\beta_i!
\right)^2}\prod_{i>j}^k\,
\frac{(\alpha_i-\alpha_j)(\beta_i-\beta_j)}{(\alpha_i+\alpha_j)(\beta_i+\beta_j)}
\prod_{i=1}^k \frac{1}{\alpha_i+\beta_i+1}
  \ee

\paragraph*{ (iv) Ensemble \eqref{measure-collection3}:
$W_\lambda(\bt,\bt',U(\bt^*))$}
 where for $U$ and $\bt=\bt'=\bt_\infty $ see \eqref{choice1}. It is defined on
 partitions $\lambda$ whose length do not exceed a given number $N$. Below
 $h_i=\lambda_i-i+N$.
 \be
W^{(6)}_\lambda(\bt^*) =\prod_{1\le i<j\le N} \,
\frac{(h_i-h_j)^2}{h_i+h_j}\,\prod_{i=1}^N\left( \frac{1}{h_i!}
\right)^2e^{-U_{h_i}(\bt^*)}
  \ee
\[
 Z^{(6)}(\bt^*)= \tau_N(\bt,U,A)=
\sum_{h_1>\cdots >h_N}\, \prod_{i<j} \,
\frac{(h_i-h_j)^2}{h_i+h_j}\,\prod_{i=1}^N\left( \frac{1}{h_i!}
\right)^2e^{-U_{h_i}}
 \]
 \be\label{Bures-density}
=\, \frac{1}{N!}\sum_{h_1,\cdots,h_N=0}^\infty\, \prod_{i<j} \,
 \frac{(h_i-h_j)^2}{h_i+h_j}\,\prod_{i=1}^N\left( \frac{1}{h_i!}
\right)^2e^{-U_{h_i}}
  \ee
Actually this choice of $A$ in (\ref{lBKP-tau-Schur}) is exactly
related to the Example \eqref{restricted-tau+NBKP U} which is the
tau function of mixed lBKP-sDKP tau function. Indeed if
$\,\bt'=\bt_\infty:=(1,0,0,\dots)\,$ then it is known that
 \[
Q_{\lambda}({\tfrac12\bt}_\infty)=
2^{\frac{\ell(\lambda)}{2}}\prod_{i=1}^k \frac
{1}{\lambda_i!}\prod^k_{i<j}\frac{\lambda_i-\lambda_j}
{\lambda_i+\lambda_j}
 \]
 see \eqref{stinftyB} in the Appendix.

In the spirit of discrete-continuous duality one may expected that
there is a continuous counterpart to \eqref{Bures-density}. Indeed
BKP-sDKP tau function may be chosen as follows
  \be\label{Bures-density-continues1}
\tau^{lBKP-sDKP}_N(\bt,\bt',U)=\langle
N+l,0|\,\g(\bt)\g'(\bt')\,e^{\int \,
\psi(x)\phi(x)d\mu(x)}\,\g(\bt)\g'(\bt')\,|l,0\rangle
  \ee
  \be
= \,\int\cdots\int\,\frac{(x_i-x_j)^2}{x_i+x_j}\prod_{i=1}^N\,
d\mu(x_i,l,\bt,{\bar \bt},\bt',{\bar \bt}')
  \ee
  where $\,l,\,$ is the Dirac sea level of lBKP vacuum vector $,\bt\,$ and
  $\,{\bar \bt}\,$ are higher times of the coupled lBKP, and $\,\bt',{\bar \bt}'\,$
  are higher times for coupled sBKP. These "times" play the role of deformation
  parameters:
 \[
d\mu(x,l,\bt,{\bar \bt},\bt',{\bar
\bt}')=\frac{x^l}{\sqrt{2}}\, \,\exp\,{\sum_{n=1}^\infty
\,(x^nt_n\,+\,x^{2n-1}t_{2n-1}'\,- \,x^{-n}{\bar
t}_n\,-\,x^{1-2n}{\bar t}_{2n-1}' ) }
  \]

Similar expressions appear in the study of random density matrices
and Bures densities \cite{OsipovSommers}.

\,

\paragraph*{ (v) Ensembles \eqref{measure-collection4}:
$W_\lambda(\bt,\bt',{\bar\bt}',U(\bt^*))$}
 where $\bt=\bt'={\bar\bt}'=\bt_\infty $. These ensembles are
 defined  on the set of all strict partitions
 $\alpha=(\alpha_1,\alpha_2,\dots)\in\DP$.

  Ensembles related to \eqref{measure-collection4} and their continues
  versions were considered in \cite{HLO} as examples of sBKP tau functions.  Ensembles
  $ W_{\alpha}(\bt',U)$ where $\bt'=\bt_\infty$ and $W_{\alpha}(\bt',{\bar\bt}',U)$
  with $\bt'={\bar \bt'}=\bt_\infty$ are written  respectively as
 \be\label{W7}
W^{(7)}(\bt^*)=\prod_{i<j} \,
 \frac{\alpha_i-\alpha_j}{\alpha_i+\alpha_j}\,\,\prod_{i=1}^n
 \,\frac{1}{\alpha_i!}\,
e^{-U_{\{\alpha_i\}}(\bt^*)}
 \ee
 \be\label{Z7} Z^{(7)}(\bt^*)=\,
 1+\sum_{n=1}^{\infty} \frac{1}{n!}\,\sum_{\alpha_1,\cdots,\alpha_n=0}^\infty\,
\prod_{i<j} \,
 \frac{\alpha_i-\alpha_j}{\alpha_i+\alpha_j}\,\,\prod_{i=1}^n
 \,\frac{1}{\alpha_i!}\,
e^{-U_{\{\alpha_i\}}(\bt^*)}
  \ee
  where
 \be\label{U-t*-sBKP}
  U_{\{\alpha_i\}}(\bt^*)=\sum_{i=1}^{\ell(\alpha)}\,U_{\alpha_i}(\bt^*)\,
  ,\quad
  U_n(\bt^*):=U_n^{(0)}+\sum_{i=1,3,\dots} n^it_i^*,\quad
  U_n^{(0)}=U_{-n}^{(0)}
 \ee
   and ensembles related to the shifted Schur measure introduced in
   \cite{TW-shifted}
   \be\label{W8}
  W^{(8)}(\bt^*)=\prod_{i<j} \,
 \left(\frac{\alpha_i-\alpha_j}{\alpha_i+\alpha_j}\right)^2\,
 \prod_{i=1}^n\,\left( \frac{1}{\alpha_i!}
\right)^2e^{-U_{\{\alpha_i\}}(\bt^*)}
 \ee
 \be\label{Z8}
Z^{(8)}(\bt^*)=\,1+\sum_{n=1}^{\infty} \frac{1}{n!}\,\sum_{
\alpha_1,\cdots,\alpha_n =0}^\infty \,\, \prod_{i<j} \,
 \left(\frac{\alpha_i-\alpha_j}{\alpha_i+\alpha_j}\right)^2\,\prod_{i=1}^n\,
 \left( \frac{1}{\alpha_i!} \right)^2e^{-U_{\{\alpha_i\}}(\bt^*)}
  \ee
The last expression may be identified with a KdV tau function
where $\bt^*$ are KdV higher times (compare with \cite{LS}).
Formula \eqref{Z7} is a sBKP tau function with respect to
variables $\bt^*$ introduced by \eqref{U-t*-sBKP}.

\paragraph*{ (vi)  Ensemble \eqref{measure-collection1}:
$W_\lambda(\bt,{\bar\bt},U(\bt^*))$ } where for $U$ and for
$\bt=\bt_\infty$ see \eqref{choice1}, and where
 \[
{\bar t}_m(x)=\frac 1m\,\sum_{i=1}^N\,x_i^m
  \]
The weight
 \be
W^{(9)}(\bt^*,x)=W_\lambda(\bt_\infty,{\bar\bt}(x),U(\bt^*))
 \ee
 is defined on partitions $\lambda$ whose
 length do not exceed a given number $N$. Below $h_i=\lambda_i-i+N$.

Then the normalization function is
 \be\label{Schur+source}
Z^{(9)}(\bt^*,x)=\sum_{h_1=0}^\infty \cdots \sum_{h_N=0}^\infty \,
(h_i-h_j)\,\prod_{i=1}^N\, \frac{1}{h_i!} \,
x_i^{h_i}\,e^{-U_{h_i}}
 \ee
 which may be considered as a discrete version of the so-called one-matrix model with a
 source:
   \[
\int\, e^{\sum_{m=1}^\infty\, Tr\, H^m t_m^* \,+\, Tr\, \Lambda
H}\, dH =\int_{\mathbb{R^N}}\,\Delta_N(z)\,\prod_{i=1}^N\,
e^{\sum\, z_i^m t_m^* + \sum_{i=1}^N\,\lambda_iz_i} dz_i
   \]
   where $z_i$ and $\lambda_i$ are respectively eigenvalues of the Hermitian matrix $H$
   and the normal matrix $\Lambda$.

\paragraph*{ (vii)  Ensemble \eqref{measure-collection1}:
$W_\lambda(\bt,{\bar\bt},U(\bt^*))$ } where for $U$ and for
$\bt=\bt(\infty,q)$, see \eqref{choice2}, and where
 \[
{\bar t}_m(x)=\frac 1m\,\sum_{i=1}^N\,x_i^m
  \]
The weight
 \be
W^{(10)}(\bt^*,x)=W_\lambda(\bt(q),{\bar\bt}(x),U(\bt^*))
 \ee
 is defined on partitions $\lambda$ whose
 length do not exceed a given number $N$. Below $h_i=\lambda_i-i+N$.

Then the normalization function is
 \be\label{Schur+source'}
Z^{(10)}(\bt^*,x)=\sum_{h_1=0}^\infty \cdots \sum_{h_N=0}^\infty
\, (q^{h_i}-q^{h_j})\,\prod_{i=1}^N\, \frac{1}{(q;q)_{h_i}} \,
x_i^{h_i}\,e^{-U_{h_i}(\bt^*)}
 \ee
 which in case $q\in S^1$ may be considered as a discrete version of models of a random
 unitary matrices with a source:
   \[
\oint\, e^{\sum_{m=-\infty}^\infty\, Tr\, U^m t_m^* \,+\, Tr\,
\Lambda U}\, dU =\oint \,\Delta_N(z)\,\prod_{i=1}^N\,
e^{\sum_{m\in\mathbb{Z}}\, e^{mz_i} t_m^* +
\sum_{i=1}^N\,\lambda_iz_i} \frac{dz_i}{z_i}
   \]
   where $e^{z_i}$ and $\lambda_i$ are respectively eigenvalues of the unitary matrix
   $U$ and the source $\Lambda$.

\paragraph*{ (viii) Ensemble \eqref{measure-collection3}:
$W_\lambda(\bt,\bt',U(\bt^*))$} where $\bt'=\bt_\infty $ and where
   \[
{ t}_m(x)=\frac 1m\,\sum_{i=1}^N\,x_i^m
  \]

The weight
 \be
W^{(11)}(\bt^*,x)=W_\lambda(\bt(x),\bt_\infty,U(\bt^*))
 \ee
 is defined on partitions $\lambda$ whose length do not exceed a given number $N$.
 Below $h_i=\lambda_i-i+N$.

The normalization function is
 \be\label{Q+source}
  Z^{(11)}(\bt^*,x)=\sum_{h_1,\dots,h_N=0}^\infty
\, \frac{h_i-h_j}{h_i+h_j}\,\prod_{i=1}^N\, \frac{1}{h_i!} \,
x_i^{h_i}\,e^{-U_{h_i}}
 \ee
which is a discrete version version of the following 2-sBKP tau
function \cite{HLO}.
  \[
Z^{(10)}=\int_{\mathbb{R}^N}\,
\frac{z_i-z_j}{z_i+z_j}\,\prod_{i=1}^N\, \,
\,e^{\sum_{m=1,3,\dots}^\infty \, \left(t_m^* z_i^m +{\bar t}_m^*
z_i^{-m}\right)+ \lambda_i z_i}\,d\mu(z_i)
  \]

\paragraph*{ (ix) Ensembles (\ref{weight-beta=4}). Discrete $\beta=4$ ensemble: }
Now take $W_\lambda(\bt,U(\bt^*))$ of (\ref{weight-beta=4}) where
we recall
$\lambda=(\lambda_1,\lambda_1,\dots,\lambda_n,\lambda_n)$ and
$\bt$ is either $\bt_\infty$ or $\bt(q)$. In the first case,
$\bt=\bt_\infty$, we have
 \be\label{symplectic-discrt}
W^{(12)}_{\lambda}(\bt^*)={\tilde\Delta}_n(x)^4\,
\prod_{i=1}^n\frac{e^{-U_{x_i}(\bt^*)-U_{x_i-1}(\bt^*)}}{x_i!(x_i-1)!}
 \ee
 where the set
 \be
x_i\ :=\,\lambda_i -2i +2n\,,\quad i=1,\dots,n
 \ee
 \be
{\tilde\Delta}_n(x)^4\,:=\,\prod_{1\le i<j\le
n}(x_i-x_j)^2\left((x_i-x_j)^2 -1 \right)
 \ee
 The grand partition function is a discrete version of grand
 partition function of the symplectic ensemble
 \be
Z^{(12)}=\sum_{n=0}^\infty \,\frac 1{n!}\,\sum_{x_1,\dots,x_n}
{\tilde\Delta}_n(x)^4\,
\prod_{i=1}^n\,\frac{e^{-U_{x_i}(\bt^*)-U_{x_i-1}(\bt^*)}}{x_i!(x_i-1)!}
 \ee
In the second case, $\bt=\bt(q)$, we have
 \be\label{symplectic-discrt'}
W^{(12)}_{\lambda}(\bt^*)={\tilde\Delta}_n(q^x)^4\,
\prod_{i=1}^n\frac{e^{-U_{x_i}(\bt^*)-U_{x_i-1}(\bt^*)}}{(q;q)_{x_i}(q;q)_{x_i-1}}
 \ee
 where the set of $\{x_i \}$ is the same as before while
 \be
{\tilde\Delta}_n(q^x)^4\,:=\,q^{-1}\prod_{1\le i<j\le
n}(q^{x_i}-q^{x_j})^2(q^{x_i-1}-q^{x_j})(q^{x_i}-q^{x_j-1})
 \ee
 The grand partition function is the grand partition function for
 a the following $\beta=4$ ensemble
 \be
Z^{(12)}=\sum_{n=0}^\infty \frac 1{n!}\sum_{x_1,\dots,x_n}
{\tilde\Delta}_n(q^x)^4\,
\prod_{i=1}^n\frac{e^{-U_{x_i}(\bt^*)-U_{x_i-1}(\bt^*)}}{(q;q)_{x_i}(q;q)_{x_i-1}}
 \ee

\section{Interlinks between tau functions of BKP \label{interlinks-section}}

Here we will show that for $\bt^{(i)}$ be one of  \eqref{t-infty}-\eqref{t(a;q)} (see also Appendix \ref{evaluated-section}) .
 and for
$U=U(\bt^*)$ be specified later there exist relations between
different lBKP tau functions as follows:
 \be\label{interlink}
\l l'|\Gamma(\bt^{(i)})\mathbb{T}(\bt^*)\,g\, |l\r = \l
l'-l|\Gamma(\bt^*_+)\, g^*_i\,\Gamma( {\bt}^*_-)\,|0\r
 \ee
where  $g^*_i$ is constructed in terms of a given $g$ according to a
choice of $\bt^{(i)}$, and where $\bt^*_+$ is the collection $\{
t^*_m,\, m>0\}$ while  $\bt^*_-$ is the collection $\{ t^*_m,\,
m<0\}$.

Here, in case $\bt^{(i)}$ is chosen either via
\eqref{choicetinfty'} or via \eqref{choicet(a)'}, the higher times
$\bt^*$ are related to variables $U$ by \eqref{U_n(bt^*,{bar bt^*})}, namely
 \[
U_n=U_n(\bt^*,{\bar \bt^*})=U_n^{(0)}+\sum_{m=1}^\infty\,\left(\frac{an+b}{cn+d}\right)^mt^*_m+
\, t_0^*\,\ln\,\left(\frac{an+b}{cn+d}\right)\,-\,\sum_{m=1}^\infty\, \left(\frac{an+b}{cn+d}\right)^{-m} {\bar t}^*_{m}
 \]
where $a,b,c,d$ are parameters chosen arbitrary in a way we have no singular tems in the the above sum. In particular, if
we do not want to have dependence on $\bt^*_-$ parameters the $U$-dependence may be chosen as 
 \be\label{choice1interlink}
  U_n=U_n(\bt^*,c)=U_n^{(0)} -\sum_{m > 0} \, n^m t^*_m 
 \ee
as it was done in, say, \cite{ABW}, or in \cite{NOk}, for the different case, for a TL tau function.
 while 

In case \eqref{t(q)}, or \eqref{t(a;q)} by
\be\label{choice2interlink-abcd}
   U_n=U_n(\bt^*;q,a,b,c,d)=U_n^{(0)} -\sum_{m > 0} \, \left(\frac{aq^n+b}{cq^n+d} \right)
   ^m{ t}^*_m -\sum_{m < 0} \, \left(\frac{aq^n+b}{cq^n+d} \right)^m
   { t}^*_m - t^*_0\,\ln \left(\frac{aq^n+b}{cq^n+d} \right)
 \ee
In particular case we may take

 \be\label{choice2interlink}
   U_n=U_n(\bt^*;q)=U_n^{(0)}-\sum_{m > 0} \, q^{nm}{
   t}^*_m 
 \ee
as in \cite{OR} or \cite{NakTak}.

Now we equate a lBKP tau functions depending on parameters $U$ and thus depending on the parameters $\bt^*$ to a 
certain multisoliton 2-lBKP 
tau functions where $\bt^*$ play the role of higher times. Thus we present a sort of duality between the $U$ and
$\bt$ variables\footnote{Here we convert an observation of
\cite{hypsol} to the case of lBKP hierarchy.}. This link may be
compared with \cite{NakTak} (see also references there).

Actually this section explains "dualities" between discrete and
continuous expressions for tau functions, see subsection
\ref{sums-section}. Such dualities are also considered in
\cite{HO-convol}.

Let us choose (\ref{choice1}). Now, in Frobenius  notation
$\lambda=(p_1,\dots,p_k|-q_1-1,\dots,-q_k-1)$,  we have
\eqref{stinfty}:
 \be\label{schur-vand2} s_\lambda(p,0,0,\dots)=
 \prod_{n < m\le k}\frac{(p_n-p_m)(q_n-q_m)}{(p_n-q_m)(p_m-q_n)} \left(\prod_{n=1}^k \,
\frac{(p_n-q_n)^{-1}}{ \Gamma(p_n+1)\Gamma(-q_n)}\right)
  \ee
  (where numbers $p_n$ and $q_n$ are related to the Frobenius
  coordinates as $p_n=\alpha_n$ and $q_n=-\beta_n-1$
  \footnote{or, more generally one can take
 $
 p_n=\frac{a\alpha_n+b}{c\alpha_n+d},\,q_n=\frac{a(-\beta_n-1)+b}{c(-\beta_n-1)+d}$ with
 arbitrary $a,b,c,d$ which keeps linear fraction factor in (\ref{schur-vand2})
 \cite{hypsol}.}). Written in this form the Schur function may be interpreted as a
 familiar formula for shift of solitons due to interaction where factor in
 bracket is irrelevant and may be included to the choice of initial position of solitons. Therefore
 \[
\tau(\bt,U(\bt^*))=\sum_{\lambda\in
  \Pa}e^{-U_\lambda(\bt^*)}s_\lambda(\bt)=
  \]
\be\label{soliton1}
  \sum_{k=0} \sum_{\{p,q\}_k}\prod_{n < m\le k}
  \frac{(p_n-p_m)(q_n-q_m)}{(p_n-q_m)(p_m-q_n)} \left(\prod_{n=1}^k \,
\frac{(p_n-q_n)^{-1}}{ \Gamma(p_n+1)\Gamma(-q_n)}\right)e^{
\sum_{n=1}^\infty { t}^*_n \left(p_j^n-q_j^n\right)}
  \ee
  where $\{p,q\}_k$ means summation over sets of integers
  $p_1>\cdots > p_k\ge 0>q_k>\cdots >q_1$
which may be interpreted as a  lDKP multisoliton tau function,
$\tau^*(\bt^*)$, where higher times, $ \bt^*$, are related to
$U=U( \bt^*)$ of original lBKP tau function. The
fermionic expression for this multisoliton tau function is
 \be\label{equality}
\tau^*(\bt^*):=\langle 0| \,e^{J( \bt^*)}\,g^{-}({\bar A},{\bf
p})\,g^{+}({\bar B},{\bf q})\, |0\rangle
=\tau(\bt,U)
 \ee
 where operators $ g^{-}({\bar A},{\bf
p}), g^{+}({\bar B},{\bf q})$ coincide respectively with $
g^{-}({\bar A}), g^{+}({\bar B})$ if in the last group we replace each
fermionic Fourier mode $\psi_n,\psi^\dag_n$ respectively by
$\psi(p_n),\psi^\dag(q_n)$:

\section{Partition functions for certain random processes \label{partition for
random-section}}

A number of random processes may be described via a sort of
partition function $Z$ as follows. We have an un-normalized weight
$W_{\mu\to\nu}(\t)$ for a transition from a state $\mu$ to a state
$\lambda$ during a time interval $\t$. To get the probability for
this transition, $p_{\mu\to\nu}(\t)$, we divide this weight by the
normalization function $Z_\mu(\t)$ which is equal to the sum of
weights to get any state during elapse of time starting from the
state $\mu$, then,
 \[
p_{\mu\to\lambda}(\t)=\frac{1}{Z_\mu(\t)}W_{\mu\to\lambda}(\t),
\quad Z_\mu(\t)=\sum_{\lambda}W_{\mu\to\lambda}(\t)
 \]
 where sum runs over all possible states $\lambda$ which may be achieved
 from initial state $\mu$ during a time interval $\texttt{T}$. The
 normalization function $Z$ provides the condition that
 the sum of probabilities is equal to unity.

 In case the state of a system may be identified with basis Fock
 vectors $\langle \mu|$ and the transition weight $W_{\mu\to\lambda}(\texttt{T})$ may be
 written as a matrix element of some given operator $o(\texttt{T})$ which acts in Fock
 space:
  \[
W_{\mu\to\lambda}(\t):=\langle \mu|o(\t)|\lambda\rangle
  \]
  then, to get the partition function $Z_\mu(\t)$ we
  need to evaluate
   \[
Z_\mu(\t)=\langle \mu|o(\t)|\Omega_0\rangle
   \]

Examples include random turn vicious walkers model \cite{F} (model
(B) in section 4) which may be treated as a modification of
  `	exclusion processes. Versions of this model were considered in
\cite{BF}, \cite{BaikRains-involution},\cite{Baik}. Relations to

\cite{AvMSchur} (section 4.1), see also \cite{Forr1}. In these
version walkers move in one direction, then

Recently, we found a mentioning of the fermionic approach to this
problem in the paper by Okounkov \cite{Ok-uses} (Section 1.4.4.).
Below we present an explicit evaluation of the probability to get
a given final position of the walkers in the simplest version of
this model where we shall use lDKP tau function \eqref{the
simplest}.

   {\bf Example 1}.
We are interested in creating a Young diagram (YD) $\lambda$ by
gluing box by box in a way that each intermediate figure is a
Young diagram. It means that at each step we can glue a box only
to a certain number of admissible places on the boundary of a
diagram. A consequence of Young diagrams may be called the path
connecting initial and final Young diagrams. The number of YD
along a path will be called the length of the path. For simplicity
we  take an empty YD (YD without nodes) as the initial one. Now
one can address few questions: (1) what is the number of paths of
length $\t$ starting at empty YD and ending at a given YD
$\lambda$, (this number will be denoted by $W_{0\to\lambda}(\t)$).
(2) what is the number of paths starting at empty  YD (empty
state relates to $\mu=0$) of a length $\t$ (this number will be
denoted by $Z_0(\t)$). If we consider this creation of a YD as a
random process describing the gluing of the boxes, then the
probability $p_{0\to\lambda}(\t)$ to achieve in $\t$ steps a given
configuration $\lambda$ is defined as the ratio of these numbers:
 \[
p_{0\to\lambda}(\t)=\frac{W_{0\to\lambda}(\t)}{Z_0(\t)}
 \]

   Let the initial state be given by the vacuum vector $\langle 0|$ and
   $o(\t)=J_{1}^\t$, $\t=0,1,2,\dots$ is a discrete time. Then the random process
   describes the process of creating a Young diagram by gluing boxes to a boundary of
   Young diagram in a way that at each step we get a Young diagram. A single action
   action of  $J_1$ on $\lambda$ glues one box uniformly to any admissible  place of the
   Young diagram of $\lambda$ ("admissible place" means that a gluing here a box we get
   a figure which is Young diagram again). $J_1^\t$ glues $\texttt{T}$ boxes one by one.
   Now
    \be\label{W1'}
W_{0\to\lambda}(\t)=\langle 0|J_1^\t|\lambda\rangle
    \ee
    is an integer which is equal to the number of ways
    to create a Young diagram of a given shape $\lambda$ by
    gluing boxes one by one in a way that each intermediate figure
    is a Young diagram. It is clear that the weight of the
    partition is equal to the duration of time:
     \be\label{T=weight}
T=|\lambda|
     \ee
     otherwise the transition weight vanishes.

    The partition function (normalization function) is the sum of all these numbers over
    final  states (Young diagrams) $\lambda$:
    \[
Z_{0}(\t)=\langle 0|J_1^\t|\Omega_0\rangle
    \]

KP tau function \eqref{KP-schur} evaluated at
$\bt=t_1\bt_\infty:=(t_1,0,0,\dots)$ generates all
$\{W_{0\to\lambda}(\t),\t=0,1,\dots\}$:
 \[
\tau^{KP}(t_1\bt_\infty)=\langle 0|e^{t_1J_1}
|\lambda\rangle=\sum_{\t}\, \frac{t_1^\t}{\t !} \langle 0|J_1^\t
|\lambda\rangle=\sum_{\t}\, \frac{t_1^\t}{\t !}
W_{0\to\lambda}(\t)
 \]
 Let us notice that $W_{0\to\lambda}(\t)=0$ in case $\t \neq |\lambda|$ vanish. Taking
 into account \eqref{KP-schur} we obtain
  \be
W_{0\to\lambda}(\t)=\t !
s_\lambda(\bt_\infty)\delta_{\t,|\lambda|}
  \ee
  where  $\delta$ is the Kronecker symbol.

 lDKP tau   function (\ref{the simplest}) evaluated at
 $\bt=t_1\bt_\infty:=(t_1,0,0,\dots)$ generates all
    $\{Z(\t):=Z_0(\t),\t=0,1,\dots\}$:
     \be
\tau_o(t_1\bt_\infty)=\langle 0|e^{t_1J_1}
|\Omega\rangle=\sum_{\t\ge 0}\, \frac{t_1^\t}{\t !}\sum_\lambda
\,W_{0\to\lambda}(\t)=\sum_{\t\ge 0}\, \frac{t_1^\t}{\t !}Z_0(\t)
     \ee
On the other hand thanks to the right-hand side of \eqref{the
simplest} we have $\tau_o(t_1\bt_\infty)=e^{\frac 12 t_1^2+t_1}$
which in turn is equal to
$\sum_{\t=0}^\infty\,t_1^\t\,s_{(\t)}(\bt_2)$ where $\bt_2:=(
1,\frac 12,0,0,\dots)$ and where $s_{(n)}$ denotes the elementary
Schur function known also
 as $n$-th completely symmetric function \cite{Mac}. Therefore we obtain
 \be
Z_0(\t)=\t ! s_{(\t)}(\bt_2)=\sum_{n=0}^{\left[\frac \t 2
\right]}\,\frac{\t ! 2^{2n-\t}}{ n!(\t -2n)!}
 \ee
Via saddle point method we find that in $\t\to\infty$ limit the
main contribution in the sum over $n$ is due to
$n\approx\frac{\t}{2}-\frac 14 \sqrt{\frac{\t}{2}}$ which yields
for large $\t$
 \[
Z_0(\t)\, = \, e^{\frac{\t}{2}\log\,{\t}+\frac {\t}{2}(\log
2-1)-\frac 14 \sqrt{\t}\log \t +O(\sqrt{\t})}
 \]

  At last we obtain
 the answer for the probability to achieve a configuration
 $\lambda$ in $\t$ steps:
 \be\label{standard-tableaux-number}
p_{0\to\lambda}(\t)=
\frac{s_\lambda(\bt_\infty)}{s_{(\t)}(\bt_2)}\delta_{\t,|\lambda|},\quad
\bt_\infty:=(1,0,0,\dots),\quad \bt_2:=(1,\frac 12,0,0,0,\dots)
 \ee
 As one can see the probability to achieve the state $\lambda=(\t)$ is given by
 \[
p_{0\to (\t)}(\t)=\frac{1}{Z_0(\t)}
 \]

One can ask, given \t, what is the configuration $\bar{\lambda}$
which maximizes the number $W_{0\to\lambda}(\t)$ and thereby the
probability $p_{0\to\lambda}(\t)$.  The answer is known \cite{HO}
and is given by Kerov-Vershik formula for the so-called limit
shape of Young diagram, see \cite{Kerov-Vershik,Kerov-Vershik-2}
\footnote{The reason of this coincidence is the following. The
Vershik-Kerov limit shape YD maximizes Plancheral measure on
partitions, and Plancheral measure is basically
$\left(s_\lambda(\bt_\infty)\right)^2$.}.

  If we modify our random process and admit both creation and
 elimination of a box we obtain basically the same shape of
 $\bar{\lambda}$ however now the weight of this configuration will be less than $\t$
 \cite{HO}:

 {\bf Example 2}.  Starting from the vacuum zero Young diagram,
 at each time step we either add or remove a
 box at random in a way that a figure we obtain during at each time step is a Young
 diagram, see the figure below. This model is equivalent is a model of random turn
 walk suggested in \cite{F} where initial configuration of walkers
 is the step function.

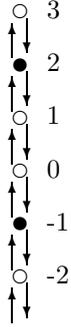
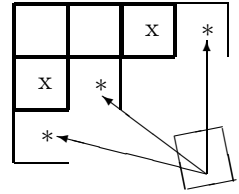
\begin{picture}(100,170)
 \put(37,30){\vector(0,1){13}}\put(42.5,45){\vector(0,-1){13}}
 \put(37,50){\vector(0,1){13}}\put(42.5,65){\vector(0,-1){13}}
 \put(37,70){\vector(0,1){13}}\put(42.5,85){\vector(0,-1){13}}
 \put(37,90){\vector(0,1){13}}\put(42.5,105){\vector(0,-1){13}}
 \put(37,110){\vector(0,1){13}}\put(42.5,125){\vector(0,-1){13}}
\put(37,130){\vector(0,1){13}}\put(42.5,145){\vector(0,-1){13}}

  \put(40,47.5){\circle{5}}
   \put(40,67.5){\circle*{5}}
 \put(40,87.5){\circle{5}}
  \put(40,107.5){\circle{5}}
   \put(40,127.5){\circle*{5}}
    \put(40,147.5){\circle{5}}
 \put(50,45){-2}
 \put(50,65){-1}
 \put(50,85){0}
 \put(50,105){1}
 \put(50,125){2}
 \put(50,145){3}

  \put(0,00){1.  Random turn walk}
 \put(0,-10){of particles}
 \put(0,-20){on a Maya diagram.}

\end{picture}
\begin{picture}(100,170)

 \put(149.7,50){\line(0,1){60}}\put(150.4,70){\line(0,1){40}}

 \put(169.7,70){\line(0,1){40}}\put(170.4,70){\line(0,1){40}}

 \put(189.7,70){\line(0,1){40}}\put(190.4,90){\line(0,1){20}}

 \put(209.7,90){\line(0,1){20}}\put(210.4,90){\line(0,1){20}}

 \put(230,90){\line(0,1){20}}

\put(150,50){\line(1,0){20}}

 \put(150,69.7){\line(1,0){20}}\put(150,70.4){\line(1,0){20}}

\put(150,109.7){\line(1,0){60}}\put(150,110.4){\line(1,0){80}}

\put(150,89.7){\line(1,0){60}}\put(150,90.4){\line(1,0){60}}

\put(210,60){\line(5,1){18}}\put(215,40){\line(5,1){18}}
\put(210,60){\line(1,-5){4}}\put(228,63){\line(1,-5){4}}

 \put(222,46){\vector(-4,3){39}}
\put(222,46){\vector(-4,1){56}} \put(222,46){\vector(0,3){50}}

\put(180,74){*} \put(160,55){*}\put(220,95){*}

\put(159,78){x}\put(199,98){x}

 \put(70,00){2.   Random adding/removing a box to a Young diagram}
 \put(70,-10){related to the up/downward hops of particles on Maya}
 \put(70,-20){diagram. At unit time instant either a box has to be added}
 \put(70,-30){at any  of vacant places marked by star, or a box marked}
 \put(70,-40){by x has to be removed.}

\end{picture}

{\vskip 40 pt}

\[
\]

 This model describes hard core particles ("walkers", "hard core"
 means that two particles can not occupy the same site) situated at the sites of 1D
 lattice. The model implies that at each tick of clock one chosen at random particle
 hops either to the left or to the right. In our picture particles are
 fermions which we placed on the vertical lattice. The step
 function is the vacuum state, $\langle 0|$, describing the Dirac
 sea where all sites downward to the sea level are occupied.
 We count particles from the top. Excitations may be described by partitions
 $\lambda=(\lambda_1,\lambda_2,\dots)$ where $\lambda_1$ describes
 the upward shift of the up-most particles (the particle number one) with respect to its
 original position in the Dirac seas, the shift of the  particle number $i$ is equal to
 $\lambda_i$. It is clear that $\lambda_1\ge\lambda_2\cdots$.
 Such configuration is denoted by $\langle \lambda|$. Each upward
 step of a particle from a configuration $\lambda$ may be described as gluing a box
 to the Young diagram $\lambda$, while each downward step is
 described as removing one box from the Young diagram of $\lambda$.

 The number of paths of a length $\t$ which start at the vacuum
 configuration and end at a given configuration $\lambda$ divided
 by the number of all paths of the length $\t$ which start at the
 vacuum configuration defines the transition probability
 $p_{0\to\lambda}(\t)$.

 Now the number of ways to achieve a given configuration $\lambda$
 during a lapse of time $\t$ starting from the vacuum (step function)
 configuration is
\be\label{W2'} W_{0\to\lambda}(\t)=\langle
0|(J_1+J_{-1})^\t|\lambda\rangle
    \ee
  where $\t$ is not necessarily equal to $|\lambda|$.

  The number of ways to achieve any configuration in $\t$ time
  steps starting from the vacuum configuration is
    \[
Z_{0}(\t)=\langle 0|(J_1+J_{-1})^\t|\Omega_0\rangle
    \]
   A usage of Baker-Campbell-Hausdorff formula
   $e^{tJ_1+tJ_{-1}}=e^{\frac12 t^2}e^{tJ_{-1}}e^{tJ_1}$
   may be considered as an advantage of the fermionic approach. After some algebra we
   obtain \cite{HO}
   \bp \em We have
\be\label{time-duration} W_{0\to\lambda}(\t)={\t}!\cdot
s_{\lambda}({\bt}_\infty)\,\delta(\t,|\lambda|),\quad
\delta(\t,|\lambda|):=2^{ \frac{|\lambda|-{\t}}{2} }
\frac{1}{(\frac{{\t}-|\lambda|}{2})!}
 \ee
 where $\delta(\t,|\lambda|)$ replaces the Kronecker symbol
 $\delta_{\t,|\lambda|}$. In \eqref{time-duration} $\t -|\lambda|$ is an even
 number otherwise $W_{0\to\lambda}(\t)$ vanishes.
  \ep

The following lDKP tau function generates partition functions
    \be
\langle 0|e^{tJ_1+tJ_{-1}} |\Omega\rangle=\sum_{\t\ge 0}\,
\frac{t^\t}{\t !}\sum_\lambda \,W_{0\to\lambda}(\t)=\sum_{\t\ge
0}\, \frac{t^\t}{\t !}Z(\t)
     \ee
     On the other hand
     \be
\langle 0|e^{tJ_1+tJ_{-1}} |\Omega\rangle=e^{\frac 12 t^2}\langle
0|e^{tJ_1} |\Omega\rangle=e^{t^2+t}=\sum_{\t =0}^{\infty}\,
t^{\t}s_{(\t)}(\bt_2')
     \ee
     where $\bt_2':=(1,1,0,0,\dots)$.
Therefore we get
 \bp \em The number of paths of the length $\t$ which start at the vacuum configuration
 is given by
   \be
Z_0(\t)=\t ! s_{(\t)}(\bt_2'),\quad \bt_2':=(1,1,0,0,\dots)
   \ee
   where $s_{(\t)}(\bt_2')$ is the elementary Schur function
   related to the partition $(\t)$:
   \be\label{schur(t_2')}
s_{(\t)}(\bt_2')=\sum_{n=0}^{\left[\frac \t 2
\right]}\,\frac{1}{n!(\t -2n)!}
 \ee
 The number of paths of a length $\t$ which start at the vacuum
 configuration and end at a given configuration $\lambda$ divided
 by the number of all paths of the length $\t$ which start at the
 vacuum configuration is given by
\be
p_{0\to\lambda}(\t)=\frac{s_\lambda(\bt_1)}{s_{(\t)}(\bt_2')}\delta(\t,|\lambda|),\quad
\bt_1:=(1,0,0,\dots),\quad \bt_2':=\left(1,1,0,0,0,\dots \right)
 \ee
 \ep

As one can see
 \[
p_{0\to (\t)}(\t)=\frac{1}{Z_0(\t)}
 \]

For large $\t$ one may apply the saddle point method to evaluate
the sum \eqref{schur(t_2')}. The saddle point is related to
$n=\frac \t 2-\frac 12 \sqrt{\frac\t 2}+O(1)$. This yields
 \bp \em
 In large $\t$ limit we obtain
  \be\label{}
Z_0(\t)=e^{\frac{\t}{2}\log \t +\frac{\t}{2}\log
\frac{2}{e}+O(\sqrt{\t})}
  \ee
  \be\label{}
p_{0\to \lambda}(\t)=s_\lambda(\bt_1)e^{-\frac{\t}{2}\log
{2}+O(\sqrt{\t})}
  \ee
  \ep

\paragraph{'One dimensional dimer' target configurations}
At last let us focus on the following problem. Let us evaluate the number of paths of a given length $\t$ 
which end on a configuration related to a fat partition $\lambda\cup\lambda$ : 
 \be
N_{\FP}(\t)\,:=\,\sum_{\lambda\in\Pa} \,W_{0\to \lambda\cup\lambda}(\t)
 \ee 
The number $N_{\FP}(\t)$ vanishes in case $\t$ is odd.
Let us recall that each configuration $\lambda\cup\lambda$ describes a configuration of pairs of particles
('one dimensional dimers').
\bp \em
\be
N_{\FP}(\t)=\begin{cases} 2^{\frac12\t}(\t-1)!!   &\mbox{ iff } \t \mbox{ is even }\\
0  &  \mbox{ iff } \t \mbox{ is   odd }
\end{cases}
\ee
\ep
Indeed, on the one hand
    \be
\sum_{\lambda\in\Pa}\, \l 0|e^{tJ_1+tJ_{-1}} |\lambda\cup\lambda\r = \sum_{\t\ge 0}\,
\frac{t^\t}{\t !}\sum_\lambda \,W_{0\to\lambda\cup\lambda}(\t)=\sum_{\t\ge
0}\, \frac{t^\t}{\t !}N_{\FP}(\t)
     \ee
     On the other hand (see \eqref{SchurSum1})
     \be
\sum_{\lambda\in\Pa}\, \l 0|e^{tJ_1+tJ_{-1}} |\lambda\cup\lambda\r =e^{\frac 12 t^2}
\sum_{\lambda\in\Pa}\, \l 0|e^{tJ_1} |\lambda\cup\lambda\r =e^{t^2}=\sum_{\t =0,2,4,\dots}\,
\frac{t^{\t}}{\left(\frac 12 \t\right)!}
     \ee

\section*{Acknowledgements}
We are grateful to John Harnad, Johan van de Leur, Vladimir Osipov
and  and most of all to Eugene Kanzieper for discussions of the
topic. One of the authors (A.O) thanks E. Kanzieper and Holon Technology Institute for
hospitality (July 2008) where a part of this work (a main part of
the section "Asymmetric two-matrix ensemble" in \cite{OST-II}) was done. We thank Andrei
Mironov for the reference \cite{Mironov}. The work was supported by RFBR grants .... and by Japanese-RFBR
grant 10-01-92104 JF and also by RAS Program "Fundamental Methods
in Nonlinear Dynamics".
This work is also partly supported by Grant-in-Aid for Scientific Research
No.~22540186 from the Japan Society for the Promotion
of Science and by the Bilateral Joint Project ``Integrable Systems,
Random Matrices, Algebraic Geometry and Geometric Invariants''
(2010--2011) of the Japan Society for the Promotion of Science and the
Russian Foundation for Basic Research.

\appendix
\section{Appendices}\subsection{Pfaffnians. Partitions. Schur functions \label{tools-section}}

\paragraph{(A) Pfaffians.} We need the notion of Pfaffian. If $A$ an
anti-symmetric matrix of an odd order its determinant vanishes.
For even order, say $k$, the following multilinear form in
$A_{ij},i<j\le k$
 \be\label{Pf'}
\Pf [A] :=\sum_\sigma
{\sgn(\sigma)}\,A_{\sigma(1),\sigma(2)}A_{\sigma(3),\sigma(4)}\cdots
A_{\sigma(k-1),\sigma(k)}
 \ee
where sum runs over all permutation restricted by
 \be
\sigma:\,\sigma(2i-1)<\sigma(2i),\quad\sigma(1)<\sigma(3)<\cdots<\sigma(k-1),
 \ee
 coincides with the square root of $\det A$ and is called the
 Pfaffian of $A$, see, for instance \cite{Mehta}. As one can see the Pfaffian  contains
 $1\cdot  3\cdot 5\cdot \cdots \cdot(k-1)=:(k-1)!!$ terms.

The following equality is known as Schur identity
\begin{equation}
\label{PfSchurA}
 \Pf\left(\left(
\frac{x_i-{x_j}}{x_i+{x_j}} \right)_{1\le i,j\le
2n}\right)=\Delta_{2n}^*(x)
\end{equation}
where
 \be
\Delta_{k}^*(x):= \prod_{1\le i<j\le k}\frac{x_i-{x_j}}{x_i+{x_j}}
 \ee
Let us mark that a special case of this relation is obtained if
$x_{2n}$ vanishes. In this case we write
 \be\label{PfSchur-odd}
\Pf(A)=\Delta^*_{2n-1}(x)
 \ee
 where $A$ is an antisymmetric $2n\times 2n$ matrix defined by
 \be\label{PfSchurA'}
A_{ij}=\begin{cases} \frac{x_i-{x_j}}{x_i+{x_j}} &{\mbox if}\quad
 {1<i<j < 2n} \\
 1 &{\mbox if}\quad i<j=2n,
 \end{cases}
 \ee

\paragraph{Hafinans} The {\em Hafnian} of  a symmetric matrix $A$ of even order $N=2n$
is defined as
 \be\label{Hf}
\Hf (A) :=\sum_\sigma
\,\,A_{\sigma(1),\sigma(2)}A_{\sigma(3),\sigma(4)}\cdots
A_{\sigma(2n-1),\sigma(2n)}
 \ee
where sum runs over all permutation restricted by
 \be
\sigma:\,\sigma(2i-1)<\sigma(2i),\quad\sigma(1)<\sigma(3)<\cdots<\sigma(2n-1),
 \ee
  As one can see the this sum  contains $1\cdot  3\cdot 5\cdot \cdots
  \cdot(2N-1)=:(2N-1)!!$ terms.

\br \em \label{Hfdiag} Let us note that entries on the diagonal of the
matrix $A$ does not contribute the sum \eqref{Hf}.

\er

The following equality was found in \cite{IKO}
\begin{equation}
\label{pfaffhaf}
 \Pf\left(\left(
\frac{x_i-{x_j}}{\left(x_i+{x_j}\right)^2} \right)_{1\le i,j\le
2n}\right)= \prod_{1\le i<j\le 2k}\frac{x_i-{x_j}}{x_i+{x_j}}
\mbox{Hf}\left(\left(\frac{1}{x_i+x_j}\right)_{1\le i,j\le
2n}\right)
\end{equation}
Another proof of this relation was presented in \cite{LOS}. Let us
mark that a special case of this relation is obtained if $x_{2n}$
vanishes. In this case we write
 \be\label{pfaffhaf-odd}
\Pf(B)=\Delta^*_{2n-1}(x)\Hf(C)=:\Delta^{**}_N(x)
 \ee
 where $B$ and $C$ are respectively antisymmetric and symmetric $2n\times 2n$
 matrices whose relevant entries (see Remark \ref{Hfdiag}) are given by
 \be\label{pfaffhafBC}
B_{ij}=\begin{cases} \frac{x_i-{x_j}}{(x_i+{x_j)^2}} &{\mbox
if}\quad
 {1\le i<j < 2n} \\
 \frac 1{x_i} &{\mbox if}\quad i<j=2n,
 \end{cases}
 \qquad
C_{ij}=\begin{cases} \frac 1{x_i+{x_j}} &{\mbox if}\quad
 {1\le i < j < 2n} \\
 \frac 1{x_i} &{\mbox if}\quad i<j=2n,
 \end{cases}
 \ee

\paragraph{(B) Partitions.} Polynomial functions in many variables, like the Schur
functions, are parameterized by partitions.

Let us remind that a partition of certain number $n$ is an ordered set of integers
$\lambda=(\lambda_1,\dots,\lambda_i)$ where $\lambda_1\ge \dots
\ge \lambda_i\ge 0$ such that
$n=\sum_{k=1}^i \lambda_i$. Then, $n$ is called the weight of $\lambda$ and
commonly denoted by $|\lambda|$), see \cite{Mac}. Integers
$\lambda_k$ are called parts of the partition $\lambda$. The number of
non-vanishing parts of $\lambda$ is called the length of $\lambda$
and will be denoted by $\ell(\lambda)$.

Almost everywhere throughout the paper we will denote partitions
by Greek characters.

Strictly ordered sets $\alpha=(\alpha_1,\dots,\alpha_i)$,
$\alpha_1>\dots>\alpha_i\ge 0$ are called the strict partitions,
see \cite{Mac}. In this paper we write  $\{\alpha\}$ to denote
strictly ordered sets $\alpha_1>\dots >\alpha_i$ where numbers
$\alpha_k$ are not necessarily positive.

We basically use notations adopted in \cite{Mac}.

\paragraph{Young diagrams} 

The {\em (Young) diagram} of a partition $\lambda$ is defined as
the set of points (or nodes) $(i,j) \in {\mathbb Z}^2$, such that
$1\le j \le \lambda_i$. Thus, it is a subset of a rectangular
array with $\ell(\lambda)$ rows and $\lambda_1$ columns. We denote
the diagram of $\lambda$ by the same symbol $\lambda$. For
example,
\begin{equation}\label{YD331}
\YD3310
\end{equation}
is the diagram of $(3,3,1)$. The weight of this partition is $7$,
the length is equal to $3$.

The partition whose diagram is obtained by the transposition of
the diagram $\lambda$ with respect to the main diagonal is called
the conjugated partition and denoted by $\lambda^t$.

In the Frobenius notations (see \cite{Mac}) we write
$\lambda=(\alpha_1,\dots,\alpha_k|\beta_1\dots,\beta_k)$ or just
$\lambda=(\alpha|\beta)$. For instance the partition \eqref{YD331} is written as $(2,1|2,0)$.  The partition
$\lambda^{tr}:=(\beta|\alpha)$ is called transposed to
$\lambda=(\alpha|\beta)$.

\paragraph{Hook polynomials, Pochhammer symbols}

The {\em product of hook lengths} $H_\lambda$ is defined as
\begin{equation}\label{Hlambda}
H_\lambda=\prod_{i,j\in \lambda}h_{ij},\quad h_{ij}= \lambda_i-i
+\lambda_j^t-j+1 \ ,
\end{equation}
where the product ranges over all nodes of the diagram of the
partition $\lambda$.

Given number $q$, the so-called hook polynomial $H_\lambda(q)$ is
defined as:
\begin{equation}\label{Hlambda(q)1}
H_\lambda(q)=\prod_{i,j\in\lambda}(1-q^{h_{ij}}),\quad h_{ij}=
\lambda_i-i +\lambda_j^t-j+1
\end{equation}
In what follows, we also need notations:
\begin{equation}\label{n(lambda)}
n(\lambda) :=\sum_{i=1}^k (i-1)\lambda_i \ ,
\end{equation}
\begin{equation}\label{Poch}
(a)_\lambda :=
(a)_{\lambda_1}(a-1)_{\lambda_2}\cdots(a-k+1)_{\lambda_k} \ ,
\quad (a)_m :=\frac{\Gamma(a+m)}{\Gamma(a)} \ ,
\end{equation}
\begin{equation}\label{Pochq}
(q^a;q)_\lambda :=
(q^a;q)_{\lambda_1}(q^{a-1};q)_{\lambda_2}\cdots(q^{a-k+1};q)_{\lambda_k}
\ , \quad (q^a;q)_m :=(1-q^a)\cdots(1-q^{a+m-1}) \ ,
\end{equation}
where $k=\ell(\lambda)$. We set $(a)_0=1$ and $(q^a;q)_0=1$.

Useful relations are
 \be\label{a-lambda-h}
(a)_\lambda=c_k(a)\,\prod_{i=1}^k\,(a)_{h_i+1-k},\quad
(q^a;q)_\lambda=c_k(a,q)\,\prod_{i=1}^k\,(q^a;q)_{h_i+1-k}
 \ee
 where
  \be\label{c-k-a-q}
c_k(a)=\prod_{i=1}^k\,(a-i)^{k-i}\, ,\quad
c_k(a,q)=\prod_{i=1}^k\,(1-q^{a-i})^{k-i}\, ,\quad
h_i=\lambda_i-i+k
  \ee

\paragraph{Schur functions.} 

We now consider a semi-infinite set of variables ${\bf
t}=(t_1,t_2,t_3,\dots)$. Given partition $\lambda$, the Schur
function $s_\lambda({\bf t})$ is defined by
\begin{equation}\label{Schurt}
s_\lambda({\bf t})=\det\bigl(h_{\lambda_i-i+j}({\bf
t})\bigr)_{1\le i,j\le \ell(\lambda)}\ ,\quad\hbox{where}\quad
\sum_{k=0}^\infty z^kh_k({\bf t}) = \exp\sum_{m=1}^\infty z^mt_m \
,
\end{equation}
and, for $k<0$, we put $h_k=0$ . The $h_k({\bf t})$ is called the
elementary Schur function.

There is another definition of the Schur function; it is the
following symmetric function in the different variables
$x:=x^{(n)}:=(x_1,\dots,x_n)$, where $n\ge\ell(\lambda)$:
\begin{equation}\label{detSc}
\underline
s_\lambda(x)=\frac{\det\bigl(x_i^{\lambda_j-j+n}\bigr)_{1\le
i,j\le n}} {\det\bigl(x_i^{n-j}\bigr)_{1\le i,j\le n}} \ ;
\end{equation}
for the zero partition one puts $\underline s_{\bf0}(x)=1$.
If
$$ 
{\bf t}={\bf t}(x^{(n)})=(t_1(x^{(n)}),t_2(x^{(n)}),\dots),\quad
t_m(x^{(n)})=\frac 1m \sum_{i=1}^n x_i^m,
$$
then definitions (\ref{Schurt}) and (\ref{detSc}) are equivalent
\cite{Mac}:
\begin{equation}
s_\lambda({\bf t}(x^{(n)}))=\underline s_\lambda(x^{(n)}).
\end{equation}
\br \em\label{Schurvanish} From definition (\ref{Schurt}) it follows
that $s_\lambda({\bf t}(x^{(n)}))=0$ if $\ell(\lambda)>n$.
 \er
 The
Schur functions $\underline s_\lambda(x_1,\dots,x_n)$, where
$\ell(\lambda)\le n$, form a basis in the space of symmetric
functions in $n$ variables. We use the underline in $\underline
s_\lambda$ only to distinguish the two definitions. If an $n\times
n$ matrix $X$ has eigenvalues $x_1$, \dots, $x_n$, we may denote
$\underline s_\lambda(x_1,\dots,x_n)$ by $s_\lambda(X)$, without
underline, since in this paper the Schur function with uppercase
argument is used only in this sense.

One of the wonderful results of Kyoto school is the formula
 \be\label{Schur=VEV}
s_\lambda({\bf t})=\l 0|\g(\bt)|\lambda \r
 \ee
 which may be obtained by the direct calculation using
 \eqref{Schurt} and
 $\g(\bt)\psi_n\g(\bt)^{-1}=\sum_{m=0}^\infty\psi_{n-m}h_m(\bt)$.

 We want to
mark out that apart of traditional notation $s_\lambda$ we shall
use also notation $s_{\{h\}}$ widely used in physical literature,
say for instance \cite{KazakovW}. Here $h_i=\lambda_i-i+N$ where
$N$ is the length of a partition $\lambda$.

\section{ Charged and neutral free fermions. \label{fermions-section}} For the charged fermions  we shall
 use notations and conventions adopted in \cite{JM}. In particular for
charged fermions we have
  \be\label{vacuum'}
\l 0|\psi^\dag_{-n-1}| =\l 0|\psi_{n}=0,\quad \psi_{-n-1}|0\rangle
=\psi^\dag_{n}|0\rangle=0,\quad n \ge  0
  \ee
  \be\label{anticom-psi''}
\psi^\dag_m\psi_n+\psi_n\psi^\dag_m=\delta_{nm}\, ,\ \quad
n\in\mathbb{Z}
  \ee
Such fermions are used to construct tau functions of KP, Toda
lattice (TL) and the large lDKP hierarchies.

For "small" BKP hierarchy and "small" BKP hierarchy coupled to
lDKP we need neutral fermions, $\{ \phi_n\}_{n\in {\bf Z} }$, see
\cite{JM}, defined by the following property:
\begin{equation}\label{canonical}
    \phi_n\phi_m +\phi_m\phi_n =(-1)^n\delta_{n,-m}\, ,\ \quad
n\in\mathbb{Z}
\end{equation}
 \be\label{phi-psi}
\phi_n\psi_m +\psi_m\phi_n = 0,\quad \phi_n\psi^\dag_m
+\psi^\dag_m\phi_n = 0
 \ee

 It results from \eqref{canonical} that $\left(\phi_0\right)^2=\frac12$.

The action of neutral fermions on vacuum states are defined
 by
 \be\label{phi-on-vacuum}
\phi_n|0\rangle =0, \qquad \langle 0|\phi_{-n}=0,\qquad  n<0,
  \ee
 \be  \label{zero-phi-on-vacuum} \phi_0|0\rangle=\frac{1}{\sqrt
2}|0\rangle, \qquad \langle 0|\phi_0=\frac{1}{\sqrt 2}\langle 0|
 \ee
 Note that the action of $\phi_0$ on the vacuum vectors is different from the action
 defined in \cite{JM}. This causes a modification in formulation of the Wick's
 relations, see below.
Here we follow \cite{KvdLbispec}, where the choice of the
corresponding Fock space is different from the suggested in
\cite{JM}. See also the Appendix in the arxiv version of \cite{LO}
for some details.

For $n,m\in\mathbb{Z}$ we have
 \[
\l 0|\psi_n\phi_m|0 \r=\l 0|\psi^\dag_n\phi_m|0 \r=0
 \]

\paragraph{}

The right vacuum vector of a charge $l$ is defined via
 \be
\label{l-right-vacuum} |l\r=
\begin{cases} \psi_{l-1}\cdots\psi_0|0\r &\mbox{ if }\, l>0 \\
\psi^\dag_{l}\cdots\psi^\dag_{-1}|0\r &\mbox{ if }\, l<0
\end{cases}
 \ee
The dual vector is defined as
 \be \label{l-left-vacuum} \l l|=
\begin{cases} \l 0|\psi^\dag_{0}\cdots\psi^\dag_{l-1} &\mbox{ if }\ l>0 \\
\l 0|\psi_{-1}\cdots\psi_{l} &\mbox{ if } l<0
\end{cases}
 \ee

\paragraph{Basis Fock vectors.}
We shall use the following notation
 \be\label{basic-lambda-l-Frob}
 |\lambda,l\rangle =(-)^{\beta_1+\cdots +\beta_k} \,\psi_{\alpha_1+l}\cdots
 \psi_{\alpha_{k}+l}\psi^\dag_{l-\beta_k-1}\cdots
 \psi^\dag_{l-\beta_1 -1} \,|l\rangle
  \ee
  where $\lambda=(\alpha_1,\dots,\alpha_k|\beta_1,\dots,\beta_k)$.
  Dual vectors may be defined via $\l \lambda',l'|\lambda,l\r=
  \delta_{\lambda,\lambda'}\delta_{ll'}$.
  The vector \eqref{basic-lambda-l-Frob} may be also written as
\be\label{basic-lambda-l-h}
 |\lambda,l\rangle \,= \,\psi_{h_1+l-N}\cdots
 \psi_{h_{N}+l-N} \,|l-N\rangle
  \ee
  where $\lambda=(\lambda_1,\dots,\lambda_N)$ and
   \be\label{shifted-parts}
h_i\,:=\,\lambda_i-i+N,\quad i=1,\dots,N
   \ee
   which are called shifted parts of $\lambda$.
   Here $N>\beta_1$
   .

\paragraph{ Wick's relations.} Let each of $w_i$ be a linear
combination of Fermi operators:
  \[
w_i=\sum_{m\in\mathbb{Z}}\,w^-_{im}\psi_m\,+\,
\sum_{m\in\mathbb{Z}}\,w^+_{im}\psi^\dag_m\,
+\,\sum_{m\in\mathbb{Z}}\,w_{im}\phi_m\, ,\quad i=1,\dots,n
 \]
  If $n$ is even
  \be\label{Wick}
\l l|w_1\cdots w_n |l\r=\Pf\left[ A \right]
  \ee
  where $A$ is $n$ by $n$ antisymmetric matrix with entries
 \[
 A_{ij}\, = \,\l l|w_i w_j|l\r\, ,\quad i<j
  \]

Now turn to the case where $n$ is odd. In case zero mode operator
$\phi_0$ is absent in series for $w_i$, namely each $w_{i0}=0$, we
have the standard situation where $\l l|w_1\cdots w_n |l\r $
vanishes\footnote{We exclude $\phi_0$ because we use Kac-de Leur
Fock space \cite{KvdLbispec} where $\l0|\phi_0|0\r=1$ non-vanishes,
see Appendix \ref{remarks-BKP-section}}.

\section{ Wick's rule and Pfaffian representations \label{pfaffian-Wick-section}}

From the fermionic expression for tau functions we
obtain its Pfaffian representations as a result of application of
the Wick theorem.

First we suppose that $g$ may be factorized into the product
$g=g^-g^+$ such that
 \be
g=g^-g^+:\quad \l l'|g^-=\l l'|,\quad g^+|l\r=|l\r\
 \ee

 Let each $w_i,\, i=1,\dots,2m$ is a linear combination
  \[
w_i=\sum_{n\in\mathbb{Z}}\,(\alpha_{in}\psi_n +
\beta_{in}\psi^\dag_n)\,+\,\sum_{n
\in\mathbb{Z}}\,\gamma_{in}\phi_n
  \]
If $n+m=2k$ by Wick theorem we have
 \be
\l l|w_1\cdots w_n \,g \, w_{n+1}\cdots w_{2m} |l\r =
\Pf[W_{ij}]|_{i,j=1,\dots,2k}
 \ee
 where $W$ is the $2k\times 2k$ antisymmetric block-structured matrix with entries
 given as follows
  \be
 W_{ij}=\l l|\,w_iw_i\,g\,| l\r, \quad W_{ip}=\l l|w_i\,g\,w_p|
 l\r, \quad W_{pq}=\l l|\,g\,w_pw_q| l\r
  \ee
  where $1\le i <j <p < q \le n$.

\section{ On the central extension in Lie algebra of $\Psi DO$\label{PsiDO}}

The central extension in the algebra of $\Psi DO$ may be chosen via the choice of nontrivial 2-cocycle in this algebra.
It is known that there two independent cocycles found in \cite{KrHes}:
 \[
\omega_1(A,B)=\,\res_x\res_{\partial_x}\, \left(\,A[\log\, x\,,\, B]\,\right)\,,\quad 
\omega_2(A,B)=\,\res_x\res_{\partial_x}\, \left(\,A[\log \,\partial_x\,,\, B]\,\right)
 \] 
In our case the choice of the cocycle should reproduce the Japanese cocycle in the algebra of infinite Jacobian matrices
\cite{JM} and it should be chosen as follows 
 \[
\omega(A,B)=\res_{x}\res_{D_x}\,\left(x^{-1} A[\log \,D_x\,,\,B]\,\right)\,, \quad A,B\in \Psi DO
 \] 
where the commutator of $log D_x$ with a $\Psi DO$ is defined via relations
 \[
[\,\ln \,D_x\,,\, f(D_x)\,]=0\,, \quad [\,\log\, D_x\,,\,f(x)\,]:=
\sum_{n=1}^\infty(-)^{n+1}\frac{\left(D_x^n\cdot f(x) \right)D_x^{-n}}{n}
 \] 
Up to coboundary term $\omega = \omega_1+\omega_2$.

\section{Hirota equations} Fermionic form of Hirota equations was invented in the papers of Kyoto school, see
for instance \cite{JM} and references therein.

Introduce
  \be
S_1:=\sum_{ n\neq 0}\,(-)^n\phi_n \otimes \phi_{-n}\, ,\quad
S_2:=\sum_{ n\neq 0}\,(-)^n{\hat\phi}_n \otimes {\hat\phi}_{-n}
 \ee
  \be
 S_0:=\,\phi_0 \otimes \phi_0
  \ee
  \,
  \be
S_3:=\sum_{n\in\mathbb{Z}}\,\psi_n \otimes \psi^\dag_n\, ,\quad
S_4:=\sum_{n\in\mathbb{Z}}\,\psi^\dag_n \otimes \psi_n\, ,\quad
 \ee

 Hirota equations in the fermionic form are:\\
For KP hierarchy \cite{JM}:
   \be
\left[ g\otimes g, S_3  \right]=0
   \ee
 For large (the same fermionic) DKP hierarchy \cite{KvdLbispec}:
   \be
\left[ g\otimes g, S_3 +S_4 \right]=0
   \ee
   For large (fermionic) BKP hierarchy \cite{KvdLbispec}:
   \be
\left[ g\otimes g, S_3 +S_4 + S_0\right]=0
   \ee
 For small DKP hierarchy \cite{DJKM-B1} (see also \cite{KvdLbispec}):
   \be
\left[ g\otimes g, S_1  \right]=0  \quad {\mbox or} \left[
g\otimes g, S_1  \right]=0
   \ee
 For small BKP hierarchy \cite{DJKM-B1} (see also \cite{KvdLbispec}):
   \be
\left[ g\otimes g, S_1 + S_0  \right]=0
   \ee
   For two-component BKP hierarchy in form \cite{KvdLbispec}:
     \be\left[ g\otimes g, S_1 + S_2 + S_0  \right]=0
      \ee

\subsection{A remarks on BKP hierarchies \cite{KvdLbispec} and
\cite{JM} and  related vacuum expectation values
\label{remarks-BKP-section}}

Let us note that different vacuum states were used in the
constructions of BKP hierarchy in versions \cite{KvdLbispec} and
\cite{JM}. If we denote the left and right vacuum states used in
\cite{JM} respectively by $'\langle 0|$ and $|0\rangle'$ then
 \be
\langle 0|=\frac{1}{\sqrt{2}}\; {'\langle 0|}+{'\langle 0|}\phi_0
 ,\qquad
|0\rangle=\frac{1}{\sqrt{2}}|0\rangle' + \phi_0 |0\rangle'
 \ee
Introduce also
 \be '\langle 1|=\sqrt2\; '\langle 0|\phi_0,\qquad
|1\rangle'=\sqrt2 \phi_0|0\rangle', \ee then $'\langle
0||0\rangle'=\, '\langle 1||1\rangle'=1$ and instead of
(\ref{phi-on-vacuum}) we have \be\label{phi-on-vacuum-DJKM}
\phi_n|0\rangle' =\phi_n|1\rangle' =0, \qquad '\langle
0|\phi_{-n}= \, '\langle 1|\phi_{-n}=0,\qquad  n<0
 \ee
 see \cite{JM} for details.

Correspondingly Fock spaces used \cite{KvdLbispec} and \cite{JM}
are different. From the representational point of view this
definition is somewhat more convenient, since each Fock module
remains irreducible for the algebra $B_\infty$ which is the
underlying algebra for KP equations of type B (BKP), see
\cite{KvdLbispec}.

The vacuum states $'\langle 0|$ and $|0\rangle'$ are more familiar
objects in physics. In particular any vacuum expectation value of
an odd number of fermions vanishes, while, for instance,  $\langle
0|\phi_0|0\rangle =\frac{1}{\sqrt{2}}$.

Let $F$ be a product of even number of fermions. Then it is easy
to see that
 \be
\langle 0|F|0\rangle = {'\langle 0|}F{|0\rangle '}
 \ee

\subsection{A remark on formulae containing $Q_\lambda$ functions \label{Remark-Q-section}}

From \cite{You} it is known that
 \be
{'\langle
0|}e^{H(s)}\phi_{\lambda_1}\phi_{\lambda_2}\cdots\phi_{\lambda_N}|0\rangle'=
\begin{cases}2^{-\frac{N}2}Q_{(\lambda_1,{\lambda_2},\ldots, {\lambda_N})}(\frac{s}2)
\quad \mbox{for }N\quad\mbox{even},\\
0 \quad \mbox{for }N\quad\mbox{odd},
\end{cases}
 \ee

 \be
\sqrt 2\,  {'\langle
0|}\phi_0e^{H(s)}\phi_{\lambda_1}\phi_{\lambda_2}\cdots
\phi_{\lambda_N}|0\rangle'=
\begin{cases}2^{-\frac{N}2}Q_{(\lambda_1,{\lambda_2},\ldots,
{\lambda_N})}(\frac{s}2)\quad \mbox{for }N\quad\mbox{odd},\\
0 \quad \mbox{for }N\quad\mbox{even},
\end{cases}\\
 \ee
where $Q_\lambda\left(\frac{s}{2}\right),\;
\lambda=(\lambda_1,\lambda_2,\dots)$ are the projective Schur
functions, see \cite{Mac}. Thus
 \[
 \langle 0|e^{H(s)}\phi_{\lambda_1}\phi_{\lambda_2}\cdots\phi_{\lambda_N}|0\rangle=\\[3mm]
 \left(\frac{1}{\sqrt{2}} {'\langle 0|}+{'\langle 0|}\phi_0
 \right)e^{H(s)}\phi_{\lambda_1}\phi_{\lambda_2}\cdots\phi_{\lambda_N}
\left(\frac{1}{\sqrt{2}}|0\rangle' + \phi_0 |0\rangle'\right)=
 \]
 \[
\frac12\, {'\langle
0|}e^{H(s)}\phi_{\lambda_1}\phi_{\lambda_2}\cdots\phi_{\lambda_N}|0\rangle'
+ {'\langle
0|}\phi_0e^{H(s)}\phi_{\lambda_1}\phi_{\lambda_2}\cdots\phi_{\lambda_N}\phi_0|0\rangle'
+
\]
 \[
 +\frac{1}{\sqrt 2}\, {'\langle 0|}\phi_0 e^{H(s)}
\phi_{\lambda_1}\phi_{\lambda_2}\cdots\phi_{\lambda_N}|0\rangle'+
\frac{1}{\sqrt 2}\, {'\langle
0|}e^{H(s)}\phi_{\lambda_1}\phi_{\lambda_2}\cdots\phi_{\lambda_N}\phi_0|0\rangle'\,
 \]
 \[
= \frac12\, {'\langle
0|}e^{H(s)}\phi_{\lambda_1}\phi_{\lambda_2}\cdots\phi_{\lambda_N}|0\rangle'
+\frac12\, {'\langle
1|}e^{H(s)}\phi_{\lambda_1}\phi_{\lambda_2}\cdots\phi_{\lambda_N}|1\rangle'
 \]
 \[
\\
+\frac{1}{\sqrt 2}\, {'\langle 0|}\phi_0 e^{H(s)}
\phi_{\lambda_1}\phi_{\lambda_2}\cdots\phi_{\lambda_N}|0\rangle'+
\frac{1}{\sqrt 2}\, {'\langle 1|}\phi_0
e^{H(s)}\phi_{\lambda_1}\phi_{\lambda_2}\cdots\phi_{\lambda_N}|1\rangle'\,
 \]
 \[
=\\[3mm]
 2^{-\frac{N}2}Q_{(\lambda_1,{\lambda_2},\ldots,
{\lambda_N})}\left(\frac{s}2\right)\, , \]
 since the role of
${'\langle 0|}$ and ${'\langle 1|}$ (resp. $|0\rangle'$ and
$|1\rangle'$) is interchangeable.

\subsection{Projective Schur functions ($Q$-functions) and neutral fermions \label{projective-section}}
 There
are two bases of \emph{neutral free fermions}
\begin{equation}\label{phi-hat-phi}
    \phi_i=\frac1{\sqrt2}(\psi_i+(-1)^i\psi^*_{-i}),\quad
    \hat\phi_i=\frac i{\sqrt2}(\psi_i-(-1)^i\psi^*_{-i}),
\end{equation}
where $i\in\mathbb{Z}$, each of which generates this subalgebra.

%

Using the results for charged free fermions, the anticommutation
relations are
\begin{equation}
    [\phi_i,\phi_j]_+=[\hat\phi_i,\hat\phi_j]_+=(-1)^i\delta_{i,-j},
    \quad[\phi_i,\hat\phi_j]_+=0,
\end{equation}
and, in particular, $\phi_0^2=\hat\phi_0^2=\frac12$. Similarly,
the vacuum expectation values of quadratic elements are given by
\begin{equation}
    '\l 0| \phi_i\phi_j |0\r'='\l 0| \hat\phi_i\hat\phi_j |0\r' =
    \begin{cases}
      (-1)^i\delta_{i,-j}&i<0\\
      \frac12\delta_{j,0}&i=0\\
      0&i>0
    \end{cases},
\end{equation}
and Wick's Theorem is used for arbitrary degree products.


The neutral free fermion generator is defined by
$\phi(p)=\sum_{n\in\mathbb{Z}}p^n\phi_n$. We have (for
$|p|\ne|p'|$)
\begin{equation}
   '\l 0| \phi(p)\phi(p') |0\r'=
  \frac12\frac{p-p'}{p+p'},
\end{equation}
and $ '\l 0|  \phi(p')\phi(p) |0\r' =- '\l 0| \phi(p)\phi(p')
|0\r' $. By Wick's Theorem we get
\begin{equation}
  '\l 0|  \phi(p_1)\phi(p_2)\cdots\phi(p_N) |0\r' =
  \begin{cases}
    \Pf\left[ '\l 0|  \phi(p_i)\phi(p_j) |0\r' \right]\\
    0
  \end{cases}=\begin{cases}
    \displaystyle 2^{-N/2}\prod_{i<j}\frac{p_i-p_j}{p_i+p_j}&N\text{ even}\\
    0&\text{otherwise}
  \end{cases}.\label{eq:phi ef}
\end{equation}

The connection between the charged and neutral free fermions can
be expressed in terms of the generators as
\begin{equation}
    -q\psi(p)\psi^*(-q)+p\psi(q)\psi^*(-p)=\phi(p)\phi(q)+\hat\phi(p)\hat\phi(q).
    \label{eq:KP->NBKP}
\end{equation}

In the sBKP reduction, even times are set equal to zero and we
define $\bt' =(t_1',0,t_3',0,t_5',\dots)$, and the hamiltonian
\begin{equation}
    H^B(\bt')=\sum_{n\ge 1,\;\text{odd}}H^B_nt_n',
\end{equation}
where
\begin{equation}
    H^B_n=\frac12\sum_{i\in\mathbb{Z}}(-1)^{i+1}\phi_i\phi_{-i-n}.
\end{equation}

For the fermion generating function one has 
\begin{equation}\label{phi(t)}
    \phi(p)(\bt')=e^{H^B(\bt' )}\phi(p)e^{-H^B(\bt' )}=
    e^{\hat H^B(\bt' )}\phi(p)e^{-\hat H^B(\bt' )}
    =e^{\xi(p,\bt')}\phi(p).
\end{equation}
Note also that
\begin{equation}
    H(\bt')=H^B(\bt')+\hat H^B(\bt'),\quad [H^B(\bt'),\hat
    H^B(\bt')]=0.
    \label{eq:KP->NBKP H}
\end{equation}

Similar to the KP case, sBKP $\tau$-functions are defined by
\begin{equation}
    \tau_B(\bt')=\langle h(\bt')\rangle,
\end{equation}
where $h$ is the Clifford algebra of the neutral free fermions
$\phi_i$. The $n$-soliton $\tau$-function is obtained by the
choice $g=\exp\bigl(\sum_{i=1}^n a_i\phi(p_{i})\phi(q_{i})\bigr)$.

The Schur $q$ polynomials are defined by
\begin{equation}
    \exp(2\xi(p,\bt'))=\sum_{k\ge0}q_k(\bt')p^k.
\end{equation}
Thus
\begin{equation}
    \phi_i(\bt')=\sum_{k\ge0}q_k(\tfrac12\bt')\phi_{i-k}.
\end{equation}
We have
\begin{equation}\label{eq: phi q_i,j}
    '\l 0| \phi_i(\bt')\phi_j(\bt')|0\r' =
    \frac12q_i(\tfrac12\bt')q_j(\tfrac12\bt')+
    \sum_{k=1}^j(-1)^{k}q_{k+i}(\tfrac12\bt')q_{j-k}(\tfrac12\bt').
\end{equation}

Since
\begin{equation}
    1=\exp(2\xi(p,\bt'))\exp(-2\xi(p,\bt'))=\sum_{i,j}q_i(\bt')q_{j-i}(-\bt')=
    \sum_{i,j}(-1)^{i-j}q_i(\bt')q_{j-i}(\bt')p^j,
\end{equation}
for all $n>0$ we have
\begin{equation}\label{eq:B othog}
    \sum_{i=0}^n(-1)^{i}q_i(\bt')q_{n-i}(\bt')=0.
\end{equation}
This is trivial if $n$ is odd and if $n=2m$ is even then it gives
\begin{equation}
    q_m(\bt')^2+2\sum_{k=1}^m(-1)^{k}q_{m+k}(\bt')q_{m-k}(\bt')=0.
\end{equation}

We can also define
\begin{equation}\label{eq: q_a,b}
    q_{a,b}(\bt')=q_a(\bt')q_b(\bt')+2\sum_{k=1}^b(-1)^k
    q_{a+k}(\bt')q_{b-k}(\bt').
\end{equation}
If follows from the orthogonality condition \eqref{eq:B othog}
that
\begin{equation}
    q_{a,b}(\bt')=-q_{b,a}(\bt'),
\end{equation}
and in particular, $q_{a,a}(\bt')=0$. Comparing \eqref{eq: phi
q_i,j} and \eqref{eq: q_a,b}, it is clear that
\begin{equation}
    q_{a,b}(\tfrac12\bt')=2\langle\phi_a(\bt')\phi_b(\bt')\rangle.
\end{equation}

Now consider $\lambda=(\lambda_1,\lambda_2,\dots,\lambda_{2n})$
where $\lambda_1>\lambda_2>\cdots\lambda_{2n-1}>\lambda_{2n}\ge0$.
Note that this is a partition with an extra trivial part 0
included if necessary to ensure that the number of parts is even.
The set of such strict, or distinct part, partitions is denoted
$\DP$. For $\lambda\in\DP$ we define
\begin{equation}\label{Q-Pfaff}
    Q_\lambda(\tfrac12\bt')
    =\Pf(q_{\lambda_i,\lambda_j}(\tfrac12\bt')).
\end{equation}
This is the Schur $Q$-function. By Wick's theorem, we come to

\bl \label{You}\em
 \[
    Q_\lambda(\tfrac12\bt')
    =\Pf(2\langle\phi_{\lambda_i}(\bt')\phi_{\lambda_j}(\bt')\rangle)
    =2^n\langle\phi_{\lambda_1}(\bt')
    \phi_{\lambda_2}(\bt')\cdots\phi_{\lambda_{2n}}(\bt')\rangle.
\]

\el

This  wonderful result was obtained in \cite{You} (see also
\cite{Nimmo} where it was independently found that $Q_\lambda$ is
an example of sBKP tau function). In particular, from this Lemma
and \eqref{phi-hat-phi} ,\eqref{Schur=VEV} it was obtained
\cite{You}
\begin{equation}\label{sQ}
2^{-\frac {\ell(\lambda)}{2}}Q_{{\lambda}}(\tfrac12\bt')= \sqrt
{s_{{\tilde \lambda}}({\bf t}' )},
\end{equation}
where $s_{ {\tilde \lambda}}({\bf t} )$ is the Schur function, and
the partition ${ {\tilde \lambda}}\in P$ is the {\em double} (see
1,I,Ex9(a) of \cite{Mac}) of the strict partition $ {\lambda }$.
$\ell(\lambda)$ is the length of $\lambda$ (the number of
non-vanishing parts of $\lambda$).

\subsection{Schur functions and projective Schur
functions evaluated at special values of its
argument.\label{evaluated-section}} We introduce the following
notations:
\begin{equation}\label{choicetinfty'}
{\bf t}_\infty=(1,0,0,0,\dots) \ ,
\end{equation}
\begin{equation}\label{choicet(a)'}
{\bf t}(a,1)=\Bigl(\frac{a}{1},\frac{a}{2},\frac{a}{3},\dots\Bigr)
\ ,
\end{equation}
\begin{equation}\label{choicetinftyq'}
{\bf t}(\infty,q)=(t_1(\infty,q),t_2(\infty,q),\dots),\quad
t_m(\infty,q)=\frac{1}{m(1-q^m)}\ ,\quad m=1,2,\dots\ ,
\end{equation}
\begin{equation}\label{choicet(a)q'}
{\bf t}(a,q)=(t_1(a,q),t_2(a,q),\dots)\ ,\quad
t_m(a,q)=\frac{1-(q^a)^{m}}{m(1-q^ m)}\ ,\quad m=1,2,\dots
\end{equation}

 For various purposes these choices
of times were used in
\cite{OSch1},\cite{OSch2},\cite{hypsol},\cite{Q},\cite{O2004},
\cite{TauFuncMI},\cite{OS}.
Note that ${\bf t}(a,q)$ tends to ${\bf t}(\infty,q)$ (resp.\
${\bf t}(a,1)$) as $a\to\infty$ (resp.\ $q\to1$). As for ${\bf
t}_\infty$, if $f$ satisfies
$f(ct_1,c^2t_2,c^3t_3,\dots)=c^df(t_1,t_2,t_3,\dots)$ for some
$d\in {\mathbb Z}$, we have $\hbar^df({\bf t}(\infty,q))\to f({\bf
t}_\infty) $ as $\hbar:=\ln q \to0$.

\bl For a partition $\lambda=(\lambda_1,\lambda_2,\dots)$, let
$ 
h_i := n+\lambda_i-i$ $(1\le i \le n)$, where
$n\ge\ell(\lambda))$. Then
\begin{equation}\label{schurhook}
s_\lambda({\bf t}_\infty)=\frac{1}{H_\lambda}=\frac{
\Delta(h)}{\prod^n_{i=1}h_i!} \ ,
\end{equation}
\begin{equation}\label{schurhookt(a)}
s_\lambda({\bf t}(a,1))=\frac{(a)_\lambda}{H_\lambda}=\frac{
\Delta(h)}{\prod^n_{i= 1}h_i!} \prod_{i=1}^n
\frac{\Gamma(a-n+h_i+1)}{\Gamma(a-i+1)}\ ,
\end{equation}
\begin{equation}\label{schurhookq}
s_\lambda({\bf t}(\infty,q))
=\frac{q^{n(\lambda)}}{H_\lambda(q)}=\frac{\Delta(q^h)}
{\prod_{i=1}^n(q;q)_{h_i}}\ ,
\end{equation}
\begin{equation}\label{schurhookqa}
s_\lambda({\bf t}(a,q))=\frac{
q^{n(\lambda)}(q^a;q)_\lambda}{H_\lambda(q)}=\,c_n(a,q)\,\frac{\Delta(q^h)}
{\prod_{i=1}^n(q;q)_{h_i}}\,\prod_{i=1}^n \,(q^{a};q)_{h_i-n+1} \
,
\end{equation}
$$ 
\Delta(h):=\prod^n_{i<j}(h_i- h_j)\ ,\quad
\Delta(q^h):=\prod^n_{i<j}(q^{h_i}- q^{h_j}) \, , \quad
c_n(a,q)=\,\prod_{i=1}^n\,(1-q^{a-i})^{n-i}
$$
where for $H_\lambda$, $H_\lambda(q)$, $n(\lambda)$, $(a)_\lambda$
and $(q^a;q)_\lambda$ see respectively (\ref{Hlambda}),
(\ref{Hlambda(q)1}), (\ref{n(lambda)}), (\ref{Poch}) and
(\ref{Pochq}). Note that those quantities
(\ref{schurhook})--(\ref{schurhookqa}) are independent of the
choice of $n \ge \ell(\lambda)$.

\el

 \bl. Let $(\alpha|\beta)=(\alpha_1,\dots
,\alpha_k |\beta_1,\dots ,\beta_k)$ be the Frobenius notation for
a partition. Then,
\begin{equation}\label{stinfty}
s_{(\alpha|\beta)}({\bf
t}_\infty)=\frac{\prod^k_{i<j}(\alpha_i-\alpha_j)(\beta_i-\beta_j)}
{\prod_{i,j=1}^k(\alpha_i+\beta_j+1)}\frac {1}{\prod_{i=1}^k
\alpha_i!\prod_{i=1}^k \beta_i!} \ ,
\end{equation}
\begin{equation}\label{st(a)}
s_{(\alpha|\beta)}({\bf
t}(a,1))=\frac{\prod^k_{i<j}(\alpha_i-\alpha_j)(\beta_i-\beta_j)}
{\prod_{i,j=1}^k(\alpha_i+\beta_j+1)} \prod_{i=1}^k\frac
{(a)_{\alpha_i+1}}{\alpha_i!} \prod_{i=1}^k
\frac{(-)^{\beta_i}(-a)_{\beta_i}}{ \beta_i!} \ ,
\end{equation}
\begin{equation}\label{stinftyq}
s_{(\alpha|\beta)}({\bf
t}(\infty,q))=\frac{\prod^k_{i<j}(q^{\alpha_i+1}-q^{\alpha_j+1})(q^{-\beta_j}-q^{-\beta_i})}
{\prod_{i,j=1}^k(q^{-\beta_i}-q^{\alpha_j+1})}\frac
{1}{\prod_{i=1}^k(q;q)_{\alpha_i}\prod_{i=1}^k (q;q)_{\beta_i}} \
,
\end{equation}
\begin{equation}
s_{(\alpha|\beta)}({\bf t}(a,q))=
\end{equation}
\begin{equation}\label{st(a,q)}
\frac{\prod^k_{i<j}(q^{\alpha_i+1}-q^{\alpha_j+1})(q^{-\beta_j}-q^{-\beta_i})}
{\prod_{i,j=1}^k(q^{-\beta_i}-q^{\alpha_j+1})} \prod_{i=1}^k\frac
{(q^a;q)_{\alpha_i +1}}{(q;q)_{\alpha_i}}\prod_{i=1}^k\frac
{(-)^{\beta_i}q^{(a-1)\beta_i }(q^{1-a};q)_{\beta_i}}
{(q;q)_{\beta_i}}
\end{equation}
\el

From \eqref{sQ} and from \eqref{stinfty} it may be derived
\begin{Lemma} \label{Q-t-infty}\em Let $\lambda=(\lambda_1,\dots,\lambda_k)$ be a
strict partition. Then
\begin{equation}\label{stinftyB}
Q_{\lambda}(\tfrac12\bt_\infty)=
2^{\frac{\ell(\lambda)}{2}}\prod_{i=1}^k \frac
{1}{\lambda_i!}\prod^k_{i<j}\frac{\lambda_i-\lambda_j}
{\lambda_i+\lambda_j} \ ,
\end{equation}

\end{Lemma}

\subsection{ "Neutral" two-component BKP hierarchy and
2-BKP hierarchy \label{"small"-section}}

\paragraph{Two-component sBKP.} Consider the following tau function of the "small" two-component
BKP hierarchy
 \be\label{2c-2-sBKP}
\tau^{(B)}(\bt',\bt'',C) =\langle 0,0|\Gamma_B(\bt'){\hat
\Gamma}_B(\bt'')\exp\, \sum_{n,m\ge 0} \, C_{nm}\phi_n{\hat
\phi_m}\,|0,0\rangle
 \ee
 \be\label{hyp-tau-sQC}
= \sum_{\lambda \in \Pa}\,2^{-\frac12
\ell(\alpha)-\frac12\ell(\beta)} \det\, C_\lambda \,
Q_\alpha\left(\frac{\bt'}{2}\right)Q_\beta\left(\frac{\bt''}{2}\right)
 \ee
 where $\lambda=(\alpha|\beta)$ and $C_\lambda:=\det\,\left( C\right)_{\alpha_i,\beta_j}$ and
 $Q_\alpha(\frac\bt 2)$
 is the projective Schur function related to a (strict) partition $\alpha$, see
 \cite{Mac}. Neutral fermions $\phi_n,{\hat
\phi_n}$ were considered in \eqref{phi-hat-phi}.
 If we take $C_{nm}=C_{nm}(\bt)=e^{U_m-U_n}s_{(n|m)}(\bt)$ we obtain
  \be\label{hyp-tau-sQQ}
\tau^{(B)}(\bt',\bt'',C(\bt))=\,\sum_{\lambda\in \Pa}\,
2^{-\ell(\lambda)} e^{U_{\{\beta \}}-U_{\{\alpha
\}}}\,Q_\alpha\left(\frac{\bt'}{2}\right)Q_\beta\left(\frac{\bt''}{2}\right)\,s_\lambda(\bt)
 \ee
Thus this tau function is a tau function of  three hierarchies at
the same time which are the lDKP and two "small" BKP ones.

 Let us present the following 'symmetric' fermionic representation of this tau
 function which looks little in common with usual notation for tau
 functions
   \be
\langle
0,0,0|\Gamma(\bt)\Gamma_B(\bt')\Gamma_B(\bt'')e^{\sum_{n,m\ge 0}
\, e^{U_m-U_n}\psi_n\psi_{-m-1}^\dag\varphi_n{\hat \varphi_m}
}\,|0,0,0\rangle
   \ee

   \paragraph*{2-sBKP tau function.} This is

   \be\label{2-sBKP}
\tau(\bt',{\bar\bt}',A) = \l 0|
\Gamma_B(\bt')e^{\sum_{n,m}\,A_{nm}\phi_n\phi_m}{\bar
\Gamma_B}({\bar \bt}') |0\r
   \ee
   where $A$ is an anti-symmetric matrix. Thanks to
 \be
\l 0| \Gamma_B(\bt')=\sum_{\alpha\in \DP}\,\l
\alpha|\,2^{-\frac12 \ell(\alpha)}\,Q_\alpha,\quad {\bar
\Gamma_B}({\bar \bt}') |0\r=\sum_{\beta\in \DP}\,|
\alpha\r\,2^{-\frac12 \ell(\beta)}\,Q_\beta,\quad
\l\alpha|\beta\r=\delta_{\alpha,\beta}
 \ee
 by Wick theorem it may be
   written in form of double series over strict partitions as
   \be\label{2-sBKP-QQ}
=\sum_{\alpha,\beta\in \DP}\,2^{-\frac12 \ell(\alpha)-\frac12
\ell(\beta)}\,A_{\alpha,\beta}\,Q_\alpha(\bt')Q_\beta({\bar\bt}')
   \ee
   where
   \be
A_{\alpha,\beta}=\det\left[ A_{\alpha_n,\beta_m}\right],\quad
A_{nm}=\l 0|\phi_{-n} e^{\sum_{i,j}\,A_{ij}\phi_i\phi_j} \phi_m
|0\r
   \ee

\section{Appendix. Simplest lDKP solitons
\label{simplest-section}} We want to present certain types of
solitonic solutions typical for lDKP  hierarchy.

I. KP hierarchy is a reduction of lDKP one, therefore KP solitons
are also lDKP ones. First let me remind a typical KP two-soliton
tau function (which is also lDKP two-soliton tau function which
will be denoted by $\tau^I_{2sol}(\bt)$)
 \be\label{tau-I-vev}
  \tau^I_{2sol}(\bt)=\langle
0|\Gamma(\bt)e^{a_1\psi(p_1)\psi^\dag(q_1)+a_2\psi(p_2)\psi^\dag(q_2)}
|0\rangle=
 \ee
 \be
1+ e^{\delta_1}e^{\sum_{m=1}^\infty t_m(p_1^m-q_1^m)} +
e^{\delta_2}e^{\sum_{m=1}^\infty t_m(p_2^m-q_2^m)}+
e^{\delta_{12}}\prod_{i=1,2}e^{\delta_i}e^{\sum_{m=1}^\infty
t_m(p_i^m-q_i^m)}
 \ee
 where
 \be\label{delta}
\delta_i=\log \frac{a_i}{p_i-q_i},\qquad
\delta_{ij}=\frac{(p_i-p_j)(q_i-q_j)}{(p_i-q_j)(p_j-q_i)}
 \ee
 This two-soliton solution exhibit resonance behavior \cite{TeorSol} when $p_1\to
 p_2$, the same occurs when $q_1\to q_2$ (such solitons has the Manakov's
 Y-form).

For lDKP one can present a larger amount of various one- and
two-soliton tau functions. Examples are written down below.

II. One of them we may obtain by replacing the right vacuum vector
$|0\rangle$ by $|\Omega\rangle$ in the expectation in the right
hand side of \eqref{tau-I-vev}:
 \be \tau^{II}_{2sol}(\bt)=\langle
0|\Gamma(\bt)e^{a_1\psi(p_1)\psi^\dag(q_1)+a_2\psi(p_2)\psi^\dag(q_2)}
|\Omega\rangle=
 \ee
 \be
\tau_0(\bt)\left(1+ e^{\delta_1+\Delta_1}e^{\sum_{m=1}^\infty
t_m(p_1^m-q_1^m)} + e^{\delta_2+\Delta_2}e^{\sum_{m=1}^\infty
t_m(p_2^m-q_2^m)}+
e^{\delta_{12}+\Delta_{12}}\prod_{i=1,2}e^{\delta_i+\Delta_i}e^{\sum_{m=1}^\infty
t_m(p_i^m-q_i^m)}\right)
 \ee
 where $\delta_{1,2}$ and $\delta_{12}$ are given by \eqref{delta} and where
 \be
\Delta_i=\log \frac{1}{(1-p_i)(1+q_i)},\qquad
\Delta_{ij}=\log\frac{(1-p_iq_j)(1-p_jq_i)}{(1-p_ip_j)(1-q_iq_j)}
 \ee
This solution describes interaction of two solitons, $i=1,2$, each
is
 \be
\tau^{II}_{1sol}(\bt)=\tau_0(\bt)\left(1+
e^{\delta_i+\Delta_i}e^{\sum_{m=1}^\infty
t_m(p_i^m-q_i^m)}\right),\quad i=1,2
 \ee
(notice the $\tau_0(\bt)$ factor in front of the right-hand side).

 Now two-soliton solution exhibit resonance behavior also in case when $p_1\to
 q_2^{-1}$, the same occurs when $p_2\to q_1^{-1}$.

One can say that we add 'KP solitons' to lBKP background solution
\eqref{the simplest}.

III. Adding of 'KP solitons' to a lBKP background solution
 $\tau_0(\bt)$ of form \eqref{lBKP-tauUA} yields

 \be
 \tau^{III}_{1sol}(\bt)=\langle
N+l|\Gamma(\bt)e^{a\psi(p)\psi^\dag(q)}
g^{--}|l\rangle=\tau_0(\bt)\left(1+ap^lq^{-l}\frac{}{}e^{} \right)
 \ee

IV. lDKP 1-soliton solution may be also as follows
  \be
 \tau^{IV}_{1sol}(\bt)=\langle
0|\Gamma(\bt)e^{a_1\psi(p_1)\psi(p_2)+a_2\psi^\dag(q_1)\psi^\dag(q_2)}
|0\rangle= 1 +
e^{\delta_{12}}\prod_{i=1,2}e^{\delta_i}e^{\sum_{m=1}^\infty
t_m(p_i^m-q_i^m)}
 \ee
This soliton is characterizes by four spectral parameters
$p_1,p_2,q_1,q_2$ instead of a pair unlike, say, typical KP
soliton.

 The corresponding '2-soliton' tau function is
\be\label{tau-I-vev'}
  \tau^{III}_{2sol}(\bt)=\langle
0|\Gamma(\bt)
e^{a_{12}\psi(p_1)\psi(p_2)+a_{12}^*\psi^\dag(q_1)\psi^\dag(q_2)+a_{34}\psi(p_3)\psi(p_4)+
a_{34}^*\psi^\dag(q_3)\psi^\dag(q_4)} |0\rangle=
 \ee
 \be
1+ e^{\varepsilon_{12,12}}\prod_{i=1,2}e^{\sum_{m=1}^\infty \,
 p_i^m t_m}\prod_{i=1,2}e^{-\sum_{m=1}^\infty \,
 q_i^m t_m}
+e^{\varepsilon_{34,34}}\prod_{i=3,4}e^{\sum_{m=1}^\infty
 p_i^m t_m}\prod_{i=3,4}e^{-\sum_{m=1}^\infty q_i^m
t_m}
 \ee
 \be
+e^{\varepsilon_{12,34}}\prod_{i=1,2}e^{\sum_{m=1}^\infty
 p_i^m t_m}\prod_{i=3,4}e^{-\sum_{m=1}^\infty q_i^m
t_m} + e^{\varepsilon_{34,12}}\prod_{i=3,4}e^{\sum_{m=1}^\infty
 p_i^m t_m}\prod_{i=1,2}e^{-\sum_{m=1}^\infty q_i^m
t_m}
 +
 \ee
 \be
+\,e^{\varepsilon_{1234}}\prod_{i=1,2,3,4} e^{\sum_{m=1}^\infty
t_m(p_i^m-q_i^m)}
 \ee
 where
 \be
\varepsilon_{ij,kl}=-\log
{(p_i-q_k)(p_i-q_l)(p_j-q_k)(p_j-q_l)}+a_{ij}+\log(p_i-p_j)+a_{kl}^*+\log(q_k-q_l)
 \ee
 and where
 \be
\varepsilon_{1234}=\frac{\prod_{i<j}^4(p_i-p_j)(q_i-q_j)}{\prod_{i,j=1}^4(p_i-q_j)}
+ \log(a_{12}a_{12}^*a_{34}a_{34}^*)
 \ee

--------------

\be \tau^{IV}(\bt)\, =\,\langle
0|\Gamma(\bt)e^{\sum_{n,m}C_{nm}\psi(p_n)\psi^\dag(q_m)}
|\Omega\rangle=
 \ee
 \be
\tau_0(\bt)\sum_{k=0}^\infty\sum_{m_1,\dots,m_k}\sum_{n_1,\dots,n_k}\,
\prod_{i<j}\,\frac {\prod_{i<j\le
k}\,(p_{m_i}-p_{m_j})(q_{n_i}-q_{n_j})}{\prod_{i,j=1}^k\,(p_{m_i}-q_{n_j})}
 \ee

V. In case $N>0$:

One-soliton
 \be
\tau^{V}_{1sol}(\bt)=\langle
N|\Gamma(\bt)e^{a\psi^\dag(q_1)\psi^\dag(q_2)} g^{--}|0\rangle
 \ee
 \be
=\sum_{\ell(\lambda)\le N}\, s_\lambda(\bt) +
a(q_1q_2)^N\sum_{\ell(\lambda)\le N+2}\,
s_\lambda(\bt+[q_1^{-1}]+[q_2^{-1}])
 \ee
Now, let us be interested in the case where
 \be
t_m =t_m({\bf x})=\frac 1m \sum_{i=1}^N \, x_i^m
 \ee
Then
 \be
\tau^{V}_{1sol}(\bt({\bf x}))=\tau_0(\bt({\bf
x}))\left(1+\frac{1}{(1-q_1)(1-q_2)(1-q_1q_2)}
\prod_{i=1}^N\frac{1}{(1-q_1x_i)(1-q_2x_i)} \right)
 \ee

\end{document}